\documentclass{article}
\usepackage[english]{babel}
\usepackage[letterpaper,top=2cm,bottom=2cm,left=3cm,right=3cm,marginparwidth=1.75cm]{geometry}
\usepackage{amsthm, bm}
\usepackage{amsmath, amssymb}
\usepackage{graphicx}
\usepackage{comment}
\usepackage[colorlinks=true, allcolors=blue]{hyperref}
\usepackage[]{todonotes}
\usepackage{tikz, pgfplots}
\usepackage{tikz-cd}
\usepackage{float}
\providecommand{\keywords}[1]
{
  \small	
  \textbf{Keywords:} #1
}
\graphicspath{{Figure/}}

\usepackage{xcolor}

\newtheorem{theorem}{Theorem}[section] 

\newtheorem{remark}[theorem]{Remark}
\newtheorem{corollary}[theorem]{Corollary}
\newtheorem{proposition}[theorem]{Proposition}

\newcommand{\R}{\mathbb{R}}

\renewcommand{\P}{\mathbb P}

\newcommand{\cF}{\mathcal F}
\newcommand{\Prob}{\mathbb P}

\title{Noise estimation of SDE from a single data trajectory}

\author{
Munawar Ali \thanks{\noindent Department of
Mathematics, Florida State University, Tallahassee, 32306; email: ma22bm@fsu.edu.}~~  
Purba Das\thanks{Department of
Mathematics, King’s College London, London, UK; email: purba.das@kcl.ac.uk.} ~~Qi Feng\thanks{\noindent Department of
Mathematics, Florida State University, Tallahassee, 32306; email: qfeng2@fsu.edu. This author is partially supported by the National Science Foundation under grant \#DMS-2420029.}
 ~~  Liyao Gao\thanks{\noindent Department of Computer Science, University of Washington, Seattle, WA 98195.}~~  Guang Lin\thanks{ \noindent Department of
Mathematics, Purdue University, West Lafayette, IN;
email: guanglin@purdue.edu. This author is partially 
supported by the National Science Foundation (NSF) (DMS-2533878, DMS-2053746, DMS-2134209, ECCS-2328241, CBET-2347401, and OAC-2311848), and DOE Office of Science Advanced Scientific Computing Research program DE-SC0023161, and DOE–Fusion Energy Science, under grant number: DE-SC0024583.} 
}

\date{}
\allowdisplaybreaks
\pgfplotsset{compat=1.18}
\begin{document}
\maketitle

\begin{abstract}
In this paper, we propose a pathwise data-driven framework for model discovery of stochastic differential equations (SDEs) from a single trajectory, without requiring the ergodicity or stationary assumption on the underlying continuous process. By combining (stochastic) Taylor expansions with Girsanov transformations, and using the drift function’s initial value as input, we construct drift estimators while simultaneously recovering the model noise. This drift is unique up to the fixed underlying Brownian sample path, in line with pathwise unique Rough-Smooth decomposition. This allows us to recover the underlying $\mathbb P$ Brownian motion increments for the observer path $\omega$. Building on these estimators, we introduce the first \textit{stochastic Sparse Identification of Stochastic Differential Equation (SSISDE)} algorithm, capable of identifying the governing SDE dynamics from a single observed trajectory without requiring ergodicity or stationarity. To validate the proposed approach, we conduct numerical experiments with both linear and quadratic drift–diffusion functions. Among these, the Black–Scholes SDE is included as a representative case of a system that does not satisfy ergodicity or stationarity.
\end{abstract}


\section{Introduction}
Stochastic differential equations (SDEs) are fundamental tools for modeling systems with inherent randomness, but estimating the underlying continuous dynamics from observed discrete data remains both challenging and important. The initial development in this area was closely tied to applications in economics and finance, including \cite{Anderson1959, Merton1980, melina1994, Sahalia1996, hansen1993, hansen1998spectral, stanton1997, Duffie2004}, among many others.
A central challenge in working with stochastic differential equations (SDEs) is that, in practice, one typically only observes the process at discrete time points, while the key quantities driving the dynamics—such as the drift and diffusion coefficients or even the underlying noise—remain unknown. Recovering both the system parameters and the latent noise process from such partial information is therefore of significant theoretical and practical interest, particularly in stochastic modeling and statistical inference \cite{kutoyants2013,kessler1997}.
\par In this paper, we study the problem of recovering the drift and diffusion functions from a high-frequency single realization of a continuous-time stochastic process $\{X_t\}_{t\geq 0}$ governed by the It\^o SDE
\begin{align}\label{eq:diff}
    dX_t = \mu(X_t)dt+\sigma(X_t) dB^\mathbb P_t,
\end{align}
where $\mu$ and $\sigma$ denote the drift and diffusion functions, respectively, and $B^{\mathbb P}$ is a canonical Brownian motion under the physical measure $\mathbb P$, i.e. the filtration is generated by the $X_t$.
Classical approaches to estimating $\mu$ and $\sigma$ include kernel-based estimators \cite{lamouroux2009,florens1989,stanton1997}, maximum likelihood methods \cite{bibbySrensen1995,dacunha1986, ait-sahalia2002}, and local polynomial regression \cite{fan2003}. These methods, however, typically rely on ergodicity or stationarity of the underlying process or the availability of repeated observations. More recent developments employ tools from empirical process theory and martingale estimation to handle high-frequency data \cite{dalalyan2007,comte2007}, but these approaches still require strong structural assumptions such as stationarity, ergodicity, or multiple sample paths. These assumptions are often unrealistic in a financial context; for example, if one assumes the classical Black-Scholes model for a price path, then the price path is not ergodic. In the paper we address this problem via a data driven pathwise approach, based on a single trajectory of  \eqref{eq:diff}. A similar pathwise approach to extract noise from single trajectory has been explored in \cite[Proposition~4.1]{ananova2017}.

On the other hand, in the absence of noise in \eqref{eq:diff}, the deterministic dynamics can be analyzed in much greater detail. Particularly, the discovery of a deterministic dynamical system, obtained by setting $\sigma \equiv 0$ in equation \eqref{eq:diff},
\begin{align}
\dot{\mathbf{x}}(t) = \mu(\mathbf{x}(t)),
\label{eq:dynamical_system_x}
\end{align}
can be formulated as a symbolic regression problem of the following form:
\begin{equation}
    \label{eq:sindy_definition}
  \dot{\mathbf{X}} = \Theta(\mathbf{X})\Xi,
\end{equation}
where $\Theta(\mathbf{X})$ denotes the function library matrix composed of $p$ candidate terms, typically including polynomials, trigonometric functions, exponentials, and sinusoidal functions. The unknown sparse matrix $\Xi = (\xi_1\ \xi_2\ \cdots\ \xi_n)\in \R^{p\times n}$ encodes the active terms and thus determines the governing dynamics. The input $\mathbf{X}$ denotes the vector of data snapshots over the discrete time grid. The state snapshots $\mathbf{X}(t)$ are observed in discrete time instances, and their temporal derivatives $\dot{\mathbf{X}}(t)$ are obtained from data. The identification/estimation of the vector $\Xi$  is called regression. Over the past decades, significant efforts have been devoted to data-driven model discovery. Early approaches include symbolic regression \cite{koza1994genetic} for uncovering the structure of ODEs \cite{schmidt2009distilling, bongard2007automated}. Subsequent advances introduced sparse regression techniques for identifying ODEs \cite{mangan2016inferring, brunton2016discovering, mangan2017model} and PDEs \cite{rudy2017data, zhang2018robust, schaeffer2017learning} with constant coefficients, along with group sparsity and sequentially grouped threshold ridge regression to capture spatial- or temporal-dependent coefficients \cite{yuan2006model, friedman2010note, casella2010penalized}. More recently, symbolic regression models \cite{ brunton2016discovering, udrescu2020ai, chen2021physics} have focused on extracting analytic expressions from large datasets, offering greater interpretability.
Recent developments in data-driven frameworks further include the use of rough path theory and signature methods \cite{semnani2025}, as well as sparse identification of nonlinear stochastic dynamics \cite{boninsegna2018, wanner2024}. Sparse learning has also proven highly effective in a broad range of applications in engineering, including fluid dynamics~\cite{bai2017data}, plasma dynamics~\cite{kaptanoglu2023sparse}, nonlinear optics~\cite{ermolaev2022data}, mesoscale ocean closures~\cite{zanna2020data}, and computational chemistry~\cite{harirchi2020sparse}.

However, the stochastic extension of Sparse Identification of Nonlinear Dynamics (SINDy) is highly nontrivial, due to the fact that SINDy fundamentally requires computing the temporal derivatives, which are known to be sensitive to noise.  Additionally, it is also mathematically unclear on how to identify the drift $\mu$ and the underlying 'rough' noise term $B^\mathbb P$ of \eqref{eq:diff}, from a single path observation on a finite time interval~\cite{stevens2024learning,boninsegna2018,wanner2024}.  In contrast to the current literature concerning these problems, our method is data-driven, does not assume strong regularity conditions, ergodic or stationary properties of the underlying process, and also uses a single path of the SDE. Our technique also extracts the noise from the finite sample data.  Our proposed new methodology combines stochastic analysis, approximation techniques, and linear regression, with the only major assumption being the knowledge of the drift coefficient at the initial time point. Building on the It\^o–Stratonovich correspondence \cite{oksendal2003}, we approximate the drift and diffusion terms using Taylor expansions and reformulate the SDE in terms of iterated integrals. These iterated integrals, also known as signatures in rough path theory \cite{lyons1998, Friz_Victoir_2010}, have recently been developed into powerful signature models in finance \cite{arribas2018derivatives,cuchiero2023signature,bayer2023optimal,cuchiero2025joint}. Our approach leverages Girsanov’s theorem \cite{girsanov1960, karatzas1991-book} to transform the dynamics into a representation involving iterated integral type terms, which admit explicit computation \cite{lyons1998, kidger2020}, and the convergence rate of such terms are well explored. Through further simplifications, inspired by the structure of the Black–Scholes model \cite{black1973}, we obtain a representation suitable for regression-based identification of drift, diffusion, and the underlying Brownian motion paths. Using the estimated increment of $\mathbb P$ Brownian motion and the data snapshot of $X$, we introduce a new stochastic symbolic regression problem, 
\begin{align}
    \arg\min_{\Xi_\sigma, \Xi_\mu} ||\Delta X - \Theta_{\mu}(X)\Xi_{\mu}\Delta t - \Theta_{\sigma}(X)\Xi_{\sigma}\Delta B^{\mathbb P}_{t}||_2^2 \;+\;
\mathcal{R}_{\alpha,\rho}(\Xi_\mu,\Xi_\sigma), 
\end{align} 
where the first loss is to fit the model, and the second promotes sparsity through regularization.

The remainder of the paper is organized as follows. Section~\ref{sec: background} formulates the problem, introduces the proposed algorithms, and reviews the necessary background on SDEs. Section~\ref{sec: algo} develops the stochastic sparse identification of SDEs (SSISDE) framework for regression-based estimation, enabling the simultaneous recovery of the drift, diffusion, and underlying noise. Section~\ref{sec: experiment} presents numerical experiments illustrating the effectiveness of the proposed method, including an application to the Black–Scholes SDE, which notably does not satisfy the ergodicity and stationary conditions.

\section{Background and problem formulation}\label{sec: background}
Let $B$ be the Brownian motion defined on the canonical probability space $(\Omega, \cF, \{\cF\}_{t\geq 0}, \Prob)$\footnote{We use the notation $B^\Prob$ to stress the underlying probability measure }.  We assume our time series data $X_{t_0}, X_{t_1},\cdots X_{t_N}$ comes from the underlying continuous  process $X_t$ which follows a stochastic differential equation (under the physical measure) of the form
\begin{align}\label{eq:diff.X_t}
   d X_t &= \mu (X_t)dt+ \sigma(X_t) dB^\Prob_t.
\end{align}

Note that the underlying process \eqref{eq:diff.X_t} does not always process a unique solution for general drift ($\mu$) and diffusion ($\sigma$) coefficients.  Nonetheless, under nominal assumptions (Lipchitz continuity and linear growth) on the drift ($\mu$) and diffusion ($\sigma$) coefficient, one can ensure the existence and uniqueness of such a solution \cite[Chapter 5]{karatzas1991-book}. 

It is a well-known pitfall of SDE calibration that, from a probabilistic perspective, one cannot uniquely recover the drift of the process from the underlying process \eqref{eq:diff.X_t}. To see this, let $\theta$ be a measurable function and define $\tilde{B}_t = B_t^{\mathbb{P}} - \int_0^t \theta(X_s)\,ds$.

By Girsanov's theorem, there exists a measure $\tilde{\mathbb{Q}}$ under which $\tilde{B}_t$ is a Brownian motion. Then
\[
dX_t = \big(\mu(X_t) + \sigma(X_t)\theta(X_t)\big)\,dt + \sigma(X_t)\,d\tilde{B}^{\tilde{\mathbb{Q}}}_t.
\]
Hence, the drift changes to
\[
\tilde{\mu}(x) = \mu(x) + \sigma(x)\theta(x),
\]
so the drift is not uniquely determined if one allows to change the underlying $\P$-Brownian motion \cite{karatzas1991-book}, i.e., the underlying filtration.
\par On the flip side, the drift is indeed unique in a pathwise sense, i.e. fixing the underlying filtration a-priory. For paths with finite quadratic variation (Brownian sample paths almost surely satisfy this condition), plus some nominal additional regularity \cite[Proposition~4.1]{ananova2017} shows that this path can be uniquely represented as a 'rough'+'smooth' component once the underdying driving rough component is fixed. i.e. in the context of SDE \eqref{eq:diff.X_t}, once we fix a sample path $\omega\in \Omega$, the drift $\mu(X(\omega,t))$ i.e. the smooth component can be uniquely identified subjected to fixing the underlying rough driver $B^\P(\omega,t)$, so more preciously we should say unique $\mu^{B^{\P}(\omega)}(X(\omega,t))$. Of course if one changes the underlying rough driver $B^\P(\omega,t)$ to $B^{\bar \P}(\omega,t)$ one gets a different (but again unique) rough-smooth decomposition resulting in another unique $\mu(t,\omega)$ under the new rough driver $B^{\bar \P}(\omega,t)$. 

The first main goal of this paper is to estimate simultaneously the underlying noise vector $(B^\mathbb P_{t_0} {(\omega)},B^\mathbb P_{t_1}{(\omega)},\\ \cdots B^\mathbb P_{t_N}{(\omega)})$, the drift vector $(\mu(X_{t_0}{(\omega)}),\mu(X_{t_1}{(\omega)}),\cdots \mu(X_{t_N}{(\omega)}))$, and the diffusion vector $(\sigma(X_{t_0}{(\omega)}),\sigma(X_{t_1}{(\omega)})\\,\cdots \sigma(X_{t_N}{(\omega)}))$ of \eqref{eq:diff.X_t} from a single discrete data vector $(X_{t_0}{(\omega)}, X_{t_1}{(\omega)},\cdots, X_{t_N}{(\omega)})$. The second goal of the paper is to recover the symbolic function $\mu$ and $\sigma$ from the above estimated vectors.  Note that apriori we do not have information about the $\mu$ and $\sigma$ function except for $\mu(X_0)$  and the fact that \eqref{eq:diff.X_t} has a unique (strong) solution (A sufficient condition for strong solution is (local) Lipschitz continuity of $\mu$ and $\sigma$). 

Under the existence and uniqueness of solution of the SDE \eqref{eq:diff.X_t}, the diffusion process of $X_t$ given by \eqref{eq:diff.X_t} possesses the integral form
\begin{equation} \label{int_form}
    X_t = X_0 + \int_0^t\mu (X_s) ds + \int_0^t  \sigma (X_s)  dB_s^{\mathbb P},
\end{equation}
where the stochastic integral is understood in the It\^o sense. 
The equation \eqref{int_form} can be rewritten such that the stochastic integral is understood in the Stratonovich sense:
\begin{equation} \label{int_form_stat}
    X_t = X_0 + \int_0^t \left( \mu (X_s)-\frac{1}{2}\sigma (X_s) \sigma' (X_s) \right) ds + \int_0^t \sigma (X_s)  \circ dB_s^{\mathbb P}.
\end{equation}
The term $\frac{1}{2}  \int_0^t\sigma (X_s) \sigma' (X_s)  ds$ is known as the It\^o-Stratnovich correction term or quadratic variation correction term corresponding to the SDE \eqref{int_form}.
If we set \(\tilde{\mu}(X_s) := \mu (X_s)-\frac{1}{2}\sigma (X_s) \sigma' (X_s)\), then the SDE becomes

\begin{equation} \label{int_form_stat1}
    X_t = X_0 + \int_0^t  \tilde{\mu}(X_s)  ds + \int_0^t \sigma (X_s)  \circ dB_s^{\mathbb P}.
\end{equation}

We will solve the problem in two-fold; firstly by estimating the diffusion vector and underlying risk-neutral Brownian noise.  Secondly, by applying a series of approximations to simplify the SDE, and then using regression to identify the noise (underlying physical Brownian motion) and the drift vector simultaneously as a solution to the ODE. The detailed algorithm is provided in Section \ref{sec: algo}. Below is the list of assumptions we need to derive the algorithm.
\begin{enumerate}
\item $\mu$ and $\sigma$ are Lipschitz continuous.
    \item $\mu\in C^3(\R)$ and $\sigma\in C^4(\R)$.
    \item  Assume the initial values of $\mu$ i.e. $\mu(X_0)$ is known (we do not need prior knowledge of  $\sigma(X_0)$).
    \item Assume the  $\mathbb P$ Brownian motion starts at $0$ i.e. $B^\mathbb P_0 = 0$. Without this assumption, the algorithm still applies; however, we can only recover the increments of the noise vector rather than the noise vector itself.   
\end{enumerate}
We summarise the key steps behind the algorithm below.   
\begin{enumerate}
\item[Step 1: ] Firstly, by taking the quadratic variation of the original data process $X_t$, we can eliminate the drift $\mu$ and the randomness $B^\mathbb P$ from the system. This enables us to estimate the diffusion vector $(\sigma(X_{t_0}),\sigma(X_{t_1}),\cdots\\ \sigma(X_{t_{N-1}}))$, of \eqref{eq:diff.X_t}. We then use Girsanov’s theorem to construct a measure $\mathbb Q$ (often known as martingale\\/risk-neutral measure\footnote{This measure $\mathbb{Q}$ can be thought of as an artificial “lens” under which the data process $X_t$ looks like fair games}) under which the drift in \eqref{eq:diff.X_t} disappears. i.e.
\begin{align}\label{martingale SDE-Stat}
    X_t &= X_0 +\int_0^t  -\frac{1}{2}\sigma(X_s)\sigma'(X_s)ds +\int_0^t \sigma(X_s)\circ dB_s^{\mathbb Q}.
\end{align}
As a consequence, the risk-neutral Brownian noise (denoted as $B^\mathbb Q$) can be calculated at time points $t_0, t_1, \cdots t_{N-1}$ using the data and the (estimated) diffusion vector.

\par By using a symbolic regression (based on a preset library), we then recover the underlying diffusion function $\sigma: \mathbb R \to \mathbb R$. This allows us to directly calculate the $\sigma'$ vector in \textbf{Step 5} without using finite difference methods. 

\item[Step 2: ] For all $i\in \{0,\cdots N-1\}$ we use a change of variable formula to rewrite $\sigma(X_{t_{i+1}})$ as an expansion around $X_{t_{i}}$. Replacing this approximation of  $\sigma$ in the (Stratonovich) integral form of SDE \eqref{martingale SDE-Stat} and combining the like terms to represent the SDE in terms of weighted sums of the (Stratonovich) iterated integral (often identified as signatures) for the $B^\mathbb Q$ up to order two\footnote{By order two we mean the error terms are of order $|t^n_{i+1}-t^n_{i}|^{2-\epsilon}$ for any given $\epsilon>0$.}, we identify the weights. 
\item[Step 3: ] We approximate both the drift and diffusion coefficients $\mu(X_{t_{i+1}}), \sigma(X_{t_{i+1}})$ by a second-order Taylor expansion around $(X_{t_{i}})$ and replace the approximation into \eqref{int_form_stat1}. This way we can represent the SDE \eqref{int_form_stat1} in terms of weighted sums of the (Stratonovich) iterated integral wrt the $B^\mathbb P$ up to order two, and we identify the weights.
\item[Step 4: ] We now apply the Girsanov theorem to transform our approximated SDE in \textbf{Step 3} into an approximated SDE written as a sum of weighted (Stratonovich) iterated integral with respect to the martingale Brownian motion $B^\mathbb Q$ up to order two. 
\item[Step 5: ] For each time interval $[t^n_i, t^n_{i+1})$ by comparing the weights from \textbf{Step 2} and \textbf{Step 4}, we write $\mu$  as a solution of ODE where the coefficients of the ODE are written using  $\sigma$, $\sigma'$ and the weights from \textbf{Step 2}.
\item[{Step 6: }] We use a regression method to solve the weights from \textbf{Step 3} to identify the coefficients of the ODE in \textbf{Step 5}. Finally, we use an ODE solver on the time grid ($t_1,\cdots, t_N$) to identify the drift vector $(\mu(X_{t_0}),\mu_(X_{t_1}),\cdots \mu(X_{t_N}))$ and recover the increments of the physical Brownian motion $\{\Delta B^{\mathbb P}_{t_i}:=B^{\mathbb P}_{t_{i+1}}-B^{\mathbb P}_{t_i}\}_{i=1}^{N-1}.$
\item[{Step 7: }] We propose \textit{Stochastic Sparse Identification of Stochastic Differential Equation} (SSISDE) to recover the exact symbolic form of the SDE that
\begin{align}
dX_t=\mu(X_t)dt+\sigma(X_t)dB_t^\mathbb{P}. 
\end{align}

We introduced a stochastic symbolic regression problem inspired by SINDy, which allowed us to obtain the precise functional form of the SDE by optimizing the following target
\begin{align}
    \arg\min_{\Xi_\sigma, \Xi_\mu} \sum_{i=i}^N ||\Delta X_{t_{i}} - \Theta(X_{t_{i}})\Xi_{\mu}\Delta t_i- \Theta(X_{t_{i}})\Xi_{\sigma}\Delta B^{\mathbb P}_{t_i}||_2^2 \;+\;
\mathcal{R}_{\alpha,\rho}(\Xi_\mu,\Xi_\sigma),
\end{align}
where $\Theta(X_{t_i})$ is a library of functions (pre-)specified by the users, which commonly includes polynomials and Fourier functions. 
This concludes our algorithm to identify the stochastic governing equation from noisy, stochastic, and single-shot time-series observations. 
\end{enumerate}

\section{Main Results}\label{sec: algo}
In this section, we introduce the main algorithm. The key idea is to formulate the estimator of the drift function as the solution of an ODE, which in turn enables the recovery of the underlying physical Brownian motion $B^{\mathbb P}$ given one single trajectory observation of the state process $X_t$ on a finite time interval $[0, T]$. \\

\noindent \textbf{Step 1: recovering $\mathbb Q$-Brownian motion.}
Recast the It\^o SDE \eqref{eq:diff.X_t} with stating time $s$ (may or may not be $0$) and terminal time $t>s$ below, 
\begin{equation}
    X_t = X_s + \int_s^t  \mu(X_u)  du + \int_s^t \sigma (X_u)  dB_u^{\mathbb P}.
\end{equation}
By applying Girsanov's theorem, the above SDE can be represented as a martingale with respect to the $\mathbb Q$-Brownian motion $B^{\mathbb Q}$, which has the following form
\begin{align}\label{martingale SDE}
    X_t &= X_s +\int_s^t \sigma(X_u) dB_u^{\mathbb Q}
\end{align}
where $B^{\mathbb Q}$ is the Brownian motion under the martingale measure $\mathbb Q$, which is often referred to risk-neutral measure in finance \cite{harrison1981}, following the dynamic
\begin{align}\label{eqn:change_of_measure}
 \forall u\in [s,t); \quad  dB_u^{\mathbb Q} = dB_u^{\mathbb P} + \frac{\mu(X_u)}{\sigma(X_u)} du, 
\end{align}
By definition, the underlying limiting quadratic variation of process $X$ along a given sequence of time-grid $\pi := \{\pi^N\}_{N=0}^\infty:=\{0=t^N_0<t^N_1\cdots < t^N_i<\cdots<t^N_{\#(\pi^N)}=t\}_{N=0}^\infty$ with vanishing mesh (i.e. $|t^N_{i+1}-t^N_i|\xrightarrow[]{N\to \infty} 0$), denoted by $[X]_{\pi,t} := \lim_{n\to \infty} \sum_{\pi^n\cap [0,t]} (X(t^n_{i+1}) - X(t^n_i))^2$, has the following form
\begin{align}
    d[X]_{\pi,t} = (\sigma(X_t))^2 dt. 
\end{align}
By using the high frequency estimate of quadratic variation along the time-grid $\pi^M$ for a given $M$ large enough (i.e $[X]_{\pi^M,t}:=\sum_{\pi^M\cap [0,t]} (X(t^M_{i+1}) - X(t^M_i))^2 $), we first obtain an estimate of 
$\sigma$ as follows. 
\begin{align}\label{Sigma.QV}
    (\sigma(X_t))^2 \approx \frac{1}{\delta} \sum_{\pi^M\cap [t,t+\delta]} (X(t^M_{i+1}) - X(t^M_i))^2 
\end{align}
Note that in order to approximate the diffusion vector $(\sigma(X_{t_0}),\sigma(X_{t_1}),\cdots \sigma(X_{t_{N-1}}))$ one need to have observation of the underlying process $X$ along a finer grid, i.e. $\sigma(X_{t_i})$ is calculated by talking sum square difference of $L$ many observation of $X$ between $t_i$ and $t_{i+1}$. Once the 
$\sigma$ vector has been identified, we can then construct the 
$\mathbb Q$-Brownian motion increment on any interval $[t,t+\delta]$, for $\delta>0$, 
\begin{align}\label{Q BM}
B_{t+\delta}^{\mathbb Q}-B_t^{\mathbb Q}= \int_t^{t+\delta}   \frac{dX_s}{\sigma(X_s)}\approx \frac{\sqrt\delta (X(t+\delta)-x(t))}{\sqrt{\sum_{\pi^M\cap [t,t+\delta]} (X(t^M_{i+1}) - X(t^M_i))^2}}.
\end{align}
\\\noindent\textbf{learning $\sigma$ function from 
the vector $\bm{(\sigma(X_{t_0}),\sigma(X_{t_1}),\cdots, \sigma(X_{t_{N-1}}))}$}
By considering the input-output pairs $\{(X_{t_i}, \sigma(X_{t_i}))\}$, we build a library $\Theta(X_{t_i})$ and rewrite (generalized) SINDy into a symbolic regression form as follows.
\begin{align}\label{SINDy sigma}
    \Xi_\sigma^* = \arg\min_{\Xi_\sigma} \sum_{i=0}^{N-1}||\sigma(X_{t_i})-\Theta(X_{t_i})\Xi_\sigma|| + ||\Xi_\sigma||_0,
\end{align}
where the first loss is to fit the model (i.e., the function $\sigma$), and the second term is referred to as sparsity regularization. This optimization problem can be solved using a sequential thresholding least-squares algorithm in SINDy. Since the coefficient of $\Xi_\sigma$ is sparse, this optimization problem effectively identifies the symbolic functional form of the $\sigma(X_{t_i})$ (e.g., for the BS equation $\sigma(X_{t_i})=\sigma X_{t_i}$). Note that from \eqref{Sigma.QV} we can not recover the terminal value $\sigma(X_{t_N})$, but once we estimate the $\sigma$ function from \eqref{SINDy sigma} one can estimate $\sigma(X_{t_N})$, recovering the entire diffusion vector $\left(\sigma(X_{t_0}),\sigma(X_{t_1}),\cdots \sigma(X_{t_{N}})\right)$.

\noindent\textbf{Step 2: constructing estimator for $X_t$ via Stratonovich SDE.} For the It\^o SDE \eqref{martingale SDE}, the corresponding Stratonovich SDE has the following form,
\begin{align}\label{martingale SDE stratonovich}
    X_t &= X_s +\int_s^t  -\frac{1}{2}\sigma(X_u)\sigma'(X_u)du +\int_s^t \sigma(X_u)\circ dB_u^{\mathbb Q}\nonumber \\
    & := X_s+\int_s^t A_0(X_u) du+\int_s^t \sigma(X_u)\circ dB^{\mathbb Q}_u,
\end{align}
where we denote $A_0(X_t)=-\frac{1}{2}\sigma(X_t)\sigma'(X_t)$, which is commonly known as the It\^o-Stratonovich correction term. For the Stratonovich SDE \eqref{martingale SDE stratonovich}, applying the (Stratono-\\vich) change of variable formula (i.e. $  f(X_u) = f(X_s) + \int_s^u f'(X_r)\circ dX_r$) for the $A_0$ and $\sigma$ functions around the initial point $X_s$, we get the following expansion. Under some nominal H\"older assumptions on the underlying process $X$, this approximation error is of order $|t|^{3/2}$ (i.e. $|t_{i+1}-t_i|^{\frac{3}{2}}$ on each time grid $[t_i,t_{i+1})$).
\begin{align}
    X_t= & X_s+A_0(X_s)\int_s^t du +\int_s^t \int_s^u A_0'(X_r) \circ dX_r du + \sigma(X_s)\int_s^t \circ dB_u^{\mathbb Q}\nonumber\\&+\int_s^t\int_s^u \sigma'(X_r)\circ dX_r\circ dB_u^{\mathbb Q}\nonumber.
\end{align}
Plugging the differential from of the dynamic \eqref{martingale SDE stratonovich} into the above expansion, applying the change of variable iteratively,  and finally approximating $\sigma(X_u)$ for all $u\in [t_i,t_{i+1})$ by $\sigma(X_{t_i})$ we get the following second-order expansion\footnote{by the second order we mean the error term is of the order $O(|t-s|^{2-\delta})$ for any $\delta>0$.} of the process $X_t$ in the Stratonovich form,
\begin{align}
    \label{SDE 2 order expansion}
    X_t=&X_s+\psi_1^{s,t} \int_s^t  du +\psi_2^{s,t} \int_s^t  \circ d B_u  + \psi_{12}^{s,t}\int_s^t\int_s^u  dr \circ dB_u^{\mathbb Q}\\
    &+\psi_{21}^{s,t}\int_s^t\int_s^u\circ   dB^{\mathbb Q}_r du +\psi_{22}^{s,t}\int_s^t\int_s^u \circ  d B^{\mathbb Q}_r \circ dB_u^{\mathbb Q} \nonumber \\ &+ \psi_{222}^{s,t}\int_s^t\int_s^u\int_s^r  \circ dB_v^{\mathbb Q}\circ dB_r^{\mathbb Q} \circ dB_u^{\mathbb Q} + \mathsf{\varepsilon}^{\Psi_2},\nonumber
\end{align}
where we denote $\varepsilon^{\Psi_2}$ as the error for the higher-order Taylor expansion terms (see, e.g., \cite{Kloeden1992}). The detailed derivation is provided in Appendix \ref{section proof step 2}. In the following, we will call the set $\Psi_2^{s,t}$ the second-order estimator of the diffusion process for the interval $[s,t]$, and denote it as
\begin{align}
\Psi_{2}^{s,t}:=\{\psi_1^{s,t},\psi_2^{s,t},  \psi_{12}^{s,t}, \psi_{21}^{s,t}, \psi_{22}^{s,t},\psi_{222}^{s,t}\}.
\end{align}
Note that the estimate of the set $\Psi_2^{s,t}$ does depend on the full data on the interval $[s,t]$, whereas theoretically the set $\Psi_2^{s,t}$ only depends on observations at time $s$.
Essentially, the set $\Psi_2^{s,t}$ is a collection of different combinations of $\sigma$ and all its derivatives evaluated at the initial value $X_s$. For example, $\psi^{s,t}_1=A_0(X_s)$ and $\psi^{s,t}_2=\sigma(X_s)$. Finally, we use the estimated $\mathbb Q$-Brownian motion $B^{\mathbb Q}$ from \textbf{Step 1} to find estimators in $\Psi_2^{s,t}$ for the second order model \eqref{SDE 2 order expansion}. \\

Next, instead of applying the Girsanov transformation first as done in \textbf{Step 1}, we will apply the Taylor expansion first directly for the Stratonovich SDE under Wiener measure $\mathbb P$ in \textbf{Step 3}, and then apply the Girsanov transformation in the last step (\textbf{Step 4}) to construct the estimator of the function $\mu$. The explicit algorithm is demonstrated in the following diagram.

{\footnotesize
\begin{center}
\begin{tikzcd}[column sep=2.2cm, row sep=2cm]
\text{It\^o SDE } \eqref{eq:diff.X_t}\text{ under } \mathbb P
  \arrow[r, "\text{Step 1: Girsanov }"] 
  \arrow[d, swap, "\text{Step 3: Taylor expansion}"] 
& \text{Stratonovich SDE}\; \eqref{martingale SDE}\; \text{under}\; \mathbb Q  
  \arrow[d, "\text{Step 2: Taylor expansion }"] \\[0.3cm]
\shortstack{Stratonovich SDE}\; \eqref{step 3 expansion} \; \text{under} \; \mathbb P  \arrow[r, "\text{Step 4: Girsanov} "] 
& {\shortstack{Approximation of $X_t$   \eqref{SDE 2 order expansion}}\text{ under }  \mathbb Q}
\end{tikzcd}\label{commute graph}
\end{center} 
}

\noindent\textbf{Step 3: Taylor expansion of Stratonovich SDE under Wiener measure $\mathbb P$.}
Recall the Stratonovich SDE \eqref{int_form_stat1} with a (deterministic) starting time $s$ (not necessarily $0$) with respect to the physical measure $\mathbb P$:
\begin{align*}
    X_t = X_s + \int_s^t  \tilde{\mu}(X_u)  du + \int_s^t \sigma (X_u)  \circ dB_u^{\mathbb P}.
\end{align*}

Applying stochastic Taylor expansion on \(\tilde{\mu}\) and \(\sigma\) around the initial observation \(X_s\), and truncating the Taylor expansion at level 2 (see e.g. \cite{Kloeden1992}), we have the following (up to second order derivatives of $\tilde{\mu}$ and $\sigma$) Taylor expansion approximation of $X_t$ under Wiener measure $\mathbb P$,
\begin{align}
    X_t & = X_s + \tilde{\mu}(X_s) \int_s^t du + \sigma(X_s) \int_s^t \circ dB_u^{\mathbb P} \nonumber\\
    & + \tilde{\mu}'(X_s) \tilde{\mu}(X_s) \int_s^t \int_s^u drdu +  \tilde{\mu}'(X_s) \sigma(X_s) \int_s^t \int_s^u \circ dB_r^{\mathbb P} du \nonumber  \\
    & \quad + \sigma'(X_s) \tilde{\mu}(X_s) \int_s^t \int_s^u dr \circ dB_u^{\mathbb P} +  \sigma'(X_s) \sigma(X_s) \int_s^t \int_s^u \circ dB_r^{\mathbb P} \circ dB_u^{\mathbb P}\nonumber\\
    &  \quad + \left((\tilde{\mu}'(X_s))^2 \tilde{\mu}(X_s) + \tilde{\mu}''(X_s) (\tilde{\mu}(X_s))^2\right) \int_s^t \int_s^u \int_s^r dvdrdu\nonumber\\
    &  \quad + \left((\tilde{\mu}'(X_s))^2 \sigma(X_s) + \tilde{\mu}''(X_s) \tilde{\mu}(X_s) \sigma(X_s)\right) \int_s^t \int_s^u \int_s^r \circ dB_v^{\mathbb P} drdu\nonumber\\
    &  \quad + \left(\tilde{\mu}'(X_s) \sigma'(X_s)\tilde{\mu}(X_s) + \tilde{\mu}''(X_s) \tilde{\mu}(X_s) \sigma(X_s)\right) \int_s^t \int_s^u \int_s^r dv \circ dB_r^{\mathbb P} du\nonumber\\
    &  \quad + \left(\tilde{\mu}'(X_s) \sigma'(X_s)\tilde{\mu}(X_s) + \sigma''(X_s) (\tilde{\mu}(X_s))^2 \right) \int_s^t \int_s^u \int_s^r dv dr \circ dB_u^{\mathbb P}\nonumber \\
    &  \quad + \left(\tilde{\mu}'(X_s) \sigma'(X_s) \sigma(X_s) + \tilde{\mu}''(X_s) (\sigma(X_s))^2 \right) \int_s^t \int_s^u \int_s^r \circ dB_v^{\mathbb P} \circ dB_r^{\mathbb P} du \nonumber\\
    &  \quad + \left(\tilde{\mu}'(X_s) \sigma'(X_s) \sigma(X_s) + \sigma''(X_s) \sigma(X_s) \tilde{\mu}(X_s) \right) \int_s^t \int_s^u \int_s^r \circ dB_v^{\mathbb P} dr \circ dB_u^{\mathbb P} \nonumber\\
    &  \quad + \left( (\sigma'(X_s))^2 \tilde{\mu}(X_s) + \sigma''(X_s) \sigma(X_s) \tilde{\mu}(X_s) \right) \int_s^t \int_s^u \int_s^r  dv \circ dB_r^{\mathbb P} \circ dB_u^{\mathbb P}\nonumber \\
    &  \quad + \left( (\sigma'(X_s))^2 \sigma(X_s) + \sigma''(X_s) (\sigma(X_s))^2  \right) \int_s^t \int_s^u \int_s^r \circ dB_v^{\mathbb P} \circ dB_r^{\mathbb P} \circ dB_u^{\mathbb P}
    + \varepsilon^{\tilde \mu, \sigma},
    \label{step 3 expansion}
\end{align}
where $\varepsilon^{\tilde \mu,\sigma}$ denotes the higher order approximation error term. The proof of this claim is postponed to the following Appendix \ref{section proof step 2}, along with the discussion of the error term $\varepsilon^{\tilde{\mu},\sigma}$.
\\

\noindent\textbf{Step 4: Girsanov transformation.} Applying the Girsanov theorem, substituting for all $r\in [s,t)$; \(dB_r^{\mathbb P} = dB_r^{\mathbb Q} - \frac{\mu(X_r)}{\sigma(X_r)} dr\) into the Stratonovich SDE approximation \eqref{step 3 expansion}, we have the following second order approximation expansion of $X_t$
 {\footnotesize \begin{align}\
    X_t & = X_s - \frac{1}{2} \sigma(X_s) \sigma'(X_s)  \int_s^t du + \sigma(X_s) \int_s^t  \circ dB_u^{\mathbb Q}   \nonumber \\
    & \quad + \left( \mu'(X_s) \sigma(X_s) - \mu(X_s) \sigma'(X_s) - \frac{1}{2}\sigma(X_s) \sigma'(X_s)^2 - \frac{1}{2}\sigma(X_s)^2 \sigma''(X_s) \right) \int_s^t \int_s^u   \circ dB_r^{\mathbb Q} du     \nonumber\\
    & \quad - \frac{1}{2} \sigma(X_s) \sigma'(X_s)^2 \int_s^t \int_s^r   du \circ dB_r^{\mathbb Q}   +  \sigma'(X_s) \sigma(X_s) \int_s^t \int_s^u  \circ dB_r^{\mathbb Q}  
 \circ dB_u^{\mathbb Q} \nonumber\\
 &+\left( (\sigma'(X_s))^2 \sigma(X_s) + \sigma''(X_s) (\sigma(X_s))^2  \right) \int_s^t\int_s^u\int_s^r  \circ dB_v^{\mathbb Q}\circ d B_r^{\mathbb Q}\circ dB_u^{\mathbb Q} +\varepsilon^{\tilde{\mu},\sigma}+\varepsilon^{\mu/\sigma}. \label{expansion step 4}
\end{align}}
 where the error term $\varepsilon^{\tilde{\mu},\sigma}$ is due to \eqref{step 3 expansion}, and the error term $\varepsilon^{\mu/\sigma}$ is due to the approximation of $\frac{\mu(X_u)}{\sigma(X_u)}$ by the initial condition $\frac{\mu(X_s)}{\sigma(X_s)}$ for $u\in [s,t]$. We want to remark that only when $|t-s|$ is small the error term $\varepsilon^{\mu/\sigma}$ can be controlled and is of order $|t-s|^{2-\epsilon}$ for $|t-s|$ small and any $\epsilon>0$, not for general $s,t$. In particular, following the Taylor expansion 
 \cite{azencott2006formule} and the  Castell estimates 
 \cite{castell1993asymptotic}, the error term $\varepsilon^{\tilde{\mu},\sigma}+\varepsilon^{\mu/\sigma}$ follows a Gaussian tail estimate. The expected error (under $\mathbb E^{\mathbb Q} $) can be found in \cite{lyons2004cubature}. The proof of this claim is postponed to Appendix \ref{section proof step 4}.

 \begin{remark}
As a consequence of the above approximation, we essentially restrict ourselves to the case when $\frac{\mu}{\sigma} \approx \text{constant}$ on each interval $[t_i,t_{i+1}]$.  A higher-order Taylor approximation of \(\frac{\mu(X_u)}{\sigma(X_u)}\) is also possible, which will relax the $\frac{\mu}{\sigma} \approx \text{constant}$ assumption; we leave it for future studies. We remark that by taking the proposed higher order Taylor expansion, the \eqref{ODE estimator of mu} will change accordingly.  
\end{remark}

\noindent\textbf{Step 5: constructing estimator of $\mu$.} In this step we recursively calculate the drift vector $(\mu(X_{t_1}),\mu(X_{t_2}),\allowbreak\cdots \mu(X_{t_N}))$ under the assumption of know initial value $\mu(X_{t_0})$ is known. By comparing the coefficients of same iterated integrals of \eqref{SDE 2 order expansion} from \textbf{Step 2} and \eqref{expansion step 4} from \textbf{Step 4} on the interval $[t_0,t_{1})$ and under the assumption of know initial value $\mu(X_{t_0})$ we calculate the drift $\mu(X_{t_{1}})$, than recursively calculate all the other drift terms. Under the assumption that $\sigma$ is continuous, it is enough for short time uniqueness of ODE for the drift function $\mu$ in the interval  $[t_i,t_{i+1})$ for $ |t_{i+1}-t_{i}|<<1$, and hence in the interval $[t_i,t_{i+1})$ the drift function can be written as the (unique) solution of the following ODE with given initial value $\mu(x_{t_i})$ (which is estimated from the previous step) 
 \begin{equation}\label{ODE estimator of mu}
    \mu'(x)  = a(x)\mu(x) + b(x),
\end{equation}
where \(a(x) = \frac{\sigma'(x)}{\sigma(x)}\) and \(b(x) = \frac{2\psi_{21}^{[t_i,t_{i+1})}+\sigma(x) \sigma'(x)^2 +\sigma(x)^2 \sigma''(x)}{2\sigma(x)}\), and $\psi_{21}^{[t_i,t_{i+1})}$ is defined in \eqref{SDE 2 order expansion} from \textbf{Step 2} on the interval $[t_i,t_{i+1})$. Note that we can identify the functional form of $\sigma$ and its derivatives from \textbf{Step 1}.  The proof is postponed to Appendix \ref{section proof step 5}.

We may also consider the higher order Taylor approximation of \(\frac{\mu(X_u)}{\sigma(X_u)}\) by assuming \(\left(\frac{\mu}{\sigma}\right)' \approx \text{constant}\), instead of  \(\frac{\mu}{\sigma} \approx \text{constant}\). To be precise
\begin{equation} \label{higher_order mu/sigma}
    \frac{\mu(X_u)}{\sigma(X_u)} \approx \frac{\mu(X_s)}{\sigma(X_s)} \left(X_u - X_s\right) + \frac{1}{2} \frac{\mu'(X_s)\sigma(X_s) - \mu(X_s)\sigma'(X_s)}{\sigma(X_s)^2} \left(X_u - X_s\right)^2, \quad \text{ for } u \in [s,t].
\end{equation}

In addition to this, we consider the iterated integrals in \textbf{Step 2} and \textbf{Step 4} to be truncated to the order \(|t-s|^{3 - \varepsilon}\). Using ~\eqref{higher_order mu/sigma} in these higher order expansions and then comparing the coefficients gives the following second order ODE 
\begin{equation} \label{ODE2}
    \mu''(x) = A(x) \mu'(x) + B(x)\mu(x),
\end{equation}
where \(A(x) = \frac{2\sigma'(x)}{\sigma(x)}\) and \(B(x) = \frac{\sigma''(x)}{\sigma(x)} - \frac{2\left(\sigma'(x)\right)^2}{\sigma(x)^2}\). Given \(\mu(x_{t_i})\) and \(\mu'(x_{t_i})\), there exists a unique solution of ODE ~\eqref{ODE2} on the interval  $[t_i,t_{i+1})$ for $ |t_{i+1}-t_{i}|<<1$. \\

\noindent \textbf{Step 6: estimating the noise with $\mathbb P$-Brownian motion.} By applying regression methods to find the estimator $\psi_{12}^{[t_i,t_{i+1})}$ recursively using the model constructed from the martingale Brownian motion $B^{\mathbb Q}$. Applying \textbf{Step 1-5} to the single trajectory observation of the states $\{X_{t_i}\}_{i=1}^N$, we find the second order estimator for the drift vector $\mu$ at the time grid $t_1,\cdots,t_N$ as the solution of ODE \eqref{ODE estimator of mu}. Then, applying the change of measure \eqref{eqn:change_of_measure} to recover the noise increment vectors $(\Delta B_{t_0}^{ \mathbb P}, \Delta B_{t_2}^{ \mathbb P},\cdots, \Delta B_{t_{N-1}}^{ \mathbb P} )$ by using the following discretization scheme, 
\begin{align}\label{eqn:change_of_measure discrete}
\Delta B_{t_i}^{\mathbb Q} = \Delta B_{t_i}^{\mathbb P} + \frac{\mu(X_{t_{i}})}{\sigma(X_{t_{i}})} \Delta t_i, \quad \text{for}\quad i=1,\cdots, N,
\end{align}
where we use the convection $\Delta B_{t_i}^{ \mathbb P}=B_{t_{i+1}}^{\mathbb P}-B_{t_{i}}^{ \mathbb P}$
 (similarly for $B^{\mathbb Q}$ and t).
\\

\noindent {\textbf{Step 7: SSISDE to recover drift and diffusion function.} We now present the \textit{Stochastic Sparse Identification of Stochastic Differential Equation} (SSISDE) algorithm, which is able to precisely identify the form of the governing SDE \eqref{eq:diff.X_t}
\begin{align*}
dX_t=\mu(X_t)dt+\sigma(X_t)dB_t^\mathbb{P}. 
\end{align*}
}

If $X_t$ and the noise $B^{\mathbb P}_{t}$ can be observed along the time grid $t_1<\cdots<t_N$, this can be viewed as a symbolic regression problem that 
\begin{align}
    dX_t = \Theta_\mu(X_t)\Xi_\mu dt + \Theta_\sigma(X_t)\Xi_\sigma dB_t^\mathbb{P},
\end{align}
which is motivated from the the {\em sparse identification of nonlinear dynamics} (SINDy) \cite{brunton2016discovering}. 
 Following the idea of SINDY, 
According to \textbf{Step 6}, we have constructing the original noise increment $\{\Delta B_{t_i}^{\mathbb P}\}_{i=0}^{N-1}$ from observation data $\{X_{t_i}\}_{i=0}^N$. We construct symbolic regression for the following identity 
\begin{align}\Delta X_{t_i} = \Theta_\mu(X_{t_{i}})\Xi_\mu \Delta t_i + \Theta_\sigma(X_{t_{i}})\Xi_\sigma \Delta B_{t_i}^\mathbb{P},
\end{align}
for $i=0,\cdots, N-1$. We denote $\Xi_\sigma$ ( and $\Xi_\mu$ respectively) as the unknown sparse matrix corresponding to function $\sigma$ (and $\mu$ respectively). Similarly, we denote $\Theta_\sigma(\cdot)$ and  $\Theta_\mu(\cdot)$ as the candidate models for $\sigma$ and $\mu$. We thus propose the following \textit{Stochastic Sparse Identification of Stochastic Differential Equation} (SSISDE) algorithm, which can be written as a stochastic symbolic regression problem, 
\begin{align}\label{stochastic SINDy}
    \arg\min_{\Xi_\sigma, \Xi_\mu} \sum_{i=0}^{N-1}||\Delta X_{t_{i}} - \Theta_{\mu}(X_{t_{i}})\Xi_{\mu}\Delta t_i- \Theta_{\sigma}(X_{t_{i}})\Xi_{\sigma}\Delta B^{\mathbb P}_{t_i}||_2^2\;+\; \mathcal{R}_{\alpha,\rho}(\Xi_\mu,\Xi_\sigma),
\end{align}
where the first term is the residual $L_2$-norm, and the second is a sparsity-inducing regularizer.

\begin{remark}
    The \textit{Stochastic Sparse Identification of Stochastic Differential Equation} (SSISDE) \eqref{stochastic SINDy} only takes input data $\Delta X_{t_i}$ and the estimated noise $\Delta B^{\mathbb P}_{t_i}$, which is different from the SINDy form \eqref{SINDy sigma} in \textbf{Step 5}. In the end, one can compare the estimated symbolic forms for the function $\sigma$. In general, we only use \eqref{SINDy sigma} in \textbf{Step 5} to construct the estimate of the drift vector $\mu(X_{t_i})$ and the desired noise $\Delta B^{\mathbb P}_{t_i}$.
\end{remark}

The main algorithm is described below. 
{\footnotesize
\begin{center}
\begin{tikzcd}[column sep=.9cm, row sep=2cm]
\text{Observations}: \{X_{t_i}\}_{i=1}^N\arrow[r, "\text{Step 1: find} \{\sigma(X_{t_i})\}_{i=1}^N"] \arrow[dr, swap, "\text{Step 3+Step 4: find ODE for}\quad \mu"] &  \text{construct} \{\Delta B^{\mathbb Q}_{t_i}\}_{i=1}^N  \arrow[d, "\text{Step 2: find parameters $\{\Psi_2(t_i)\}_{i=1}^N$}"] \\
  & {\text{Calculate} \{\mu(X_{t_i})\}_{i=1}^N \text{ using Euler scheme}}  \arrow[d, "\text{Step 6: construct $\{\Delta B^{\mathbb P}_{t_i}\}_{i=1}^N$}"] \\
  & {\text{\small Step 7: apply SSISDE to identify the two functions}\quad  \mu,\sigma. } 
\end{tikzcd}\label{commute graph: algorithm}
\end{center}}

\section{Numerical Experiments}\label{sec: experiment}
\subsection{Model description}
In this section, we demonstrate the performance of our method for estimating the drift, diffusion function,  and noise vector across several numerical examples. We assume the process \(X_t\) satisfies the dynamics as given in Eq.~\eqref{eq:diff.X_t}. We simulate sample paths using a standard time-discretization scheme, such as the Euler-Maruyama scheme. With access to the data, we approximate its quadratic variation $[X]_{\pi,t_i}$ for each time index \(t_i: 0 \le i \le n\) where \(t_0=0, t_n=T\), \(n=100000\), and \(T=1\) of a chosen uniform partition \(\pi\). The quadratic variation satisfies the ordinary differential equation \(d[X]_{\pi,t} = (\sigma(X_t))^2 dt\). On estimated $Y_t=[X]_{\pi,t}$, we solve the symbolic regression problem ~\eqref{SINDy sigma} to get the functional form of \((\sigma(X_t))^2\) and hence \(\sigma(X_t)\). The $L_0$ penalty $\|\Xi_\sigma\|_0$ in Eq.~\eqref{SINDy sigma} induces sparsity, but since we focus only on estimating $\sigma(X_t)$ accurately, no regularization coefficient is included. We know that under risk-neutral probability $\mathbb Q$, the dynamics of \(X_t\) satisfy \(dX_t=\sigma(X_t)dB_t^\mathbb{Q}\), which can be used to get the \(B_t^\mathbb{Q}\) vector assuming it starts at zero. Now we have access to the primary process \((t_i, B_{t_i}^\mathbb{Q})\) for \(0 \le i \le N\). We extend the features by using the idea of the iterated Stratonovich integral of the primary process up to order  \(|t-s|^{2-\epsilon}\), for any $\epsilon>0$, i.e., the integrals corresponding to the following coefficients
\begin{align*}
\Psi_{2}:=\{\psi_1,\psi_2,\psi_{12},\psi_{21},\psi_{22},\psi_{222}\}.
\end{align*}
We choose a sub-partition \(\{j_k\}_{k=0}^{M}, M <<N\) of the partition \(\pi\) such that \(t_{j_0} = t_0, t_{j_M} = t_N\) with \(M=1000\). On each sub-interval \([t_{j_k}, t_{j_{k+1}}]\) of this sub-partition, \(X\)  is regressed against above features to estimate the values of coefficients in the \(\Psi_{2}\) at \(t=t_{j_k}\) for \(k=0,\dots, M\).
We would also like to remark that using \eqref{eq29} one can estimate $\Psi_2$ using the symbolic form of $\sigma$ recovered in \textbf{Step-1}, without using the regression method used in the classical signature model. 

The ODE~\eqref{ODE estimator of mu} is then solved to get the vector of \(\mu(X_{t_{j_k}})\) for each \(t_{j_k}\). Furthermore, we use the discretization scheme ~\eqref{eqn:change_of_measure discrete} to predict the noise increment vectors $(\Delta B_{t_{j_0}}^{ \mathbb P}, \Delta B_{t_{j_1}}^{ \mathbb P},\cdots, \Delta B_{t_{j_M}}^{ \mathbb P} )$ hence the noise \(B^{\mathbb{P}}\) by assuming it starts at zero. Finally, we use the \textit{SSISDE} ~\eqref{stochastic SINDy} to recover the symbolic form of drift \(\mu(X_t)\) and diffusion \(\sigma(X_t)\). We describe the model-selection procedure, i.e., $k$-fold time–series cross-validation, the CV error $\delta$, and the choice of the optimal sparsity level, before presenting the numerical examples. The code required to reproduce all experiments is available at \href{https://github.com/munawarali93/SSISDE}{https://github.com/munawarali93/SSISDE}.

\subsubsection{Model selection via \texorpdfstring{$k$}{TEXT}-fold time–series cross-validation}

We adopt an elastic-net penalty \cite{hastie2009elements} in the penalized identification problem \eqref{stochastic SINDy}, namely
\[
\mathcal{R}_{\alpha,\rho}(\Xi_\mu,\Xi_\sigma)
=\alpha\!\left(\rho\,\|\beta\|_{1}+\tfrac{1-\rho}{2}\,\|\beta\|_{2}^{2}\right),
\qquad
\beta:=\begin{bmatrix}\Xi_\mu\\[2pt]\Xi_\sigma\end{bmatrix},
\]
with $\|\cdot\|_{1}$ and $\|\cdot\|_{2}$ the usual vector norms. The penalty strength $\alpha>0$ and mixing parameter $\rho\in[0,1]$ are hyperparameters that control sparsity and shrinkage in $(\Xi_\mu,\Xi_\sigma)$ ($\rho=1$ recovers Lasso; $\rho=0$ Ridge). We select $(\alpha,\rho)$ by $k$-fold \emph{time–series} cross-validation \cite{stone1974cross} (no shuffling across time): let $\{A_j\}_{j=1}^k$ partition the index set $\{1,\dots,N-1\}$ into $k$ contiguous, non-overlapping validation blocks, and define $B_j:=\bigcup_{\ell\neq j}A_\ell$ as the complementary training indices. For each candidate pair $(\alpha,\rho)$ and each fold $j$, we fit \eqref{stochastic SINDy} on $B_j$ to obtain $(\widehat{\Xi}_\mu^{(j)},\widehat{\Xi}_\sigma^{(j)})$, and compute the mean squared one-step prediction error on $A_j$:
\begin{align*}
\delta_j^2(\alpha,\rho)
\;:=\;
\frac{1}{|A_j|}
\sum_{i\in A_j}
\Bigl\|
\Delta X_{t_i}
-\Theta_\mu(X_{t_i})\,\widehat{\Xi}_\mu^{(j)}\,\Delta t_i
-\Theta_\sigma(X_{t_i})\,\widehat{\Xi}_\sigma^{(j)}\,\Delta B^{\mathbb P}_{t_i}
\Bigr\|_2^2.
\end{align*}
The $k$-fold CV score and its standard error are
\begin{align*}
\overline{\delta}(\alpha,\rho)
=\frac{1}{k}\sum_{j=1}^k \delta_j(\alpha,\rho),
\qquad
\operatorname{SE}(\alpha,\rho)
=\sqrt{\frac{1}{k(k-1)}\sum_{j=1}^k \bigl(\delta_j(\alpha,\rho)-\overline{\delta}(\alpha,\rho)\bigr)^2 }.
\end{align*}
We evaluate $\overline{\delta}$ on a logarithmic grid $\alpha\in\Lambda$ and a discrete set $\rho\in\mathcal{P}$ (including $\rho=1$ for Lasso). To reduce shrinkage bias, after solving \eqref{stochastic SINDy} on each training fold we refit an ordinary least-squares model \emph{restricted to the active set} (nonzero entries of $(\widehat{\Xi}_\mu^{(j)},\widehat{\Xi}_\sigma^{(j)})$) before scoring on $A_j$.\footnote{All feature normalizations are computed \emph{using training data of each fold only} and then applied to both training and validation rows to prevent temporal leakage.}

\subsubsection{Choice of the final model}
Let $(\alpha_\star,\rho_\star)=\arg\min_{\alpha,\rho}\overline{\delta}(\alpha,\rho)$ and define the one–standard-error \cite{hastie2009elements} threshold
\[
\varepsilon \;=\; \overline{\delta}(\alpha_\star,\rho_\star)+\operatorname{SE}(\alpha_\star,\rho_\star).
\]
Among all $(\alpha,\rho)$ with $\overline{\delta}(\alpha,\rho)\le \varepsilon$, we pick the \emph{simplest} model, i.e., the one with the smallest support size (fewest nonzeros in $(\Xi_\mu,\Xi_\sigma)$); ties are broken toward larger~$\alpha$ (and larger~$\rho$ when applicable). This yields $(\alpha^\dagger,\rho^\dagger)$ and the corresponding $(\widehat{\Xi}_\mu,\widehat{\Xi}_\sigma)$. We then refit \eqref{stochastic SINDy} on the \emph{full} dataset with $(\alpha^\dagger,\rho^\dagger)$, followed by the same restricted least-squares de-biasing on the selected support, and we report the in-sample mean squared error as a summary of fit quality.

\subsubsection{Optimally sparse size}
It is often informative to summarize CV results as a function of the active-set sizes
\[
n_\mu := \bigl\|\widehat{\Xi}_\mu\bigr\|_0, \qquad
n_\sigma := \bigl\|\widehat{\Xi}_\sigma\bigr\|_0,
\]
where $\|\cdot\|_0$ counts nonzero entries [see e.g. \cite{efron2004least}, \cite{boninsegna2018}]. Grouping all grid points $(\alpha,\rho)$ that yield the same $n\in\mathbb{N}$, define
\[
\delta(n) \;=\; \min_{(\alpha,\rho):\, \|\widehat{\Xi}\|_0=n}\; \overline{\delta}(\alpha,\rho),
\qquad \widehat{\Xi}:=(\widehat{\Xi}_\mu,\widehat{\Xi}_\sigma).
\]
Empirically, $\delta(n)$ typically exhibits three regimes: (i) underfitting at very small $n$ (large error), (ii) a flat “good” region where accuracy is stable across several $n$, and (iii) overfitting at very large $n$ (rising error). We define the \emph{optimally sparse} sizes $\tilde n_\mu,\tilde n_\sigma$ as the smallest $n$ whose error lies within one standard error of the minimum of $\delta(n)$ (a size-based analogue of the 1-SE rule). We compare them with the sizes actually selected by $(\alpha^\dagger,\rho^\dagger)$. For completeness we also provide “$\delta$ vs.\ $n$’’ plots with $\pm1$\,SE bands for both $\mu$ and $\sigma$.

\subsubsection{Practical settings used in our experiments}
Unless stated otherwise, we use $k=7$ contiguous folds, a logarithmic grid $\Lambda=\{\alpha_{\min},\dots,\alpha_{\max}\}$ spanning several decades for $\alpha$, and $\mathcal{P}=\{0.3,0.4,\\ 0.5,0.7,0.85,0.95,1.0\}$ for $\rho$ (with $\rho=1.0$ recovering Lasso). Feature blocks are built as
\[
\Theta_\mu(X_{t_i}) \;\Delta t_i
\quad\text{and}\quad
\Theta_\sigma(X_{t_i}) \;\Delta B^{\mathbb P}_{t_i},
\]
with RMS column normalization applied separately to each block on the training portion of each fold. All reported coefficients are de-biased by restricted least squares on the selected support. Alongside $(\alpha^\dagger,\rho^\dagger)$ and $(\tilde n_\mu,\tilde n_\sigma)$, we provide the final model size, validation curves/surfaces, and \emph{$\delta$ vs.\ $n$} summaries to document the sparsity–accuracy trade-off achieved by \eqref{stochastic SINDy} in each example.

\subsubsection{Methods of estimating \(\psi_{21}\) and hence \(\mu(X_t)\) }\label{subsubsec:psi12_methods}
Here we present two methods to estimate the parameter \(\psi_{21}\) from \eqref{SDE 2 order expansion}. The estimation of \(\psi_{21}\) will lead to estimating \(\mu(X_t)\) using \ref{Euler_for_mu}.
\begin{description}
  \item[\textbf{Method~1:}]\label{psi21_method1}%
  To begin with, we approximate \(\psi_{21}\) by regressing \(X_t\) against the iterated integrals in \eqref{SDE 2 order expansion} for each \(t \in [t_{j_k}, t_{j_{k+1}})\). The coefficient of \(\int_{t_{j_k}}^{t_{j_{k+1}}}\int_{t_{j_k}}^u\circ   dB^{\mathbb Q}_r du\) is then extracted to get an estimate of \(\psi_{21}\) on each interval. This procedure is repeated for \(k: k=0,\dots, M-1\). 
  \item[\textbf{Method~2:}]\label{psi21_method2}
  Secondly, to avoid the regression error from fitting \(X_t\)  against the iterated integral terms, we use the closed-form coefficient of \(\int_s^t\int_s^u\circ   dB^{\mathbb Q}_r du\) from Eq.\eqref{eq29} to compute \(\psi_{21}\). Since the coefficient only depends on \(\sigma(X_t)\), no additional data fitting is required. We simply use the value of \(\sigma(X_t)\) and its first two derivatives at \(t = t_k\) for \(k=0,\dots, M-1\) to recover  \(\psi_{21}\) on each sub-interval \([t_{j_k}, t_{j_{k+1}})\).
\end{description}
Once \(\psi_{21}\) is recovered, we compute \(b(X_{t_k}) = \frac{2\psi^{t_k,t_{k+1}}_{21}+\sigma(X_{t_k}) \sigma'(X_{t_k})^2 +\sigma(X_{t_k})^2 \sigma''(X_{t_k})}{2\sigma(X_{t_k})}\) and substitute this value into the Euler scheme \eqref{Euler_for_mu} to estimate \(\mu(X_{t_k})\) for \(k=0,\dots, M-1\).

In addition to estimating the drift \(\mu(X_t)\) using
\hyperref[psi21_method1]{Method~1} and \hyperref[psi21_method2]{Method~2} with the Euler scheme \eqref{Euler_for_mu} for the first-order ODE, we construct an estimator based on the second-order ODE \eqref{ODE2}, as described below.

\begin{description}
    \item[\textbf{Estimator of \(\mu(X_t)\) via ODE2:}]\label{psi21_method3} We estimate the drift \(\mu(X_t)\) using the second-order ODE \eqref{ODE2} and refer to this approach as the \emph{ODE2 method}. Its coefficients depend only on the estimated diffusion coefficient \(\sigma\) and its first two derivatives. Solving this ODE requires the initial values \(\mu(X_{t_0})\) and \(\mu'(X_{t_0})\). Introducing \(\lambda(x)=\mu'(x)\), we rewrite \eqref{ODE2} as the first-order system
    \begin{align} \label{ODE system}
        \mu'(x) &= \lambda(x), \nonumber\\
        \lambda'(x) &= A(x)\lambda(x)+B(x)\mu(x).
    \end{align}
    Given the initial estimates \(\mu(X_{t_0})\) and \(\mu'(X_{t_0})\), we apply the explicit Euler scheme
\begin{equation}\label{ODE_system_Euler_scheme}
    \begin{aligned}
        \mu(X_{t_k}) &= \mu(X_{t_{k-1}})+\lambda(X_{t_{k-1}})\,\Delta X_k,\\
        \lambda(X_{t_k}) &= \lambda(X_{t_{k-1}})+\bigl(A(X_{t_{k-1}})\lambda(X_{t_{k-1}})+B(X_{t_{k-1}})\mu(X_{t_{k-1}})\bigr)\,\Delta X_k,
    \end{aligned}
\end{equation}
for \(k=1,\dots,M-1\), where \(\Delta X_k=X_{t_k}-X_{t_{k-1}}\). This yields estimates \(\mu(X_{t_k})\) for \(k=1,\dots,M-1\).
\end{description}

Finally, we present and discuss the results of our numerical experiments for several classical examples.


\subsection{Time-shifted Brownian motion}
Consider the SDE
\begin{align}\label{eq:drifted_BM}
    dX_t = \mu \ dt + \sigma \ dB^{\mathbb{P}}_t,
\end{align}
where \(\mu > 0\) and \(\sigma > 0\) are constants. This SDE does not satisfy the ergodicity assumption. The choice of this particular SDE is motivated from the earliest model of stock price path first introduced by Bachelier in his thesis \cite{bachelier1900theorie}. The drift and diffusion coefficients are taken to be constant, with $\mu = 0.5$ and $\sigma = 1.0$ in our experiments. The explicit solution of the stochastic differential equation is  given by \(X_t = X_0 +\mu \ t + \sigma\ B^{\mathbb{P}}_t\).

\begin{figure}[H]
  \centering
  \includegraphics[width=.98\linewidth]{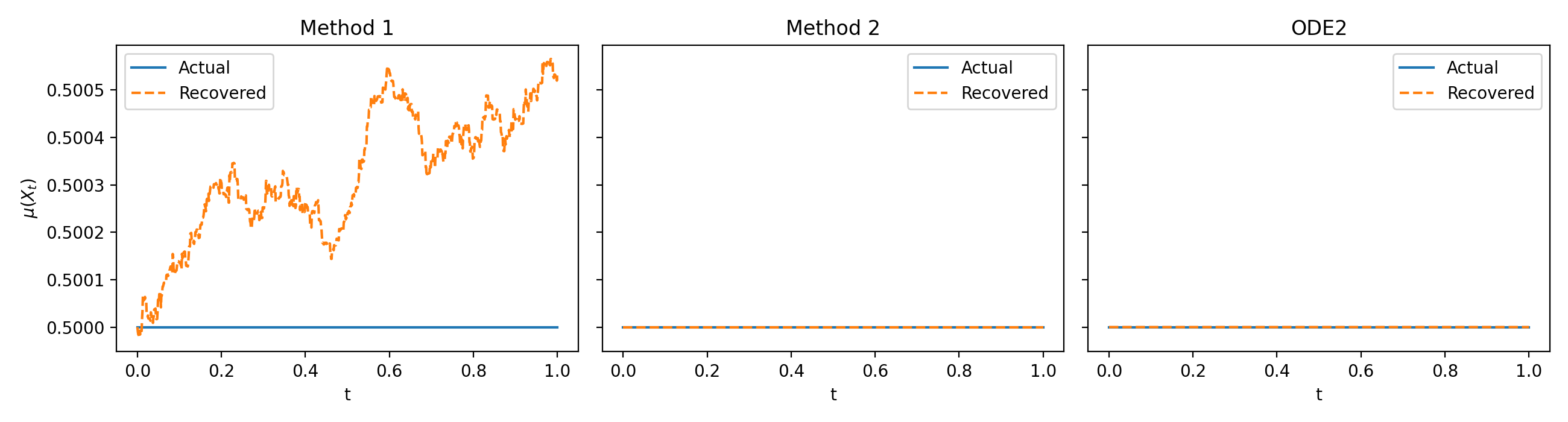}
  \caption{\textit{The true and the predicted drift \(\mu(X_t)\) using \hyperref[psi21_method1]{Method~1}, \hyperref[psi21_method2]{Method~2}, and \hyperref[psi21_method3]{ODE~2 Method}. Little deviation of the approximate value of \(\mu(X_t)\) in the \textbf{Left} figure from its true value is due to error in approximating \(\psi_{21}\) via regression. On the other hand, the approximate value of \(\mu(X_t)\) in the \textbf{middle} and the \textbf{right} figure exactly overlaps the true value of \(\mu(X_t)\).}}
  \label{drift_Euler}
\end{figure}

The solution \(X_t\) to Eq.~\eqref{eq:drifted_BM} was simulated using Euler-Maruyama scheme. It could also be achieved by using the exact solution of the SDE. Before using the penalized residual \eqref{stochastic SINDy} to identify the drift and diffusion function, the diffusion function was approximated, and the risk-neutral Brownian motion \(B_{t_i}^\mathbb{Q}\) was estimated for \(0\ \le i \le N\).
We then approximated the drift \(\mu(X_t)\) using \hyperref[psi21_method1]{Method~1}, \hyperref[psi21_method2]{Method~2}, and \hyperref[psi21_method3]{ODE~2 Method} as described in Subsection~\ref{subsubsec:psi12_methods}. The corresponding results are reported in Figure~\ref{drift_Euler}. 


We now look at how the choice of the quantity \(\psi_{21}\) affects the recovery of the drift, diffusion function, and noise from a single path trajectory. In both cases we work with the same penalized residual \eqref{stochastic SINDy} and use polynomial libraries
\[
\Theta_\mu(x)=\{1,x,\ldots,x^5\},\qquad
\Theta_\sigma(x)=\{1,x,\ldots,x^5\}.
\]
The regularization parameters \((\alpha,\rho)\) are chosen by \(k\)-fold time–series cross–validation with \(k=7\).  We search over a logarithmic grid for \(\alpha\) and a small discrete set of \(\rho\), then de–bias on the selected support and apply the \(1\)-SE rule to prefer simpler models.

\begin{figure}[H]
  \centering
\includegraphics[width=.78\linewidth]{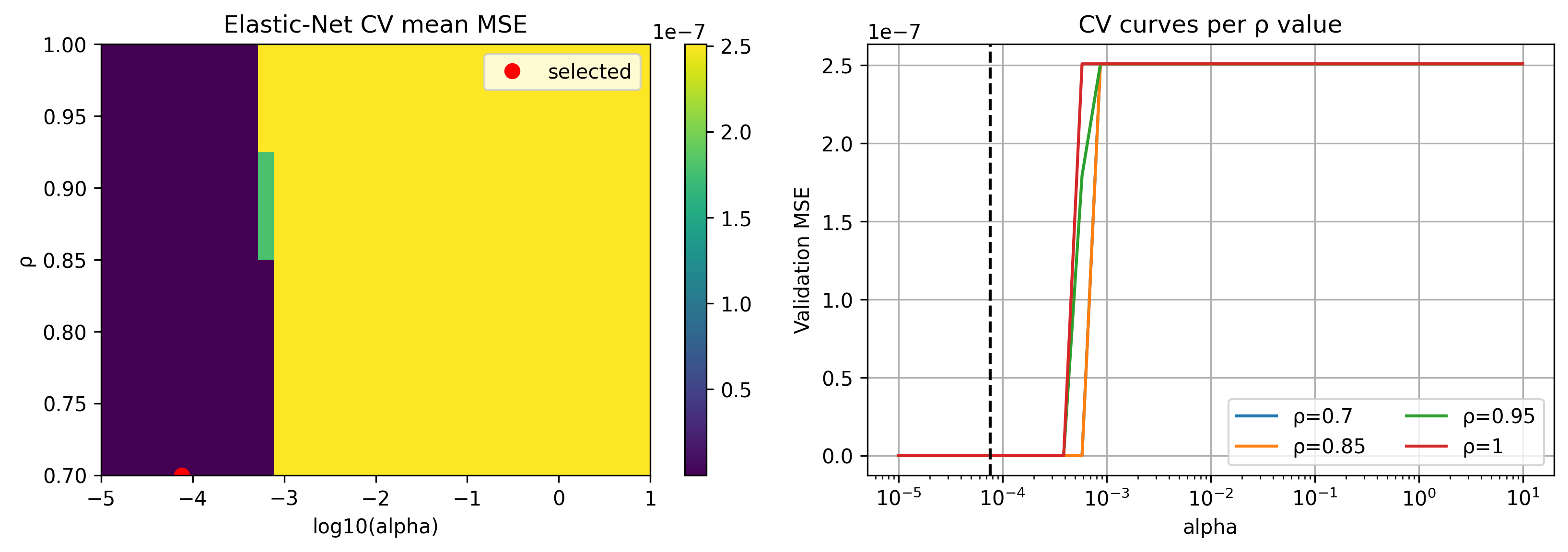}
  \caption{\textit{Time–series CV over $(\alpha,\rho)$ with de-biasing on the active set; the selected point $(\alpha^\dagger,\rho^\dagger)=(8\times 10^{-5}, 0.7)$ lies on a broad near-optimal basin.}}
  \label{fig:cv_driftedBM_reg}
\end{figure}

\begin{figure}[H]
  \centering
\includegraphics[width=.78\linewidth]{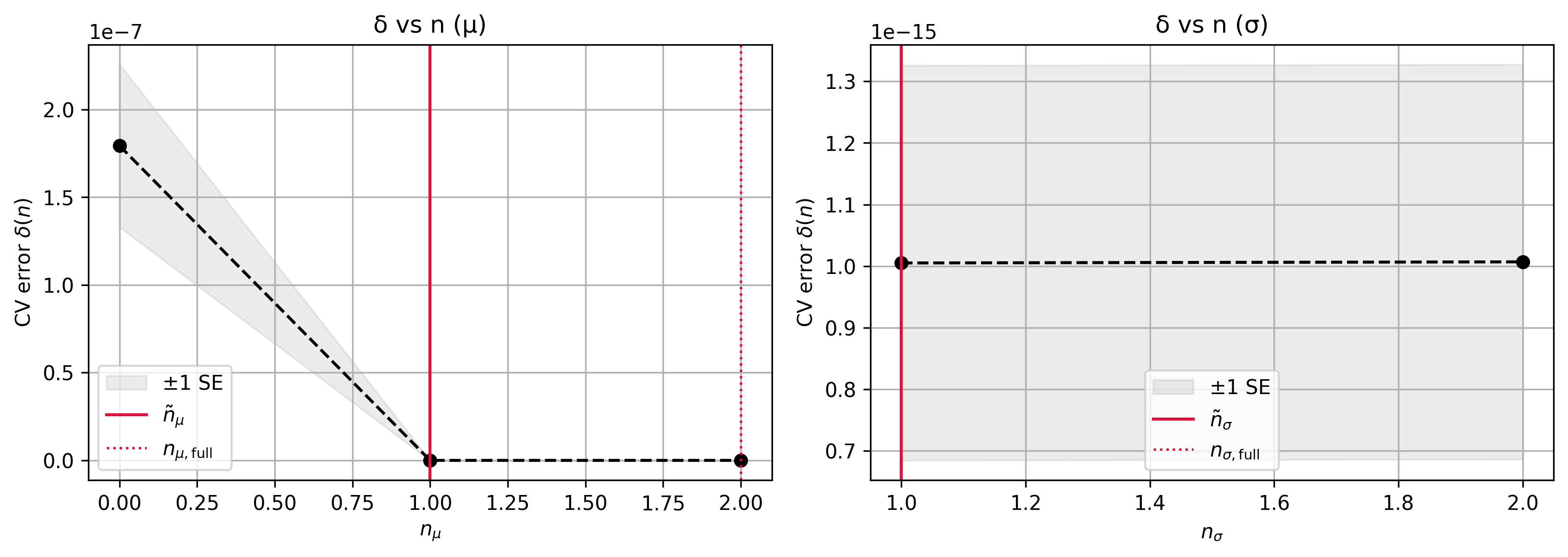}
  \caption{\textit{Validation error $\delta(n)$ vs.\ support size $n$ with $\pm1$\,SE bands for the drift and diffusion blocks. The vertical markers indicate the optimally sparse sizes and the sizes of the final full-data models.}}
  \label{fig:size_driftedBM_reg}
\end{figure}

\begin{figure}[H]
  \centering
  \includegraphics[width=.95\linewidth]{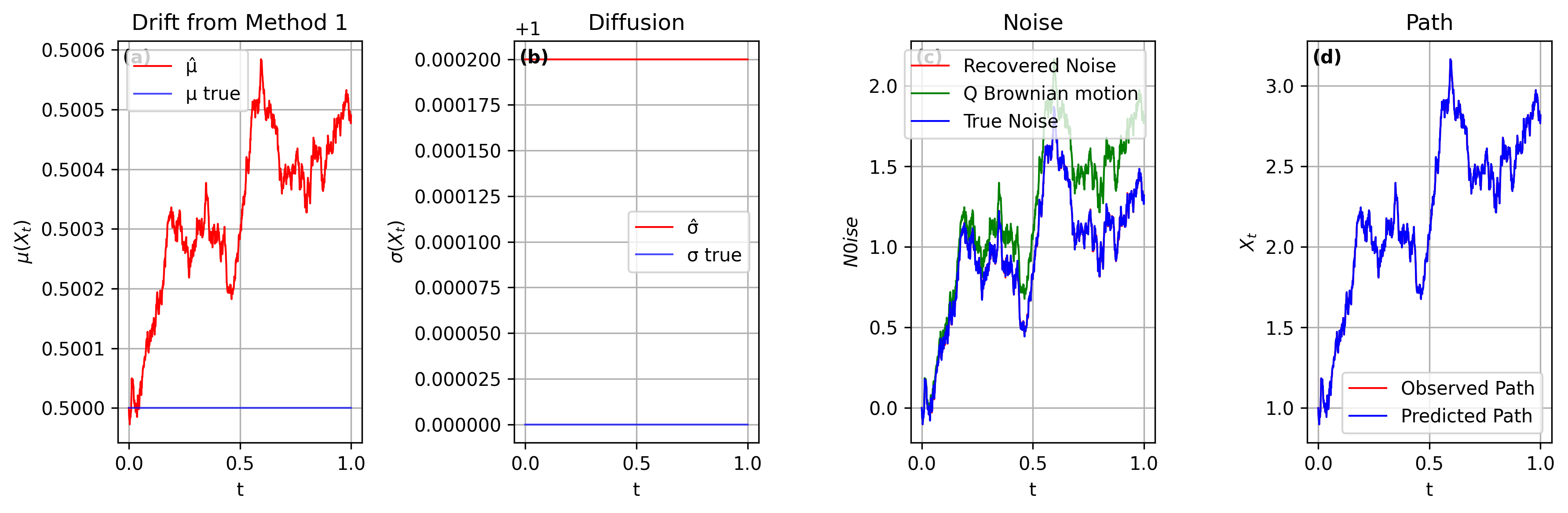}
  \caption{\textit{\textbf{(a)} The true \(\mu(X_t)\) and reconstructed  \(\widehat{\mu}(X_t)\) and \textbf{(b)} The true \(\sigma(X_t)\) and reconstructed  \(\widehat{\sigma}(X_t)\) are plotted against time on the interval \([0,1]\). \textbf{(c)} The true noise, i.e., Brownian motion \(B^{\mathbb{P}}\) that was used to simulate the actual path, the recovered noise, and the \(\mathbb{Q}\)-Brownian motion are plotted against time. \textbf{(d)} The actual path of the data \(X_t\) and the reconstructed one from the \(\widehat{\mu}(X_t)\) and \(\widehat{\sigma}(X_t)\) are plotted against time. Each vector is of length \(M\) with \(M=1000\).}}
  \label{fig:results_driftedBM_reg}
\end{figure}

\noindent\textbf{Case 1:} Here in the first experiment we use \hyperref[psi21_method1]{Method~1} to approximate the value of \(\psi_{21}\).  The cross–validation surface and curves (Fig.~\ref{fig:cv_driftedBM_reg}) show a wide region where the validation error is almost flat, and the chosen point \((\alpha^\dagger,\rho^\dagger)\) lies inside this near–optimal plateau and is \( (\alpha^\dagger,\rho^\dagger) \approx (8\times 10^{-5}, 0.7)\).  The \(\delta\)-versus-\(n\) summaries in Fig.~\ref{fig:size_driftedBM_reg} select \(n_\mu=1\) and \(n_\sigma=1\), so each block keeps only one non–zero term; adding more terms does not improve the error in any visible way. At these hyperparameters, the recovered drift and diffusion functions are
\[
\widehat{\mu}(x)=0.00027\,x + 0.49973,\qquad 
\widehat{\sigma}(x)=1.0002.
\]
They stay very close to the true constant values (see
Fig.~\ref{fig:results_driftedBM_reg}(a)–(b)), although the recovered drift shows some small fluctuations in time. In spite of this the recorded MSE was too small i.e., $\mathrm{MSE}=1.44870\times 10^{-14}$.  The reconstructed noise process tracks the driving Brownian motion quite well, and the path simulated under the learned model is practically indistinguishable from the observed trajectory (Fig.~\ref{fig:results_driftedBM_reg}(c)–(d)).  This suggests that the method can capture the main dynamics even when \(\psi_{21}\) is only approximated.


\begin{figure}[H]
  \centering
  \includegraphics[width=.78\linewidth]{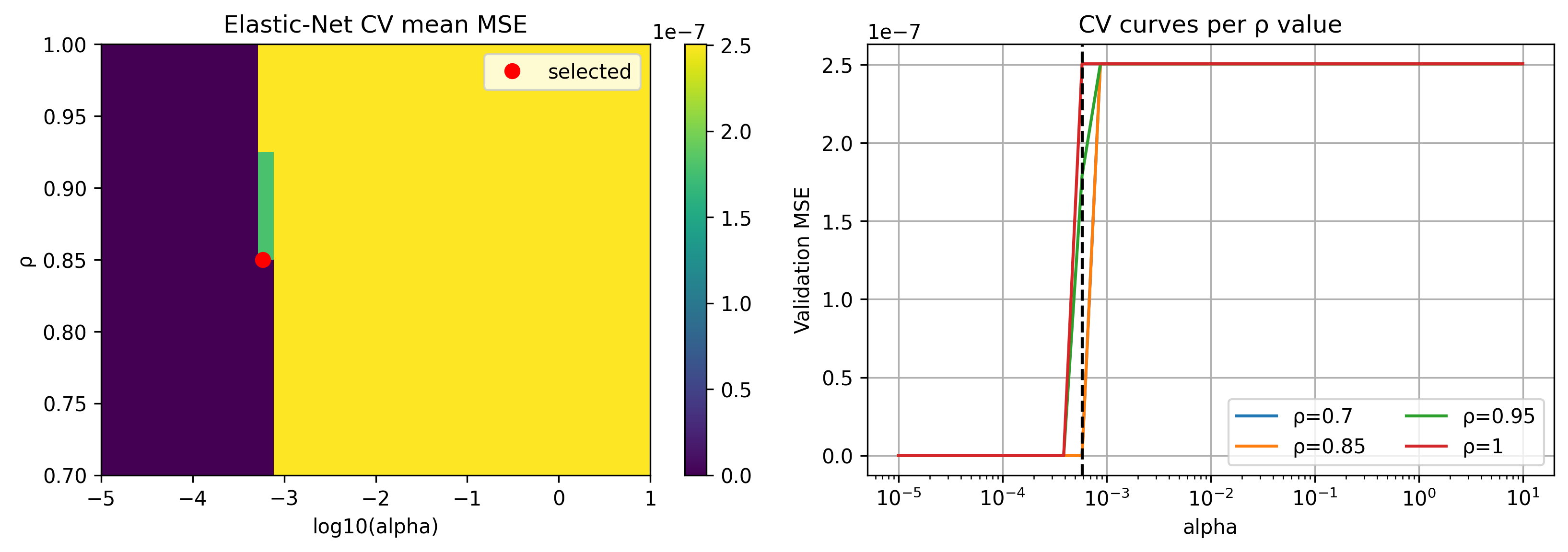}
  \caption{\textit{Time–series CV over $(\alpha,\rho)$ with de-biasing on the active set; the selected point $(\alpha^\dagger,\rho^\dagger)=(5.8\times 10^{-4},\,0.85)$ lies on a broad near-optimal basin.}}
  \label{fig:cv_driftedBM}
\end{figure}

\begin{figure}[H]
  \centering
  \includegraphics[width=.78\linewidth]{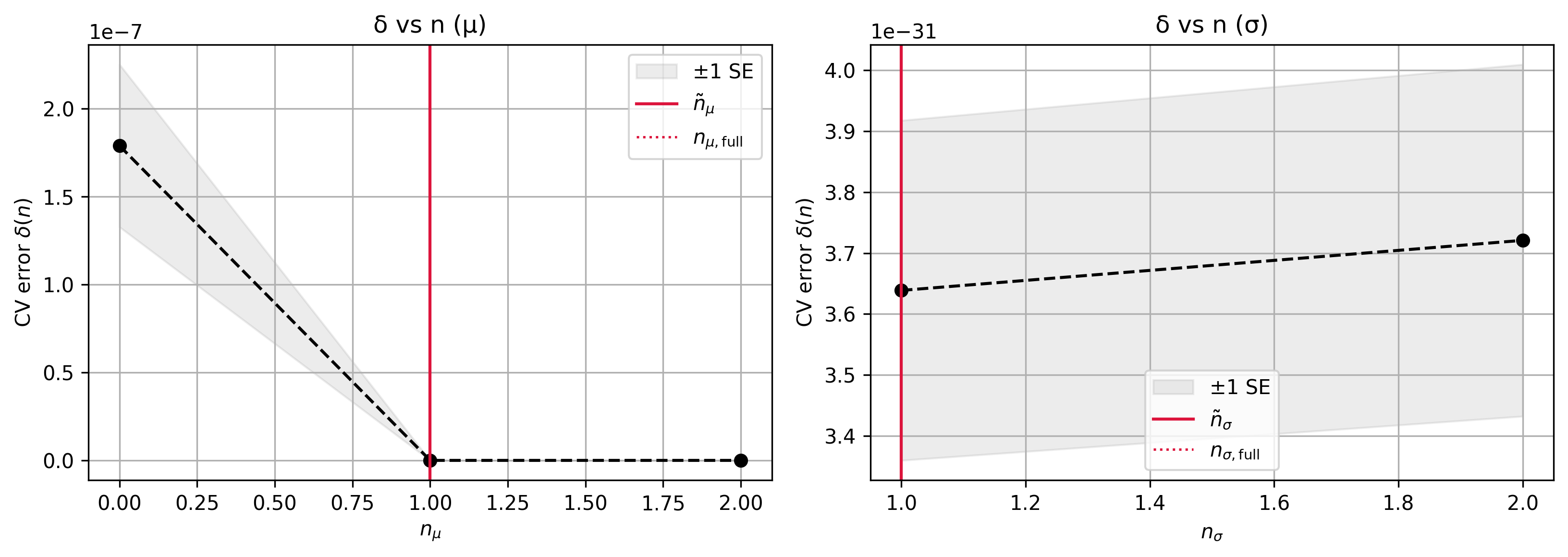}
  \caption{\textit{Validation error $\delta(n)$ vs.\ support size $n$ with $\pm1$\,SE bands for the drift and diffusion blocks. The vertical markers indicate the optimally sparse sizes and the sizes of the final full-data models.}}
  \label{fig:size_driftedBM}
\end{figure}

\noindent\textbf{Case 2:} In the second experiment, we repeat exactly the same procedure, but now we use the value of \(\psi_{21}\) computed by using \hyperref[psi21_method2]{Method~2}.  The cross–validation plots in Fig.~\ref{fig:cv_driftedBM} again display a broad flat region of low validation error, and the selected \((\alpha^\dagger,\rho^\dagger) = (5.8\times 10^{-4},\,0.85)\) is very much close to the one in the first case.  The corresponding \(\delta\)-versus-\(n\) curves in Fig.~\ref{fig:size_driftedBM} once more pick \(n_\mu=1\) and \(n_\sigma=1\).

\begin{figure}[H]
  \centering
  \includegraphics[width=.95\linewidth]{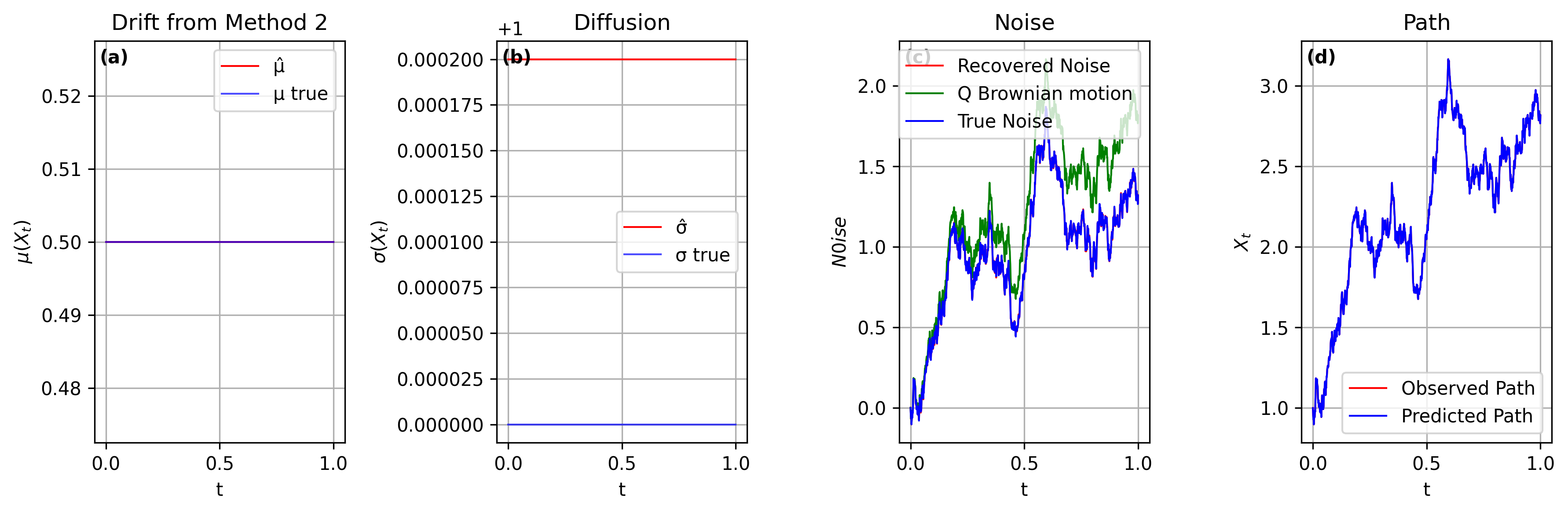}
  \caption{\textit{\textbf{(a)} The true \(\mu(X_t)\) and reconstructed  \(\widehat{\mu}(X_t)\) and \textbf{(b)} The true \(\sigma(X_t)\) and reconstructed  \(\widehat{\sigma}(X_t)\) are plotted against time on the interval \([0,1]\). \textbf{(c)} The true noise, i.e., Brownian motion \(B^{\mathbb{P}}\) that was used to simulate the actual path, the recovered noise, and the \(\mathbb{Q}\)-Brownian motion are plotted against time. \textbf{(d)} The actual path of the data \(X_t\) and the reconstructed one from the \(\widehat{\mu}(X_t)\) and \(\widehat{\sigma}(X_t)\) are plotted against time. Each vector is of length \(M\) with \(M=1000\).}}
  \label{fig:results_driftedBM}
\end{figure}

With this more accurate value of \(\psi_{21}\), the recovered drift and diffusion along the path are almost perfectly flat and lie on top of the true coefficients (Fig.~\ref{fig:results_driftedBM}(a)–(b)). The recovered drift and diffusion functions are
\[
\widehat{\mu}(x)= 0.5,\qquad 
\widehat{\sigma}(x)=1.0002,
\]
with very small $\mathrm{MSE}=3.497\times 10^{-31}$. The reconstructed noise now overlaps the driving noise very closely, and the simulated path matches the data path almost exactly (Fig.~\ref{fig:results_driftedBM}(c)–(d)). Compared with the regression case, using the exact \(\psi_{21}\) mainly smooths out the estimated coefficients; the overall picture of the dynamics remains the same.


\begin{figure}[H]
  \centering
  \includegraphics[width=.78\linewidth]{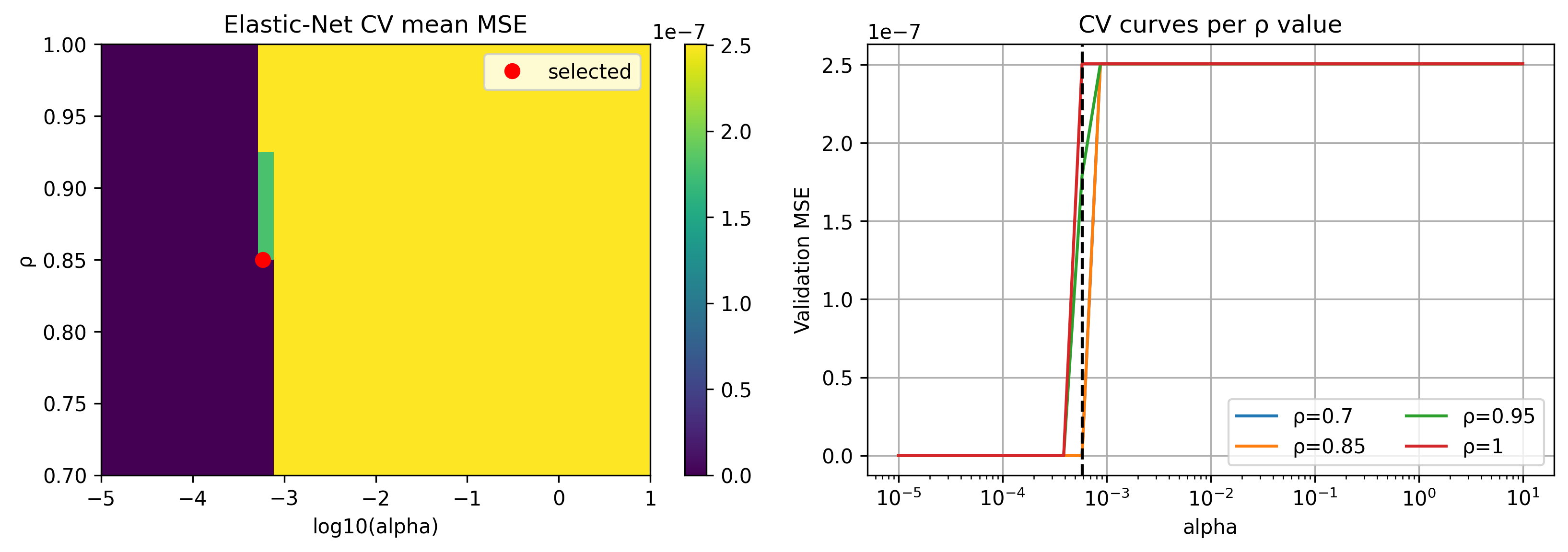}
  \caption{\textit{Time-series cross-validation over \((\alpha,\rho)\) with de-biasing on the active set, using noise reconstructed from the ODE2 drift estimate. The selected point \((\alpha^\dagger,\rho^\dagger)=(5.8\times10^{-4},\,0.85)\) lies within a broad region of low validation error.}}
  \label{fig:cv_driftedBM_ode2}
\end{figure}

\begin{figure}[H]
  \centering
  \includegraphics[width=.78\linewidth]{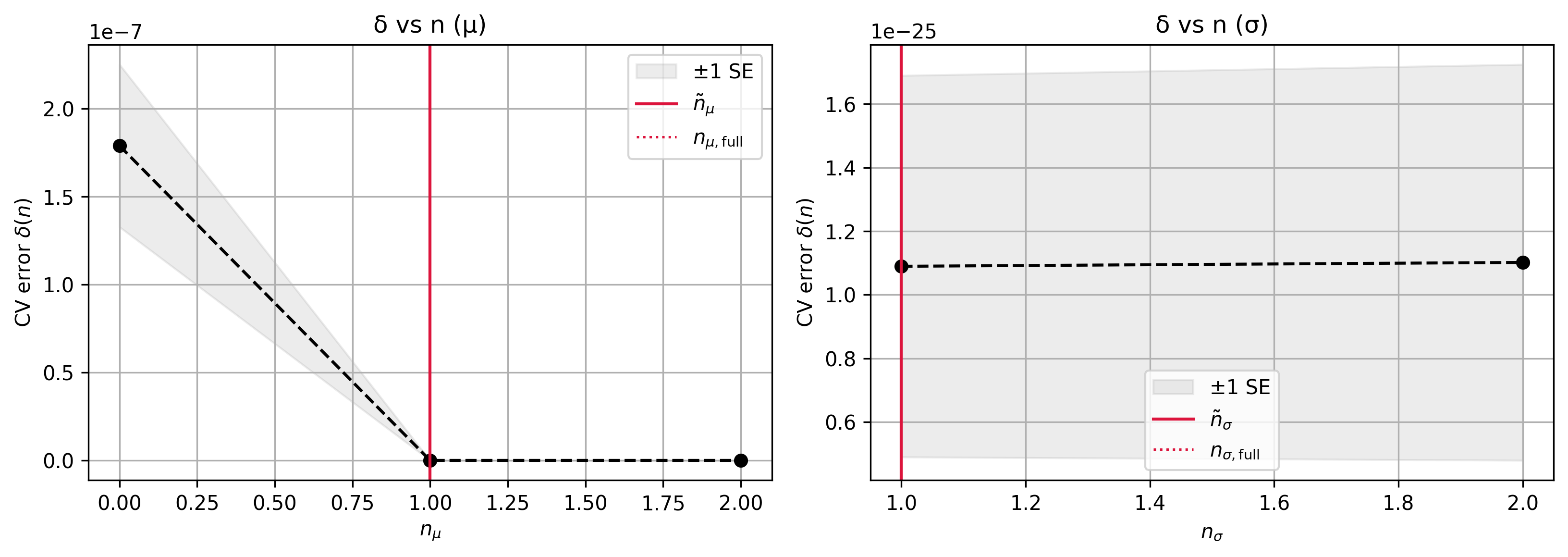}
  \caption{\textit{Validation error \(\delta(n)\) versus support size \(n\), with \(\pm1\)\,SE bands for the drift and diffusion blocks in the ODE2 case. The vertical markers indicate the selected sparse sizes and the sizes of the final full-data models. Both blocks retain one constant term: \(n_\mu=1\) and \(n_\sigma=1\).}}
  \label{fig:size_driftedBM_ode2}
\end{figure}

\noindent\textbf{Case 3:} In the third experiment, we follow the same procedure, but now use the drift recovered by the \hyperref[psi21_method2]{ODE~2 Method} to reconstruct the noise. The cross-validation plots in Fig.~\ref{fig:cv_driftedBM_ode2} again show a broad region of low validation error. The selected parameters are \((\alpha^\dagger,\rho^\dagger)=(5.8\times10^{-4},\,0.85)\), which agree with those selected in Case~2. The corresponding \(\delta\)-versus-\(n\) curves in Fig.~\ref{fig:size_driftedBM_ode2} select \(n_\mu=1\) and \(n_\sigma=1\), retaining one constant term in each recovered function.

\begin{figure}[H]
  \centering
  \includegraphics[width=.95\linewidth]{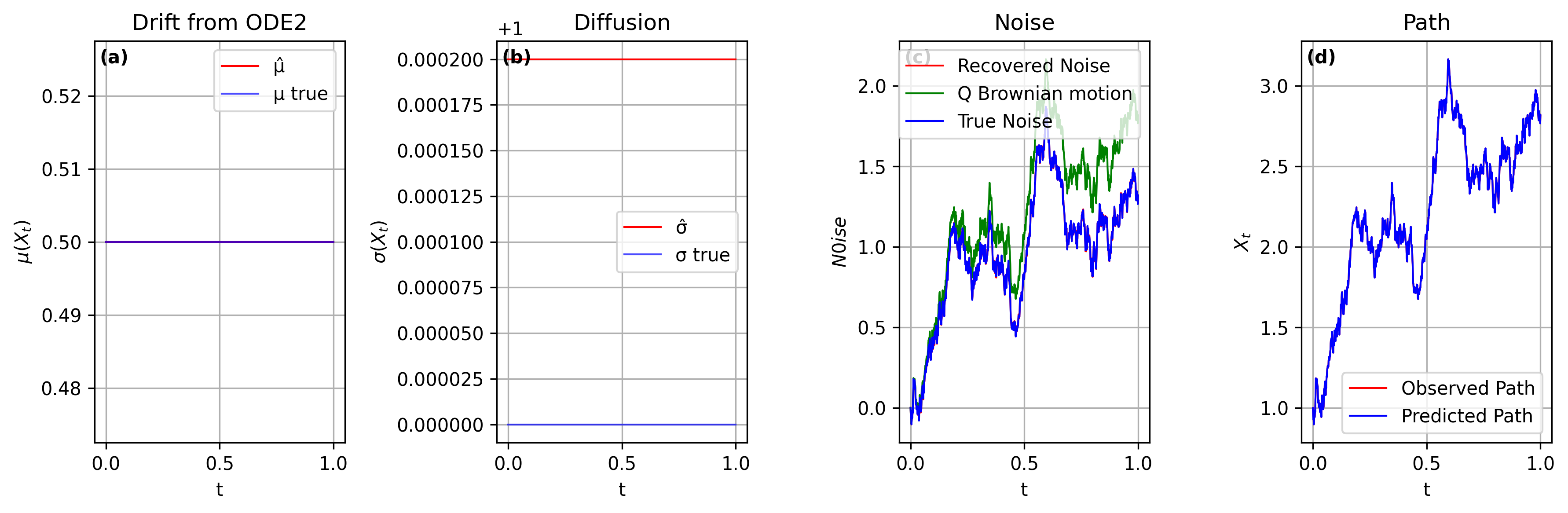}
  \caption{\textit{Recovery of the constant-coefficient model using the ODE2 method. \textbf{(a)} True drift \(\mu(X_t)\) and recovered drift \(\widehat{\mu}(X_t)=0.50000\). \textbf{(b)} True diffusion \(\sigma(X_t)\) and recovered diffusion \(\widehat{\sigma}(X_t)=1.00020\). \textbf{(c)} The driving Brownian motion \(B^{\mathbb{P}}\), the recovered noise, and the \(\mathbb{Q}\)-Brownian motion. \textbf{(d)} The observed path \(X_t\) and the path reconstructed using the recovered coefficients and noise. All quantities are plotted against time on \([0,1]\). The final prediction error is \(\mathrm{MSE}=1.35643\times10^{-14}\).}}
  \label{fig:results_driftedBM_ode2}
\end{figure}

The recovered drift and diffusion remain constant along the path, with the drift matching its true value and the diffusion lying slightly above its true value (Fig.~\ref{fig:results_driftedBM_ode2}(a)--(b)). The recovered functions are
\[
\widehat{\mu}(x)=0.50000,\qquad \widehat{\sigma}(x)=1.00020,
\]
with a final prediction error of \(\mathrm{MSE}=1.35643\times10^{-14}\). The reconstructed noise closely follows the original Brownian motion, and the predicted path is nearly indistinguishable from the observed path (Fig.~\ref{fig:results_driftedBM_ode2}(c)--(d)). At the reported precision, ODE2 therefore gives the same recovered coefficients and final MSE as Method~2, confirming consistent recovery of the constant-coefficient dynamics.

\subsection{Black--Scholes model}
Consider the Black--Scholes SDE 
\begin{align}\label{eq:BS}
    dX_t = \mu \ X_t \ dt + \sigma\ X_t \ dB^{\mathbb{P}}_t,
\end{align}
where \(\mu\) and \(\sigma>0\) are two constants, and the solution is a Geometric Brownian motion. This is an example of an SDE that does not satisfy the ergodicity assumption. It is the widely used dynamics for stock prices in finance. The idea traces back to Osborne and Samuelson’s lognormal price hypothesis \cite{osborne1959brownian} and was formalized in the Black--Scholes--Merton framework \cite{black1973} as geometric Brownian motion. The drift term is linear in the price, \( \mu X_t \), and shows a proportional (percentage) expected rate of return. The diffusion term is also linear, \( \sigma X_t\, dB^{\mathbb{P}}_t \), and \(\sigma\) is taken positive because it represents standard deviation, and any negative sign can be absorbed by flipping the Brownian motion. In short, the drift part governs the systematic trend while the diffusion part quantifies how volatile the price is around that trend.

\begin{figure}[H]
  \centering
  \includegraphics[width=.98\linewidth]{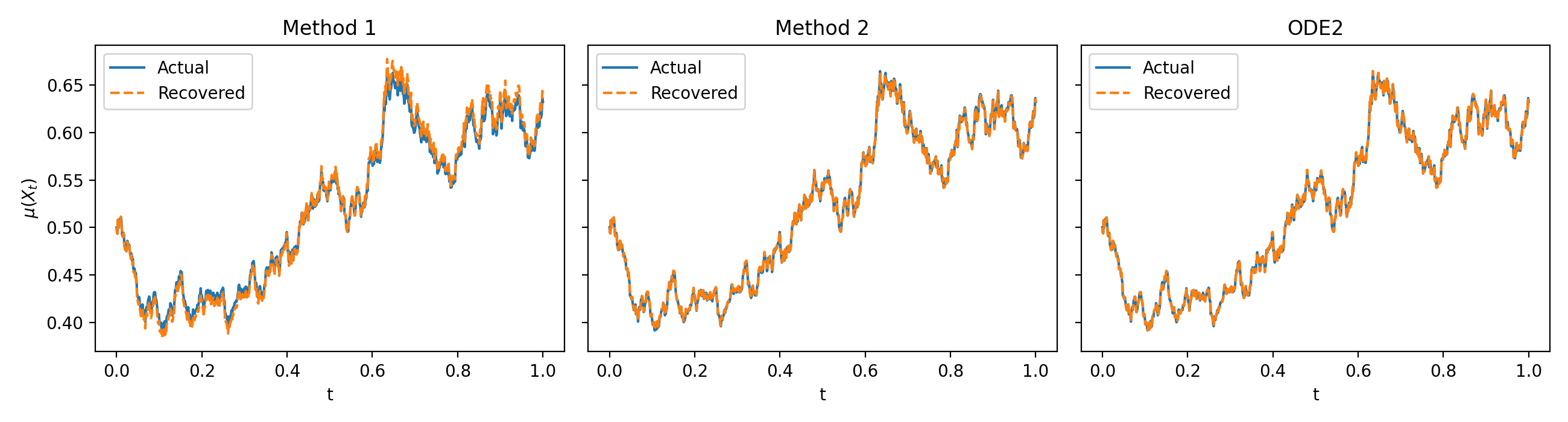}
  \caption{\textit{Comparison of the true drift \(\mu(X_t)\) with its reconstructions using two approximations of \(\psi_{21}\): (Left) \(\psi_{21}\) obtained via \hyperref[psi21_method1]{Method~1}; (Center) \(\psi_{21}\) computed using \hyperref[psi21_method2]{Method~2}, and its comparison with the drift reconstructed by using \hyperref[psi21_method3]{ODE~2 Method}.  The small discrepancy in the \textbf{Left} panel is attributed to regression error in \(\psi_{21}\). 
  In the \textbf{Center} and the \textbf{Right} panel, the recovered drift overlaps the true drift.}}
  \label{drift_Euler_BS}
\end{figure}

The path \(X_t\) solving the Black--Scholes SDE \eqref{eq:BS} was simulated using the Euler--Maruyama scheme by taking \(\mu =0.5\) and \(\sigma=0.3\).  Similar to the previous example, before applying the penalized residual \eqref{stochastic SINDy} to recover the drift and diffusion, we first approximated the diffusion term and estimated the risk-neutral Brownian motion \(B_{t_i}^{\mathbb{Q}}\) for \(0 \le i \le N\).
We then estimated the drift \(\mu(X_t)\) using the two choices of \(\psi_{21}\) provided by
\hyperref[psi21_method1]{Method~1} and \hyperref[psi21_method2]{Method~2}, and by using \hyperref[psi21_method3]{ODE~2 Method}
(see Subsection~\ref{subsubsec:psi12_methods}). The resulting drift estimates are shown in
Figure~\ref{drift_Euler_BS}.



We now examine how the choice of \(\psi_{21}\) influences the recovery of the drift, diffusion, and noise from a single observed path. In both settings, we use the same penalized residual \eqref{stochastic SINDy} and the same polynomial libraries,
\[
\Theta_\mu(x)=\{1,x,\ldots,x^5\},\qquad
\Theta_\sigma(x)=\{1,x,\ldots,x^5\}.
\]
The parameters \((\alpha,\rho)\) are selected using \(k\)-fold time--series cross--validation with \(k=7\). 
We scan a logarithmic grid for \(\alpha\) and a small discrete set for \(\rho\). 
After selecting the active terms, we re-estimate the coefficients by restricted least squares to reduce bias. 
Finally, we use the \(1\)-SE rule to favor a simpler model when its performance is close to the best one.

\begin{figure}[H]
  \centering
  \includegraphics[width=.78\linewidth]{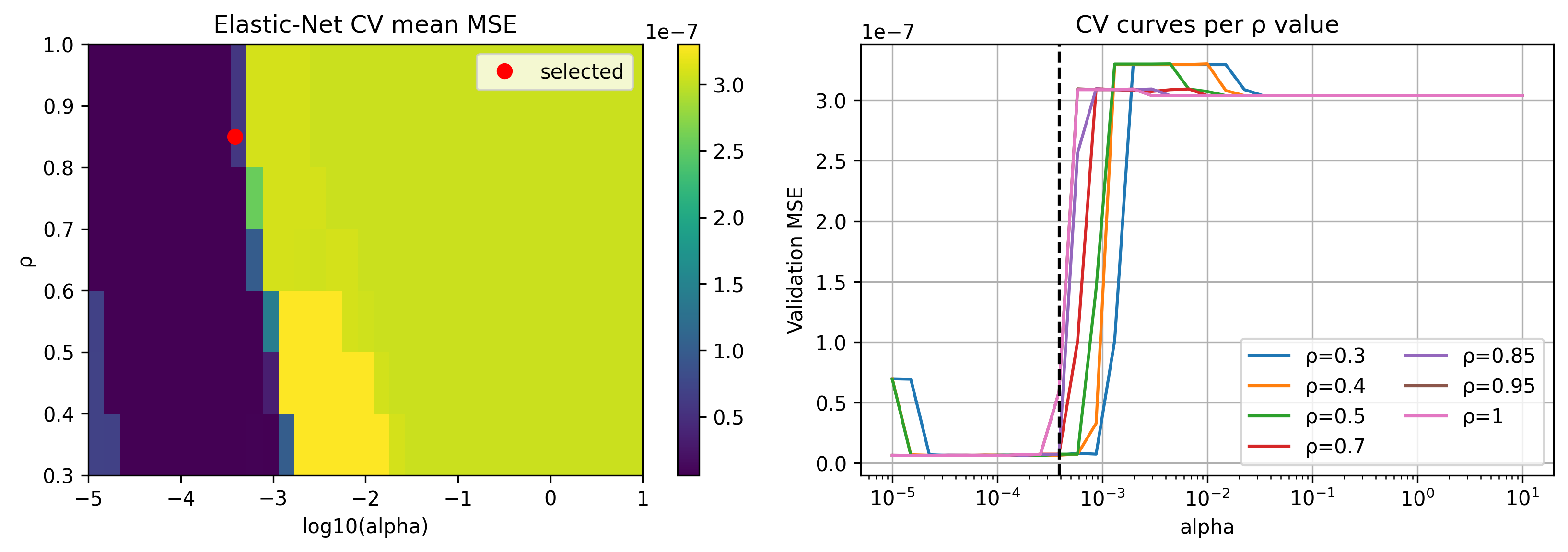}
  \caption{\textit{Time--series CV over \((\alpha,\rho)\) with de-biasing on the active set. The selected point 
  \((\alpha^\dagger,\rho^\dagger)\approx(3.9\times 10^{-4}, 0.85)\) lies in a broad near-optimal region.}}
  \label{fig:cv_BS_reg}
\end{figure}

\begin{figure}[H]
  \centering
  \includegraphics[width=.78\linewidth]{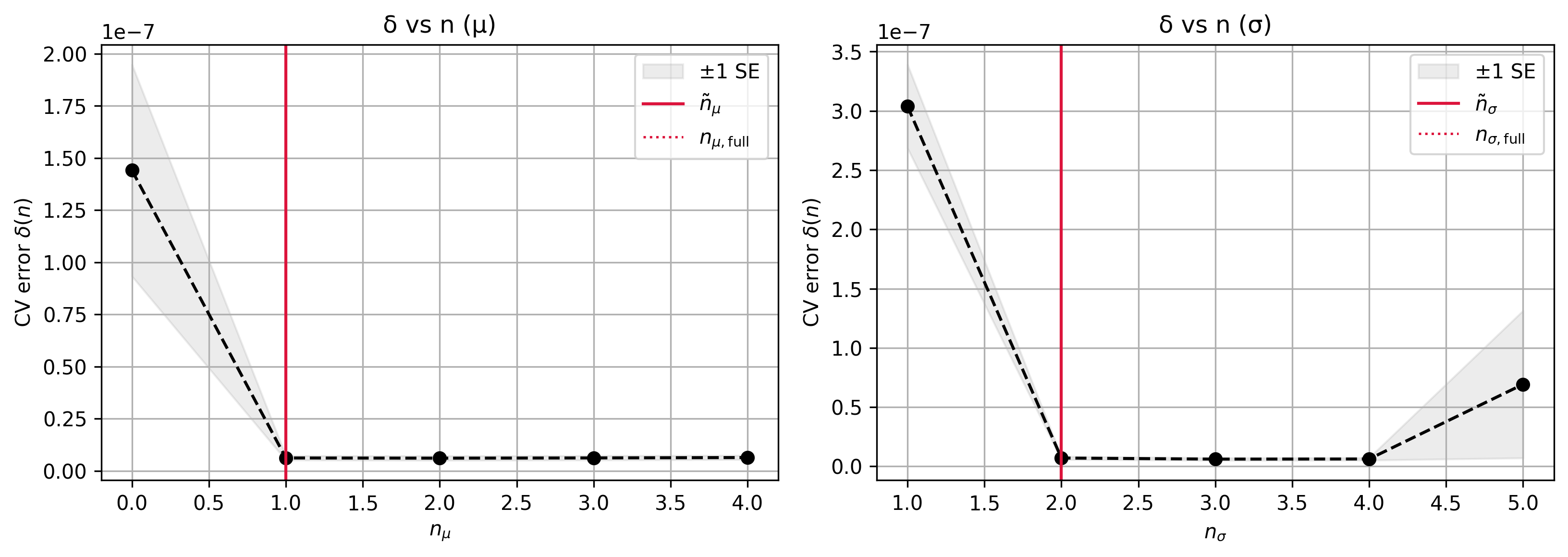}
  \caption{\textit{Validation error \(\delta(n)\) versus support size \(n\) with \(\pm 1\) SE bands for the drift and diffusion blocks. 
  The selected sizes are \(n_\mu=1\) and \(n_\sigma=2\), indicating sparse yet stable models.}}
  \label{fig:size_BS_reg}
\end{figure}

\begin{figure}[H]
  \centering
  \includegraphics[width=.95\linewidth]{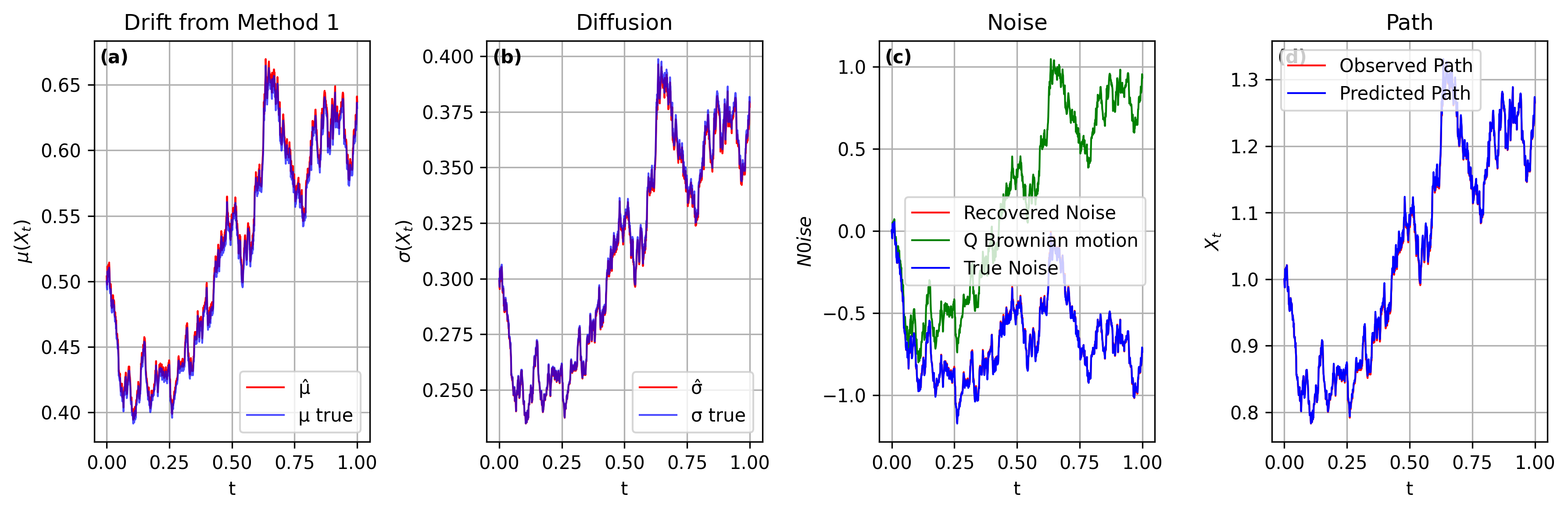}
  \caption{\textit{\textbf{(a)} True \(\mu(X_t)\) and reconstructed \(\widehat{\mu}(X_t)\). 
  \textbf{(b)} True \(\sigma(X_t)\) and reconstructed \(\widehat{\sigma}(X_t)\). 
  \textbf{(c)} True noise, recovered noise, and the \(\mathbb{Q}\)-Brownian motion. 
  \textbf{(d)} Observed path \(X_t\) and the path simulated using the recovered \(\widehat{\mu}(X_t)\) and \(\widehat{\sigma}(X_t)\). 
  Each vector is of length \(M\) with \(M=1000\).}}
  \label{fig:results_BS_reg}
\end{figure}

\noindent\textbf{Case 1:} In the first experiment, we use the value of \(\psi_{21}\) obtained by applying \hyperref[psi21_method1]{Method~1}. 
The cross--validation heatmap and the CV curves in Fig.~\ref{fig:cv_BS_reg} indicate a broad region where the validation error changes very little. 
The selected pair \((\alpha^\dagger,\rho^\dagger)\) lies inside this stable region and is approximately
\[
(\alpha^\dagger,\rho^\dagger) \approx \big(3.9\times 10^{-4},\,0.85\big).
\]
The \(\delta\)-versus-\(n\) plots in Fig.~\ref{fig:size_BS_reg} suggest \(n_\mu=1\) and \(n_\sigma=2\). 
This means the drift block keeps a single active term, while the diffusion block keeps two terms; adding more terms does not lead to a visible improvement in the validation error. 
With these hyperparameters, the recovered functions take the form
\[
\widehat{\mu}(x)=0.50377\,x,
\qquad 
\widehat{\sigma}(x)=0.00324 + 0.29576\,x.
\]
These estimates are close to the expected linear structure of the model, and the reconstructions of \(\mu(X_t)\) and \(\sigma(X_t)\) track the true curves well; see Fig.~\ref{fig:results_BS_reg}(a)--(b). 
The overall fit is strong, with a small mean squared error \(\mathrm{MSE}=5.24352\times 10^{-9}\). 
The recovered noise follows the driving signal closely, and the path generated using \(\widehat{\mu}\) and \(\widehat{\sigma}\) is almost indistinguishable from the observed trajectory; see Fig.~\ref{fig:results_BS_reg}(c)--(d). 
Overall, these results indicate that the method can still recover the main dynamics when \(\psi_{21}\) is obtained via regression.


\begin{figure}[H]
  \centering
  \includegraphics[width=.78\linewidth]{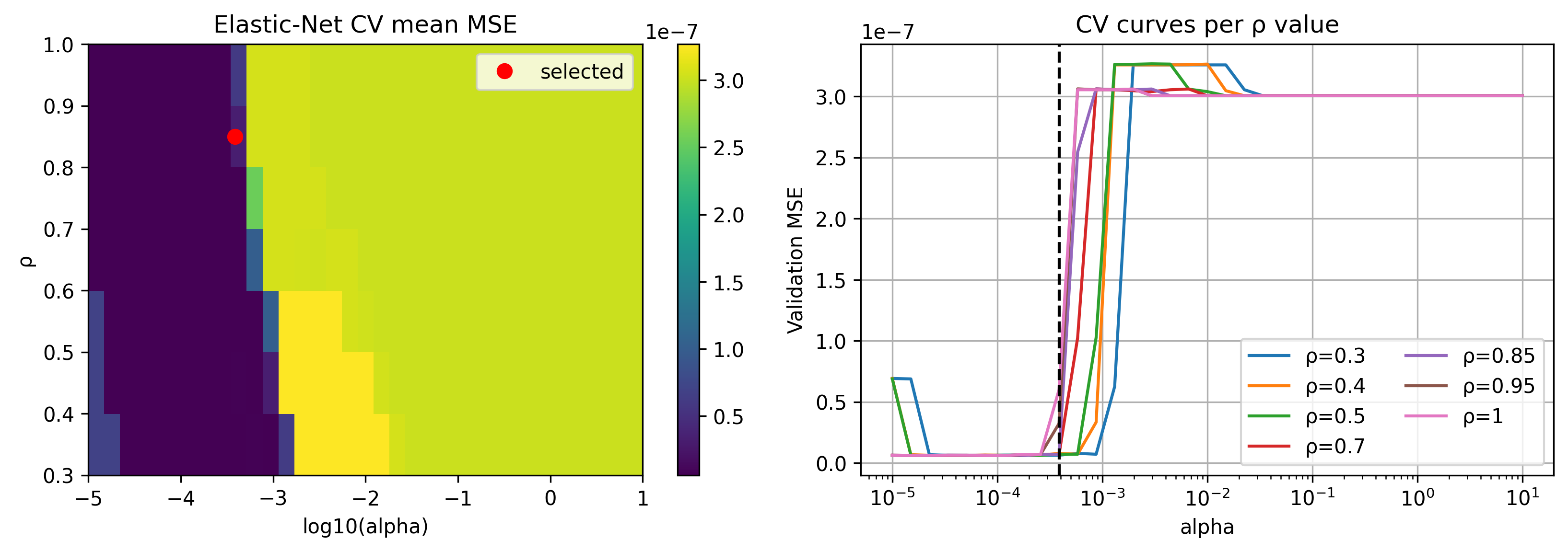}
  \caption{\textit{Time--series CV over \((\alpha,\rho)\) with de-biasing on the active set. 
  The selected point \((\alpha^\dagger,\rho^\dagger)\approx(3.9\times 10^{-4},\,0.85)\) lies in a broad near-optimal region.}}
  \label{fig:cv_BS_true}
\end{figure}

\begin{figure}[H]
  \centering
  \includegraphics[width=.78\linewidth]{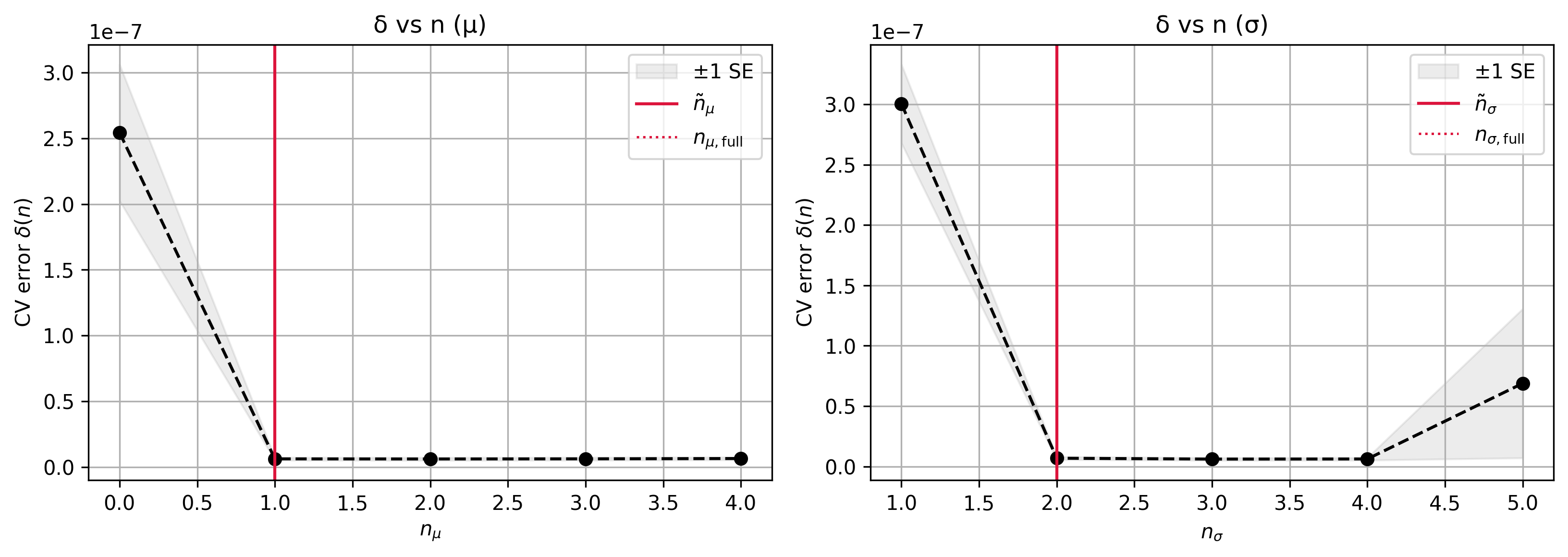}
  \caption{\textit{Validation error \(\delta(n)\) versus support size \(n\) with \(\pm 1\) SE bands for the drift and diffusion blocks. 
  The selected sparse sizes are \(n_\mu=1\) and \(n_\sigma=2\).}}
  \label{fig:size_BS_true}
\end{figure}

\noindent\textbf{Case 2:} In the second experiment, we follow the same steps as in the first one, but now we use the value of \(\psi_{21}\) computed using \hyperref[psi21_method2]{Method~2}. 
The cross--validation heatmap and CV curves in Fig.~\ref{fig:cv_BS_true} again show a wide region where the validation error is almost unchanged. 
The selected hyperparameters lie inside this stable region and are
\[
(\alpha^\dagger,\rho^\dagger)\approx\big(3.9\times 10^{-4},\,0.85\big),
\]
which is very close to the choice obtained when \(\psi_{21}\) is estimated by regression. 
The \(\delta\)-versus-\(n\) summaries in Fig.~\ref{fig:size_BS_true} select \(n_\mu=1\) and \(n_\sigma=2\), 
so the learned model remains sparse; adding more terms does not visibly reduce the CV error.

\begin{figure}[H]
  \centering
  \includegraphics[width=.95\linewidth]{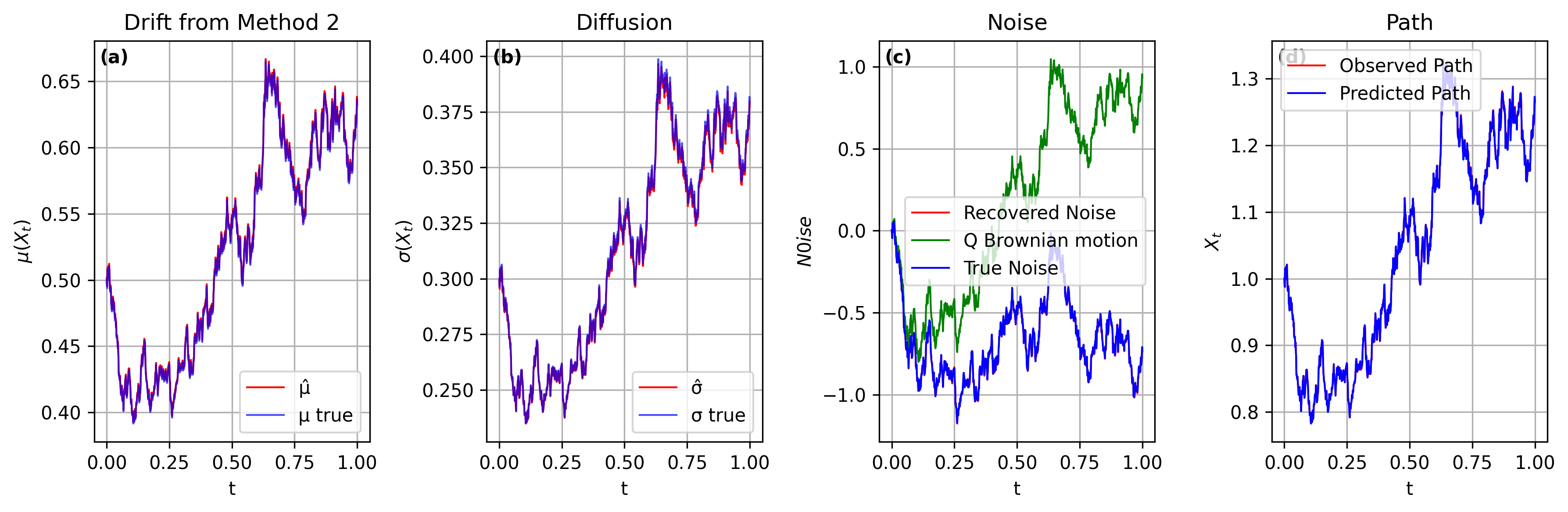}
  \caption{\textit{\textbf{(a)} True and recovered drift \(\mu(X_t)\). 
  \textbf{(b)} True and recovered diffusion \(\sigma(X_t)\). 
  \textbf{(c)} True noise, recovered noise, and the \(\mathbb{Q}\)-Brownian motion. 
  \textbf{(d)} Observed and reconstructed paths of \(X_t\). 
  Each vector has length \(M=1000\).}}
  \label{fig:results_BS_true}
\end{figure}

With this analytic \(\psi_{21}\), the recovered drift and diffusion are
\[
\widehat{\mu}(x)=0.50169\,x,
\qquad 
\widehat{\sigma}(x)=0.00318 + 0.29582\,x,
\]
with a small mean squared error \(\mathrm{MSE}=5.1917\times 10^{-9}\). 
Along the observed trajectory, the reconstructed \(\mu(X_t)\) and \(\sigma(X_t)\) track the true curves closely 
(Fig.~\ref{fig:results_BS_true}(a)--(b)). 
The recovered noise also follows the driving signal well, and the simulated path under the learned model nearly overlaps the observed path 
(Fig.~\ref{fig:results_BS_true}(c)--(d)). 
Overall, using the analytic \(\psi_{21}\) mainly refines the estimates slightly, while the main dynamics remain consistent with the regression-based case.


\begin{figure}[H]
  \centering
  \includegraphics[width=.78\linewidth]{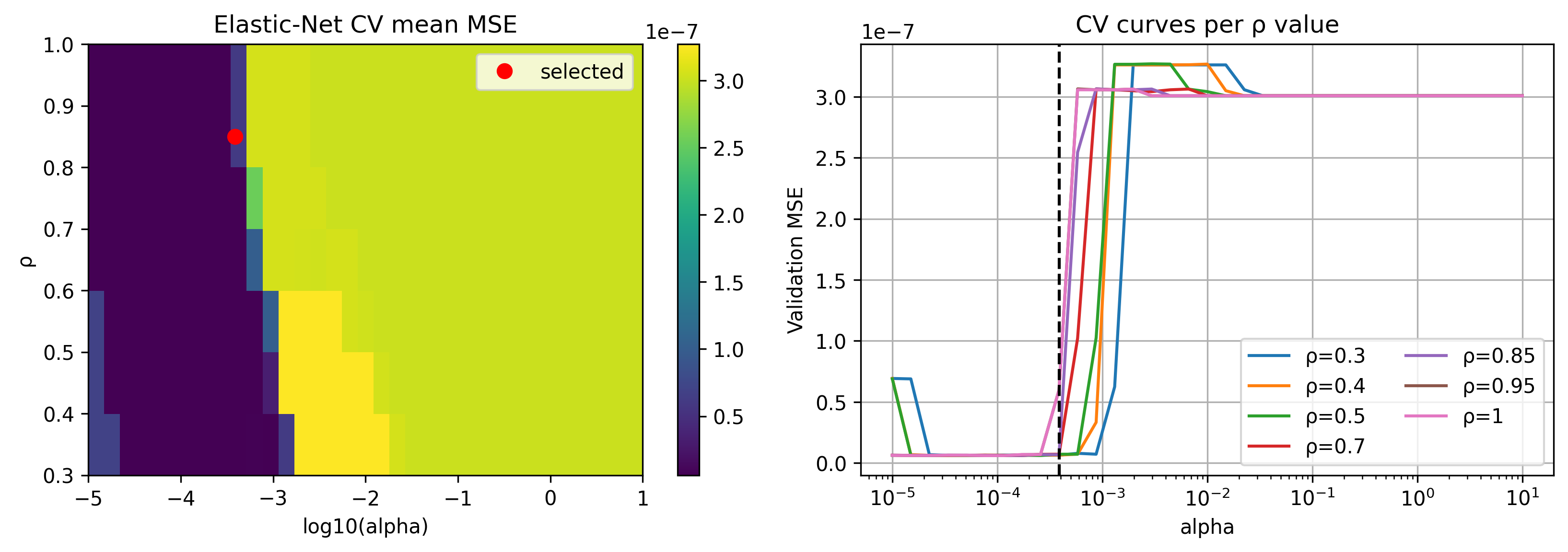}
  \caption{\textit{Time--series CV over \((\alpha,\rho)\) with de-biasing on the active set. 
  Again, the selected point \((\alpha^\dagger,\rho^\dagger)\approx(3.9\times 10^{-4},\,0.85)\) lies in a broad near-optimal region.}}
  \label{fig:cv_BS_ode2}
\end{figure}

\begin{figure}[H]
  \centering
  \includegraphics[width=.78\linewidth]{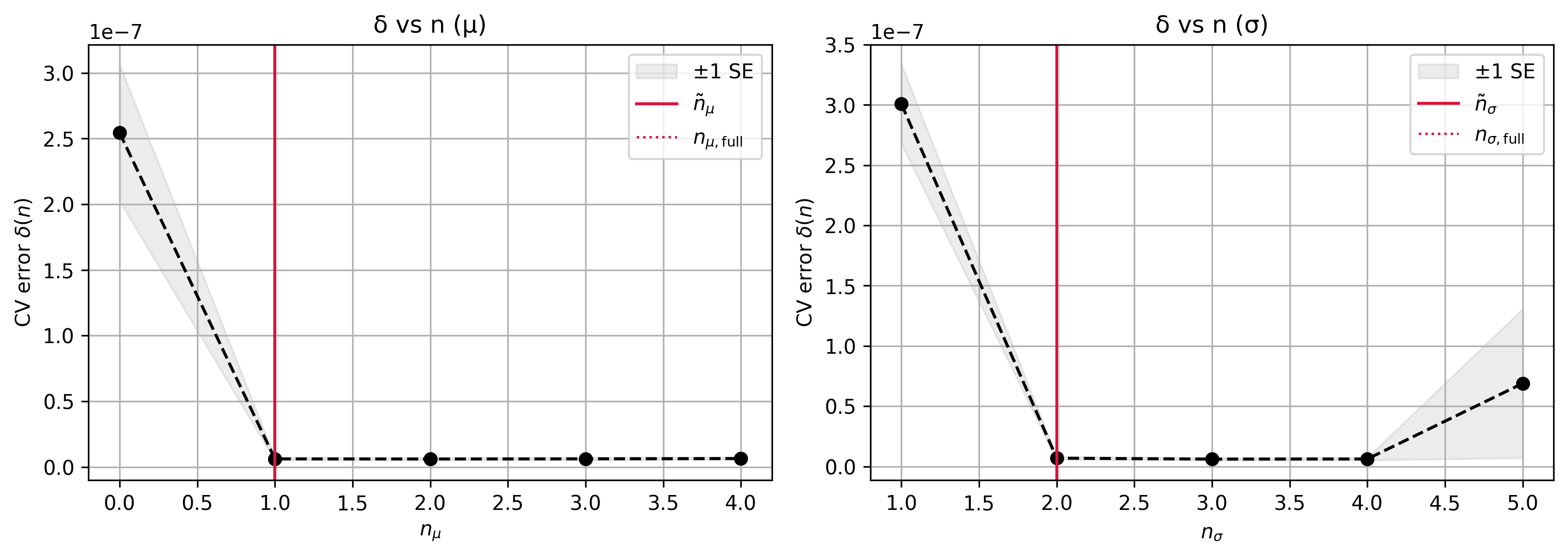}
  \caption{\textit{Validation error \(\delta(n)\) versus support size \(n\) with \(\pm 1\) SE bands for the drift and diffusion blocks. 
  The selected sparse sizes are \(n_\mu=1\) and \(n_\sigma=2\).}}
  \label{fig:size_BS_ode2}
\end{figure}

\noindent\textbf{Case 3:} In the third experiment, we reconstruct the drift function \(\mu(X_t)\) via \hyperref[psi21_method3]{ODE~2 Method}. 
The CV heatmap and curves are given in Fig.~\ref{fig:cv_BS_ode2}. A wide region can be seen where the validation error is almost constant. 
The selected values of \(\alpha\) and \(\rho\) lie inside the stable region and are
\[
(\alpha^\dagger,\rho^\dagger)\approx\big(3.9\times 10^{-4},\,0.85\big).
\]
The values are similar to the ones we recovered in \textbf{Case 1} and \textbf{Case 2}. Fig.~\ref{fig:size_BS_ode2} shows the \(\delta\)-versus-\(n\) summaries. The Model remains sparse and adding more terms does not help too much.

\begin{figure}[H]
  \centering
  \includegraphics[width=.95\linewidth]{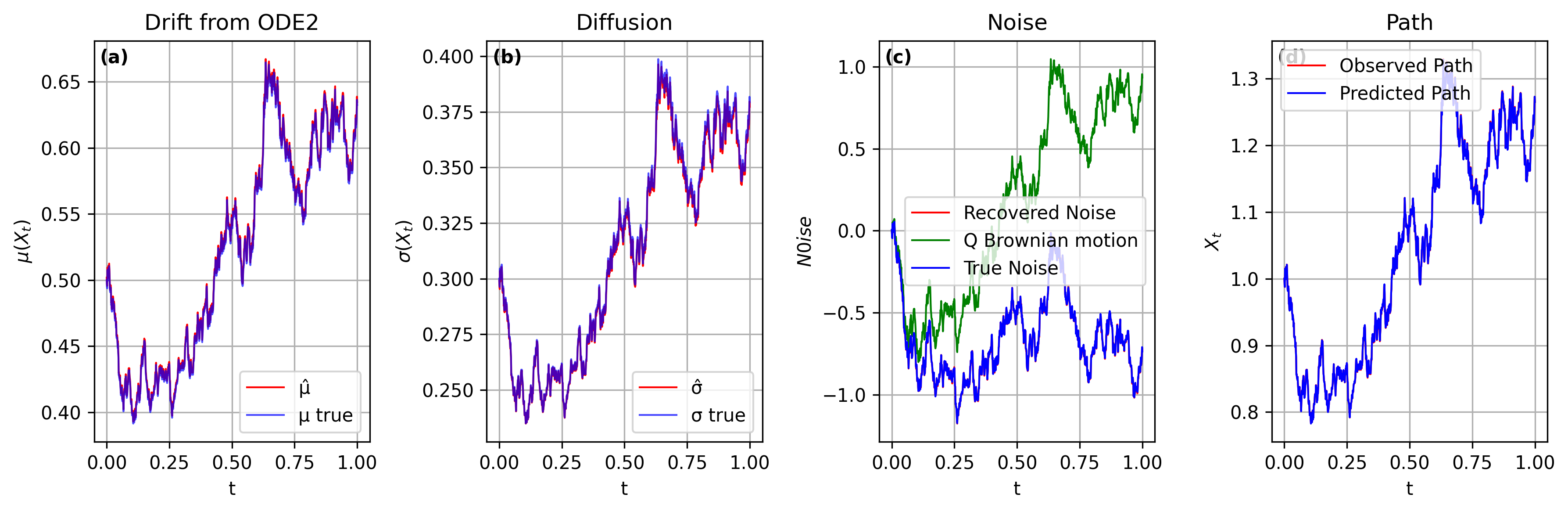}
  \caption{\textit{\textbf{(a)} True and recovered drift \(\mu(X_t)\). 
  \textbf{(b)} True and recovered diffusion \(\sigma(X_t)\). 
  \textbf{(c)} True noise, recovered noise, and the \(\mathbb{Q}\)-Brownian motion. 
  \textbf{(d)} Observed and reconstructed paths of \(X_t\). 
  Each vector has length \(M=1000\).}}
  \label{fig:results_BS_ode2}
\end{figure}

Here in this \hyperref[psi21_method3]{ODE~2 Method}, the recovered drift and diffusion are
\[
\widehat{\mu}(x)=0.50186\,x,
\qquad 
\widehat{\sigma}(x)=0.00319 + 0.29581\,x,
\]
with a small mean squared error \(\mathrm{MSE}=5.19375\times 10^{-9}\). Again, the reconstructed drift \(\mu(X_t)\) and diffusion \(\sigma(X_t)\) follow the true drift and diffusion very closely (Fig.~\ref{fig:results_BS_ode2}(a)--(b)). The recovered noise remains close to the true noise \(\mathbb{P}\)-Brownian motion too (Fig.~\ref{fig:results_BS_ode2}(c)). Finally, the path recovered by using the reconstructed drift and diffusion overlaps the observed path (Fig.~\ref{fig:results_BS_ode2}(d)). Overall, using the \hyperref[psi21_method3]{ODE~2 Method} refines the estimates slightly from \hyperref[psi21_method1]{Method~1} but remains very close to \hyperref[psi21_method2]{Method~2}.

\subsection{Quadratic coefficients SDE}
In this example, we study the SDE with quadratic drift and diffusion,
\begin{equation}\label{eq:quadSDE-ours}
dX_t \;=\; \underbrace{\bigl(0.7\,X_t + 0.3\,X_t^2\bigr)}_{\mu(X_t)}\,dt
\;+\;
\underbrace{\bigl(0.5\,X_t + 0.1\,X_t^2\bigr)}_{\sigma(X_t)}\,dB^{\mathbb{P}}_t,
\end{equation}
which belongs to the broader class of polynomial/CEV–type diffusions \cite{davydov2001pricing}. 
Unlike geometric Brownian motion, \eqref{eq:quadSDE-ours} features state–dependent nonlinearities in \emph{both} the drift and the diffusion. 

Because \eqref{eq:quadSDE-ours} has quadratic drift and diffusion with positive coefficients, the model may in principle exhibit explosive behavior over long time intervals. Since our goal is identification on the finite interval \([0,1]\), we adopt a simple safeguard in the simulation step to ensure that the generated trajectory remains finite i.e., we impose large absorbing bounds \([-L,\,L]\) such that \(-L < X_0 < L \) . Specifically, if the simulated path \(X_t\) ever hits either boundary \(\pm L\), we set \(X_t=0\) for the remainder of the time interval. This truncation acts only as a numerical protection against rare large deviations that could otherwise dominate the regression and produce unstable estimates. In our experiments, the trajectory does not reach \(\pm L\), indicating that the simulated sample is non-explosive on \([0,1]\) and that the reported identification results reflect the genuine dynamics of \eqref{eq:quadSDE-ours} over the observation interval \([0,1]\).

In line with the previous examples, we illustrate the full identification pipeline on this one-dimensional SDE and recover both the drift and diffusion function from a single trajectory via the penalized residual \eqref{stochastic SINDy}. The feature libraries for drift and diffusion are polynomial dictionaries
\[
\Theta_\mu(x)=\{1,x,x^2,x^3,x^4,x^5\},\qquad 
\Theta_\sigma(x)=\{1,x,x^2,x^3,x^4,x^5\},
\]
so that the analytic basis functions used to generate the data are the first two non-constant monomials \([x,x^2]\) in each dictionary. Hyperparameters \((\alpha,\rho)\) are selected by \(k\)-fold time–series cross-validation with \(k=7\), using a logarithmic grid \(\alpha\in\{\mathrm{geomspace}(10^{-5},10^{1},35)\}\) and \(\rho\in\{0.3,0.4,0.5,0.7,0.85,0.95,1.0\}\). After each penalized fit on the training folds, we de-bias by restricted least squares on the active support before scoring on the validation fold, and apply the 1-SE rule to favor parsimony.

\begin{figure}[H]
  \centering
  \includegraphics[width=.98\linewidth]{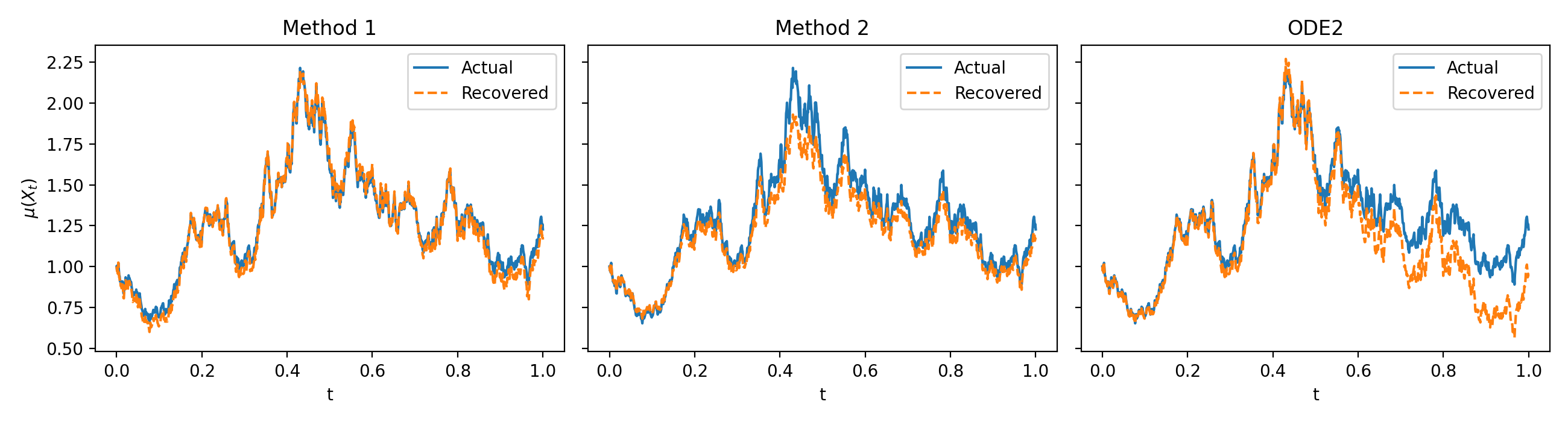}
  \caption{\textit{Comparison of the true drift \(\mu(X_t)\) with its reconstructions using two approximations of \(\psi_{21}\) and using \hyperref[psi21_method3]{ODE~2 Method}: 
  (Left) Actual and recovered \(\mu(X_t)\) when \(\psi_{21}\) obtained via \hyperref[psi21_method1]{Method~1}, (Center) \(\psi_{21}\) computed using \hyperref[psi21_method2]{Method~2}, and (Right) by using \hyperref[psi21_method3]{ODE~2 Method} .}}
  \label{drift_Euler_Quad}
\end{figure}

Following the same procedure as before, we approximate the diffusion vector and recover the \(\mathbb{Q}\)-Brownian motion \(B^{\mathbb{Q}}\). Next, we recover \(\mu(X_t)\) by plugging in \(\psi_{21}\) computed via \hyperref[psi21_method1]{Method~1} and \hyperref[psi21_method2]{Method~2}. We also recover \(\mu(X_t)\) via \hyperref[psi21_method3]{ODE~2 Method}
(see Subsection~\ref{subsubsec:psi12_methods}). Figure~\ref{drift_Euler_Quad} compares the two
drift estimates.



\begin{figure}[H]
  \centering
  \includegraphics[width=.78\linewidth]{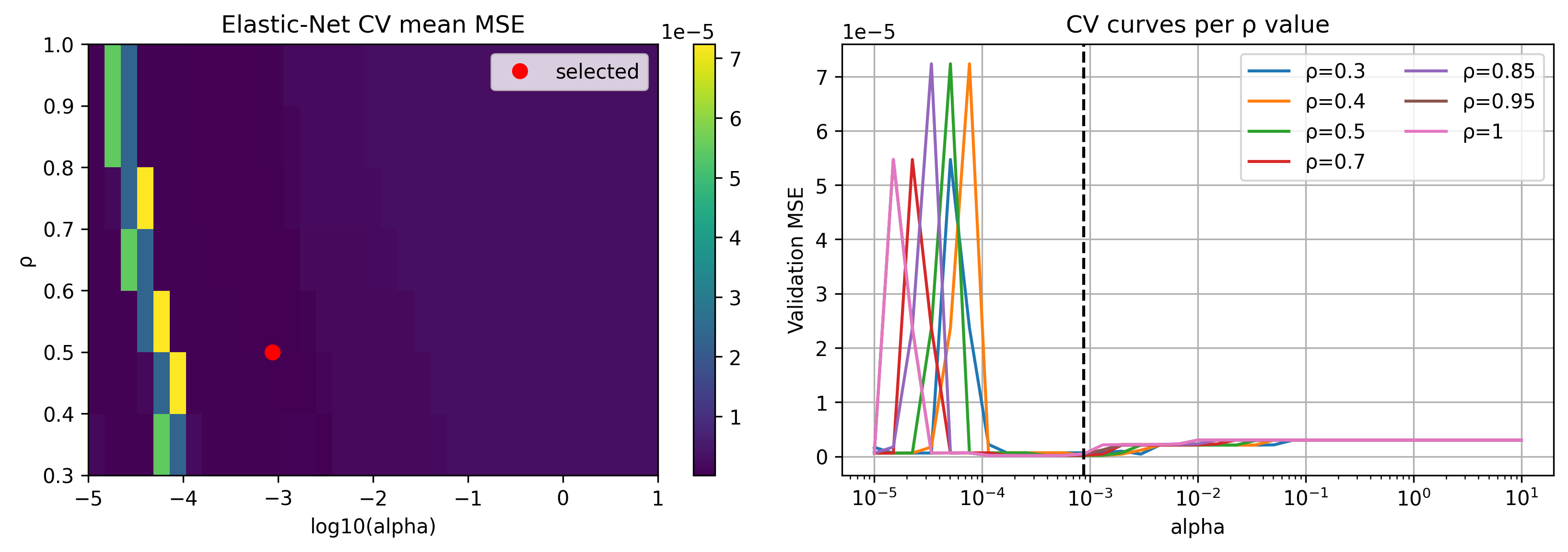}
  \caption{\textit{Time--series CV over \((\alpha,\rho)\) with de-biasing on the active set.
  The selected point \((\alpha^\dagger,\rho^\dagger)\approx(8.7\times 10^{-4}, 0.5)\) lies in a broad near-optimal region.}}
  \label{fig:cv_quad_reg}
\end{figure}

\begin{figure}[H]
  \centering
  \includegraphics[width=.78\linewidth]{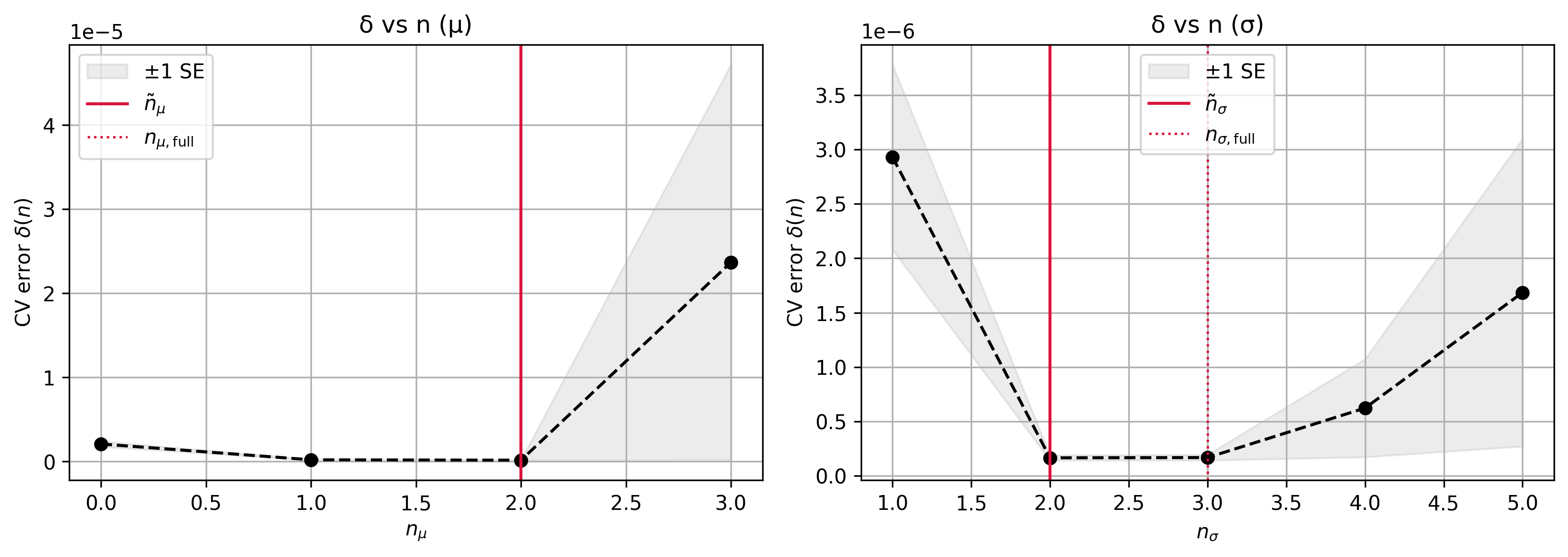}
  \caption{\textit{Validation error \(\delta(n)\) versus support size \(n\) with \(\pm 1\) SE bands for the drift and diffusion blocks.
  The selected sizes are \(n_\mu=2\) and \(n_\sigma=2\), indicating sparse yet stable polynomial models.}}
  \label{fig:size_quad_reg}
\end{figure}

\begin{figure}[H]
  \centering
  \includegraphics[width=.95\linewidth]{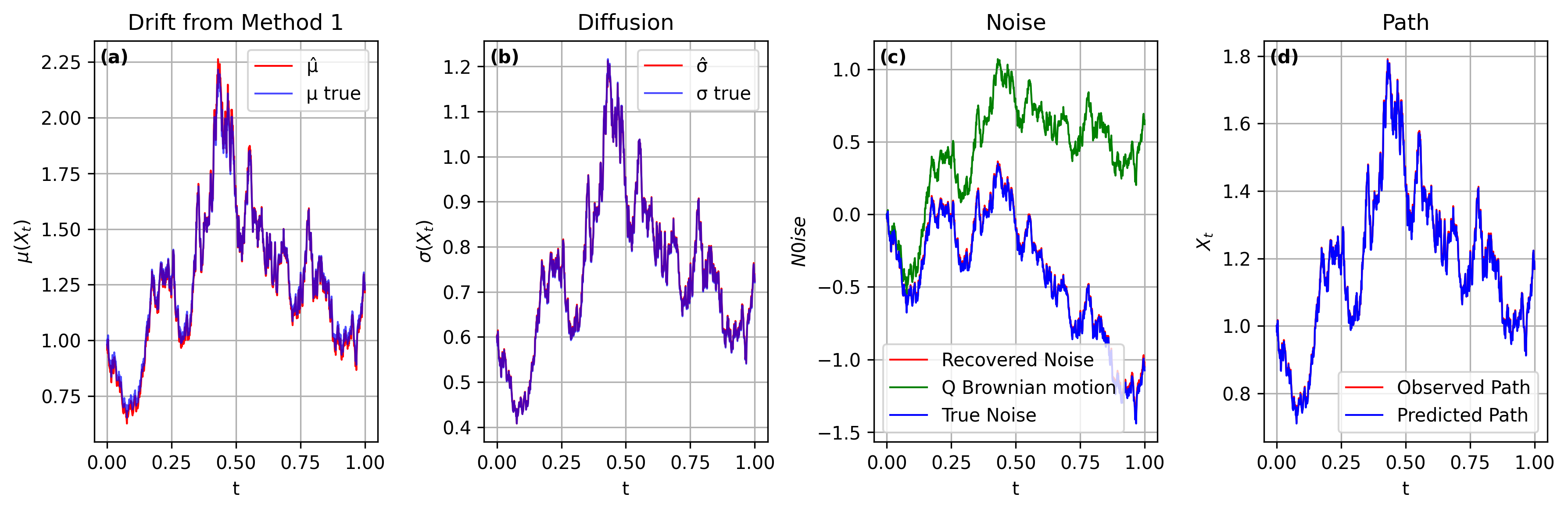}
  \caption{\textit{\textbf{(a)} True \(\mu(X_t)\) and reconstructed \(\widehat{\mu}(X_t)\).
  \textbf{(b)} True \(\sigma(X_t)\) and reconstructed \(\widehat{\sigma}(X_t)\).
  \textbf{(c)} True noise, recovered noise, and the \(\mathbb{Q}\)-Brownian motion.
  \textbf{(d)} Observed path \(X_t\) and the path simulated using the recovered
  \(\widehat{\mu}(X_t)\) and \(\widehat{\sigma}(X_t)\).
  Each vector is of length \(M\) with \(M=1000\).}}
  \label{fig:results_quad_reg}
\end{figure}

\noindent\textbf{Case 1:} To identify the drift and diffusion function, we first use the value of \(\psi_{21}\) obtained via \hyperref[psi21_method1]{Method~1} for the quadratic-coefficient SDE \eqref{eq:quadSDE-ours}. The cross--validation heatmap and the CV curves in Fig.~\ref{fig:cv_quad_reg} reveal a broad near-flat basin of low validation error around \(\alpha\sim 10^{-3}\) for moderate \(\rho\). Using the 1-SE rule, the selected pair \((\alpha^\dagger,\rho^\dagger)\) lies within this stable region and is approximately
\[
(\alpha^\dagger,\rho^\dagger) \approx \big(8.7\times 10^{-4},\,0.50\big).
\]
The \(\delta\)-versus-\(n\) plots in Fig.~\ref{fig:size_quad_reg} suggest sparse yet stable polynomial models with \(n_\mu=2\) and \(n_\sigma=2\) (the full-support choice for diffusion is larger). Thus, both the drift and diffusion blocks retain two active terms, and adding further terms does not yield a meaningful improvement in validation error. With these hyperparameters, the recovered functions are well captured by quadratic forms,
\[
\widehat{\mu}(x)=0.62051\,x + 0.35905\,x^2,  
\qquad 
\widehat{\sigma}(x)=- \, 0.01778 + 0.53598\,x + 0.08424\,x^2.
\]
These estimates align with the intended polynomial structure, and the reconstructions of \(\mu(X_t)\) and  \(\sigma(X_t)\) track the true curves closely; see Fig.~\ref{fig:results_quad_reg}(a)--(b).  The overall fit remains strong with a small mean squared error \(\mathrm{MSE}=1.44796\times 10^{-7}\). Moreover, the recovered noise follows the true driving noise well, and the path simulated using \(\widehat{\mu}\) and \(\widehat{\sigma}\) is almost indistinguishable from the observed trajectory; see Fig.~\ref{fig:results_quad_reg}(c)--(d). Overall, these results indicate that the method continues to recover the main dynamics reliably when \(\psi_{21}\) is obtained via regression, even in the presence of quadratic drift and diffusion structure.


\begin{figure}[H]
  \centering
  \includegraphics[width=.78\linewidth]{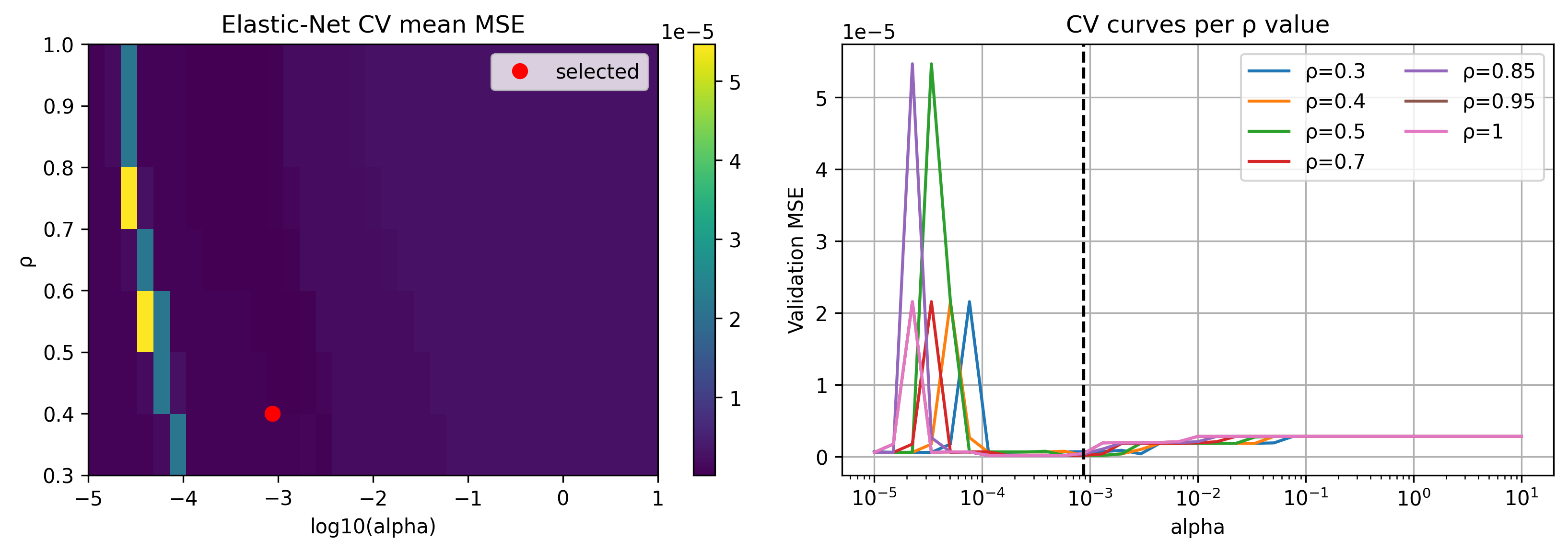}
  \caption{\textit{Time--series CV over \((\alpha,\rho)\) with de-biasing on the active set.
  The selected point \((\alpha^\dagger,\rho^\dagger)\approx(8.7\times 10^{-4},\,0.40)\) lies in a broad near-optimal region.}}
  \label{fig:cv_quad_true}
\end{figure}

\begin{figure}[H]
  \centering
  \includegraphics[width=.78\linewidth]{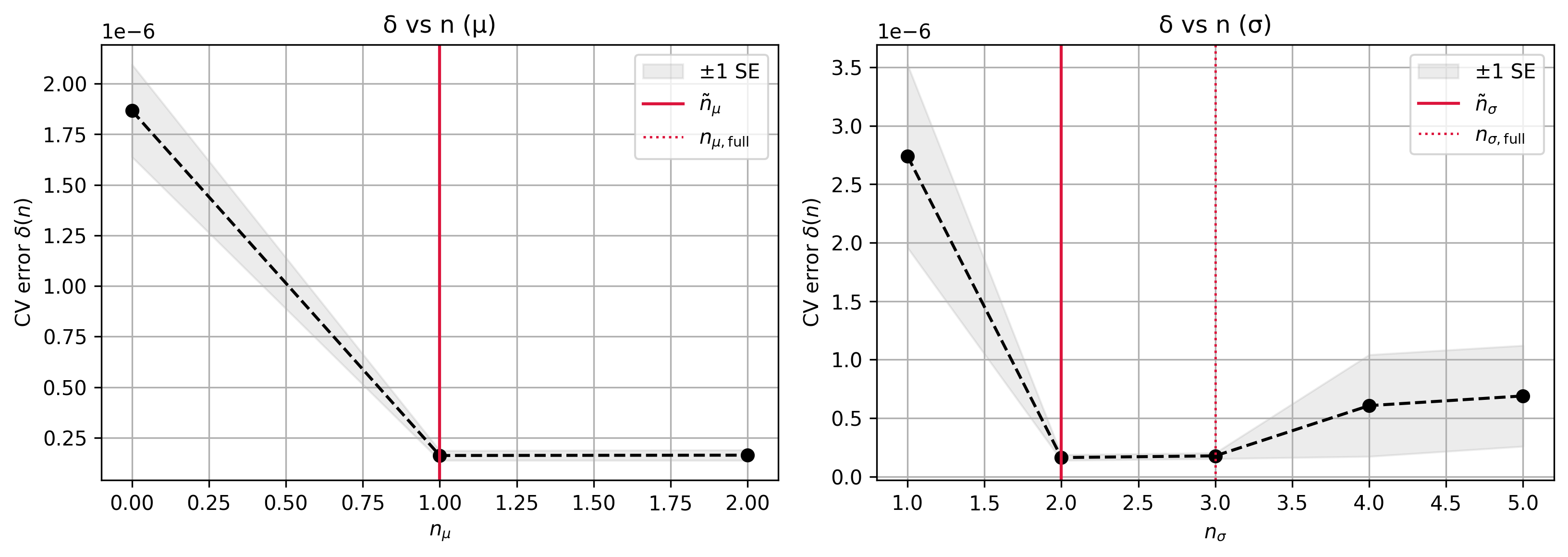}
  \caption{\textit{Validation error \(\delta(n)\) versus support size \(n\) with \(\pm 1\) SE bands for the drift and diffusion blocks.
  The selected sparse sizes are \(n_\mu=1\) and \(n_\sigma=2\).}}
  \label{fig:size_quad_true}
\end{figure}

\noindent\textbf{Case 2:} As a comparison to the first experiment, we repeat the same pipeline for the quadratic-coefficient SDE \eqref{eq:quadSDE-ours}, but now we use the value of \(\psi_{21}\) computed using \hyperref[psi21_method2]{Method~2}. The cross--validation heatmap and the CV curves in Fig.~\ref{fig:cv_quad_true} again show a wide, almost flat region of low validation error. The selected hyperparameters fall inside this stable region and are approximately
\[
(\alpha^\dagger,\rho^\dagger)\approx\big(8.7\times 10^{-4},\,0.40\big),
\]
which is of the same order as the regression-based choice. The \(\delta\)-versus-\(n\) summaries in Fig.~\ref{fig:size_quad_true} select \(n_\mu=1\) and \(n_\sigma=2\), so the learned model remains sparse; adding more terms does not noticeably improve the CV error.

\begin{figure}[H]
  \centering
  \includegraphics[width=.95\linewidth]{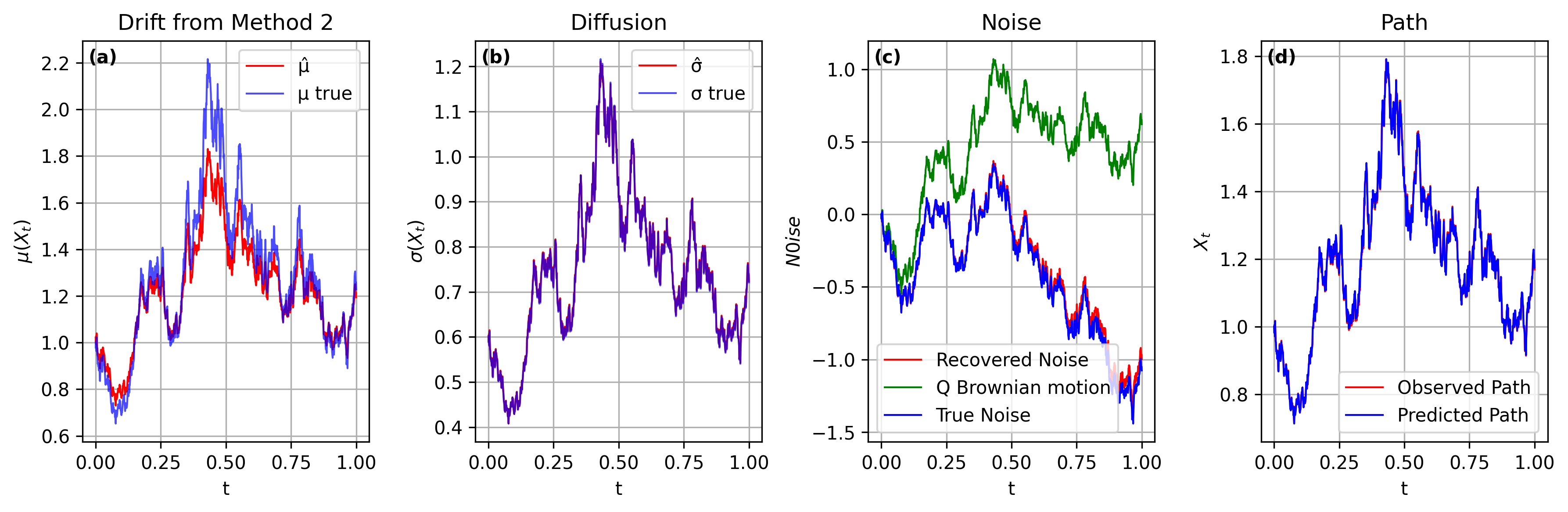}
  \caption{\textit{\textbf{(a)} True and recovered drift \(\mu(X_t)\).
  \textbf{(b)} True and recovered diffusion \(\sigma(X_t)\).
  \textbf{(c)} True noise, recovered noise, and the \(\mathbb{Q}\)-Brownian motion.
  \textbf{(d)} Observed and reconstructed paths of \(X_t\).
  Each vector has length \(M=1000\).}}
  \label{fig:results_quad_true}
\end{figure}

With this analytic \(\psi_{21}\), the recovered drift and diffusion are
\[
\widehat{\mu}(x)=1.02213\,x,
\qquad
\widehat{\sigma}(x)= - \, 0.01997 + 0.54009\,x + 0.08239\,x^2,
\]
with mean squared error \(\mathrm{MSE}=1.45359\times 10^{-7}\). Along the observed trajectory, the reconstructed \(\mu(X_t)\) and \(\sigma(X_t)\) follow the true curves closely (Fig.~\ref{fig:results_quad_true}(a)--(b)). The recovered noise also aligns well with the true driving signal, and the path simulated under the learned model nearly overlaps the observed path (Fig.~\ref{fig:results_quad_true}(c)--(d)). Overall, using the analytic \(\psi_{21}\) produces results consistent with the regression-based case, while keeping the identified model compact and stable.


\begin{figure}[H]
  \centering
  \includegraphics[width=.78\linewidth]{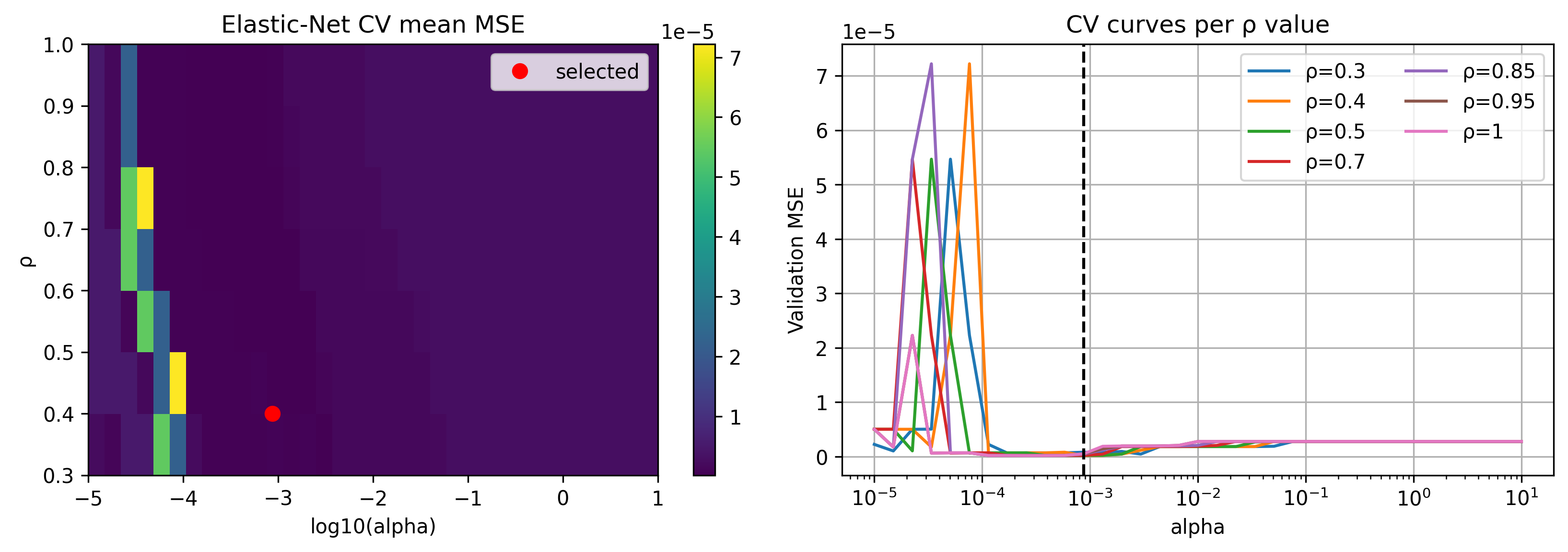}
  \caption{\textit{Time--series CV over \((\alpha,\rho)\) with de-biasing on the active set.
  The selected point \((\alpha^\dagger,\rho^\dagger)\approx(8.7\times 10^{-4},\,0.40)\) lies in a broad near-optimal region.}}
  \label{fig:cv_quad_ode2}
\end{figure}

\begin{figure}[H]
  \centering
  \includegraphics[width=.78\linewidth]{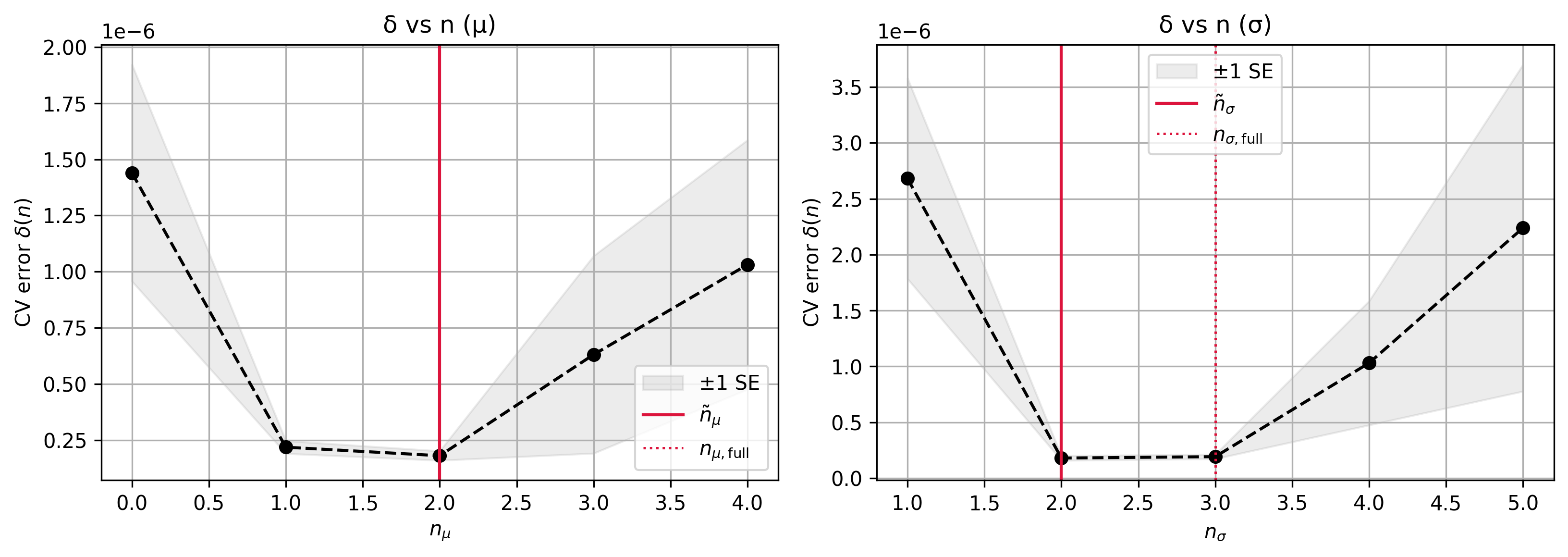}
  \caption{\textit{Validation error \(\delta(n)\) versus support size \(n\) with \(\pm 1\) SE bands for the drift and diffusion blocks.
  The selected sparse sizes are \(n_\mu=2\) and \(n_\sigma=2\).}}
  \label{fig:size_quad_ode2}
\end{figure}

\noindent\textbf{Case 3:} Again, we repeat the same procedure to recover the drift function \(\mu(X_t)\), the drift function \(\sigma(X_t)\), and the noise for the quadratic-coefficient SDE \eqref{eq:quadSDE-ours}. But now we use the \hyperref[psi21_method3]{ODE~2 Method}  to recover the drift function \(\mu(X_t)\). The CV heatmap and the CV curves given in Fig.~\ref{fig:cv_quad_ode2} show a wide, almost horizontal region of small validation error. The chosen hyperparameters are approximately
\[
(\alpha^\dagger,\rho^\dagger)\approx\big(8.7\times 10^{-4},\,0.40\big),
\]
and are same as those in \textbf{Case 2}. The \(\delta\)-versus-\(n\) plot given in Fig.~\ref{fig:size_quad_ode2} shows that the selected values of \(n_\mu=\) and \(n_\sigma\) are \(1\) and \(2\) respectively. These small values show that the learned model remains sparse and adding more terms does not improve the CV error significantly.

\begin{figure}[H]
  \centering
  \includegraphics[width=.95\linewidth]{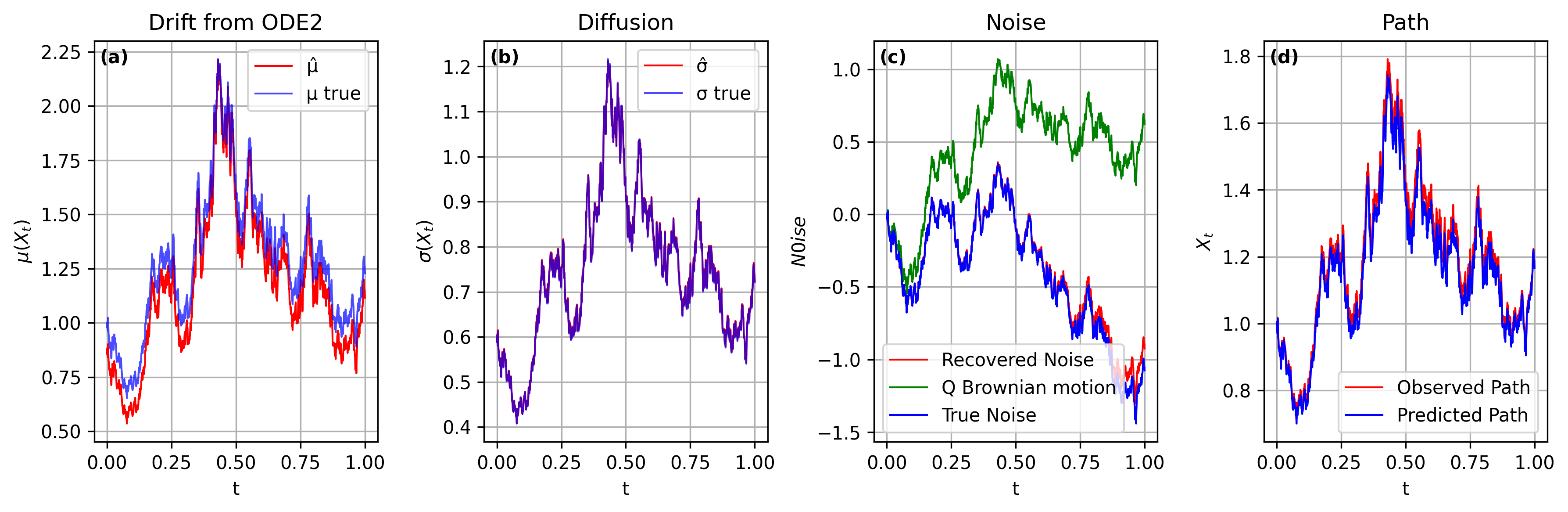}
  \caption{\textit{\textbf{(a)} True and recovered drift \(\mu(X_t)\).
  \textbf{(b)} True and recovered diffusion \(\sigma(X_t)\).
  \textbf{(c)} True noise, recovered noise, and the \(\mathbb{Q}\)-Brownian motion.
  \textbf{(d)} Observed and reconstructed paths of \(X_t\).
  Each vector has length \(M=1000\).}}
  \label{fig:results_quad_ode2}
\end{figure}

The recovered drift and diffusion are
\[
\widehat{\mu}(x)=0.42731\,x + 0.45097\,x^2,
\qquad
\widehat{\sigma}(x)= - \, 0.02019 + 0.54020\,x + 0.08250\,x^2,
\]
with mean squared error \(\mathrm{MSE}=1.56184\times 10^{-7}\). Fig.~\ref{fig:results_quad_ode2}(a) and Fig.~\ref{fig:results_quad_ode2}(b) show that the reconstructed \(\mu(X_t)\) and \(\sigma(X_t)\) follow the true curves closely, Fig.~\ref{fig:results_quad_ode2}(c) shows that the recovered noise also aligns well with the true driving signal, and Fig.~\ref{fig:results_quad_ode2}(d) shows that the path simulated under the learned model nearly overlaps the observed path. Overall, the \hyperref[psi21_method3]{ODE~2 Method} performs well on estimating the diffusion function \(\sigma(X_t)\) and has the small mean squared error similar to what we have in \textbf{Case 1} and \textbf{Case 2}. Unlike \textbf{Case 1}, this performs better on recovering the drift function as both quadratic and linear coefficients are recovered.

\subsection{Ornstein--Uhlenbeck (OU) SDE}
Consider the Ornstein--Uhlenbeck (OU) SDE
\begin{align}\label{eq:OU}
    dX_t = r(\theta-X_t)\,dt + \sigma\,dB^{\mathbb{P}}_t,
\end{align}
where \(r>0\) is the mean-reversion rate, \(\theta\in\mathbb{R}\) is the equilibrium level, and \(\sigma>0\) controls the noise level \cite{uhlenbeck1930theory}. The drift pulls \(X_t\) towards \(\theta\). This is an example of an ergodic process.

\begin{figure}[H]
  \centering
  \includegraphics[width=.98\linewidth]{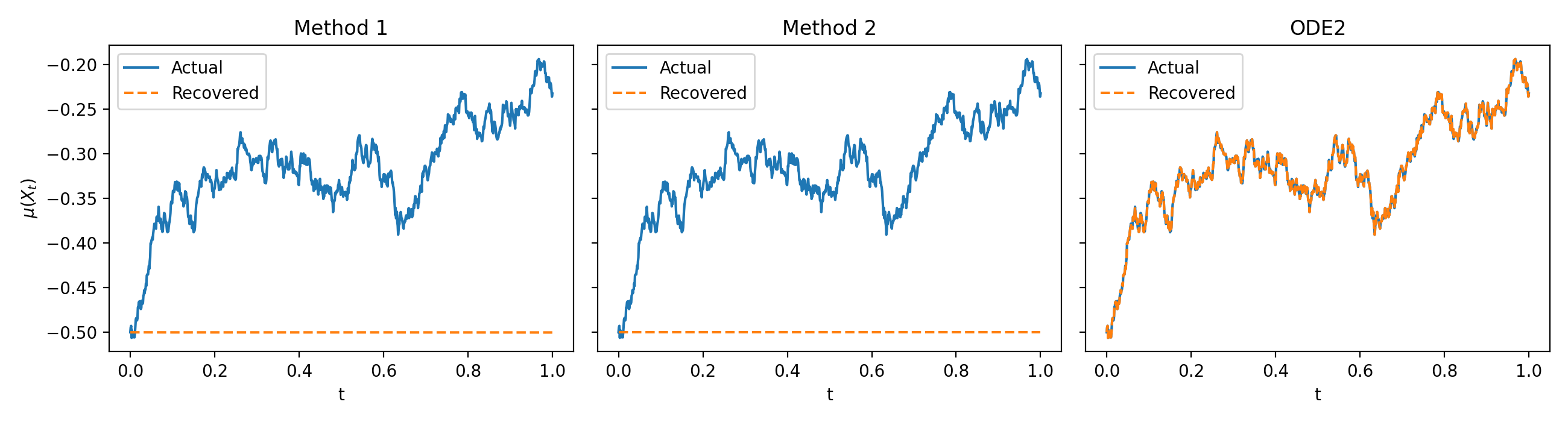}
  \caption{\textit{Comparison of the true drift \(\mu(X_t)\) with its reconstructions using two approximations of \(\psi_{21}\) and using \hyperref[psi21_method3]{ODE~2 Method}: 
  (Left) Actual and recovered \(\mu(X_t)\) when \(\psi_{21}\) obtained via \hyperref[psi21_method1]{Method~1}, (Center) \(\psi_{21}\) computed using \hyperref[psi21_method2]{Method~2}, and (Right) by using \hyperref[psi21_method3]{ODE~2 Method} .}}
  \label{drift_Euler_OU}
\end{figure}

For our experiments, we chose \(r=0.5\), \(\theta=0\), and
\(\sigma=0.3\), giving the drift function \(\mu(X_t)=-\,0.5 \, X_t\).
The path \(X_t\) was simulated using the Euler--Maruyama scheme.
Similar to the previous examples, before applying the penalized
residual \eqref{stochastic SINDy} to recover the drift and diffusion,
we first approximated the diffusion coefficient and estimated the
Brownian motion \(B_{t_i}^{\mathbb{Q}}\), \(0\le i\le N\), under
the change of measure.
We then estimated the drift using the two choices of \(\psi_{21}\)
provided by \hyperref[psi21_method1]{Method~1} and
\hyperref[psi21_method2]{Method~2}, and by using
\hyperref[psi21_method3]{ODE~2 Method}
(see Subsection~\ref{subsubsec:psi12_methods}).
The resulting drift estimates are shown in Figure~\ref{drift_Euler_OU}.

This is an example where the \hyperref[psi21_method3]{ODE~2 Method} performs better than \hyperref[psi21_method1]{Method~1} and \hyperref[psi21_method2]{Method~2}. We expect the same when we run run the whole SSISDE pipeline. Again in all settings, we use the same penalized residual \eqref{stochastic SINDy} and the same polynomial libraries,
\[
\Theta_\mu(x)=\{1,x,\ldots,x^5\},\qquad
\Theta_\sigma(x)=\{1,x,\ldots,x^5\}.
\]
The number of folds in the \(k\)-fold cross--validation and the grids for \(\alpha\) and \(\rho\) remain same. Once again, after selecting the active terms, the coefficients are re-estimated to reduce bias and \(1\)-SE rule is used to favor simpler model closer to the best one.  

The recovery of the models and the noise in all the three cases: \hyperref[psi21_method1]{Method~1}, \hyperref[psi21_method2]{Method~2}, and \hyperref[psi21_method3]{ODE~2 Method} is discussed below.

\begin{figure}[H]
  \centering
  \includegraphics[width=.78\linewidth]{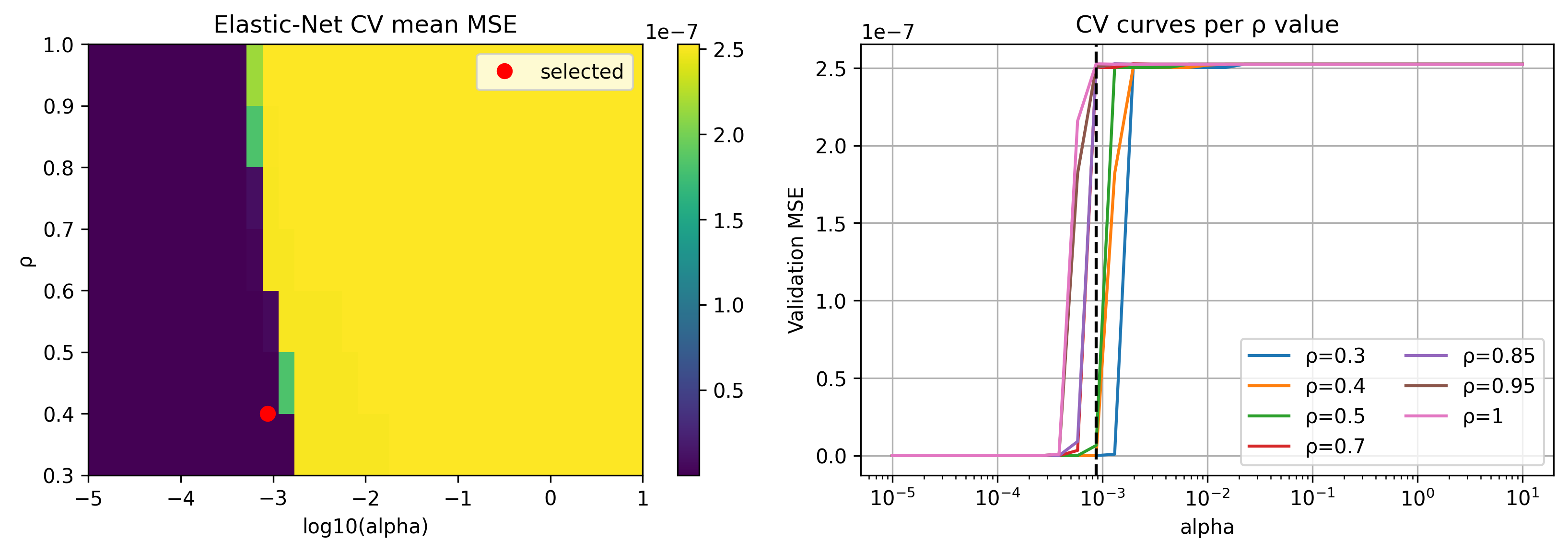}
  \caption{\textit{Time--series CV over \((\alpha,\rho)\) with de-biasing on the active set. The selected point 
  \((\alpha^\dagger,\rho^\dagger)\approx(8.7\times 10^{-4}, 0.4)\) lies in a broad near-optimal region.}}
  \label{fig:cv_OU_reg}
\end{figure}

\begin{figure}[H]
  \centering
  \includegraphics[width=.78\linewidth]{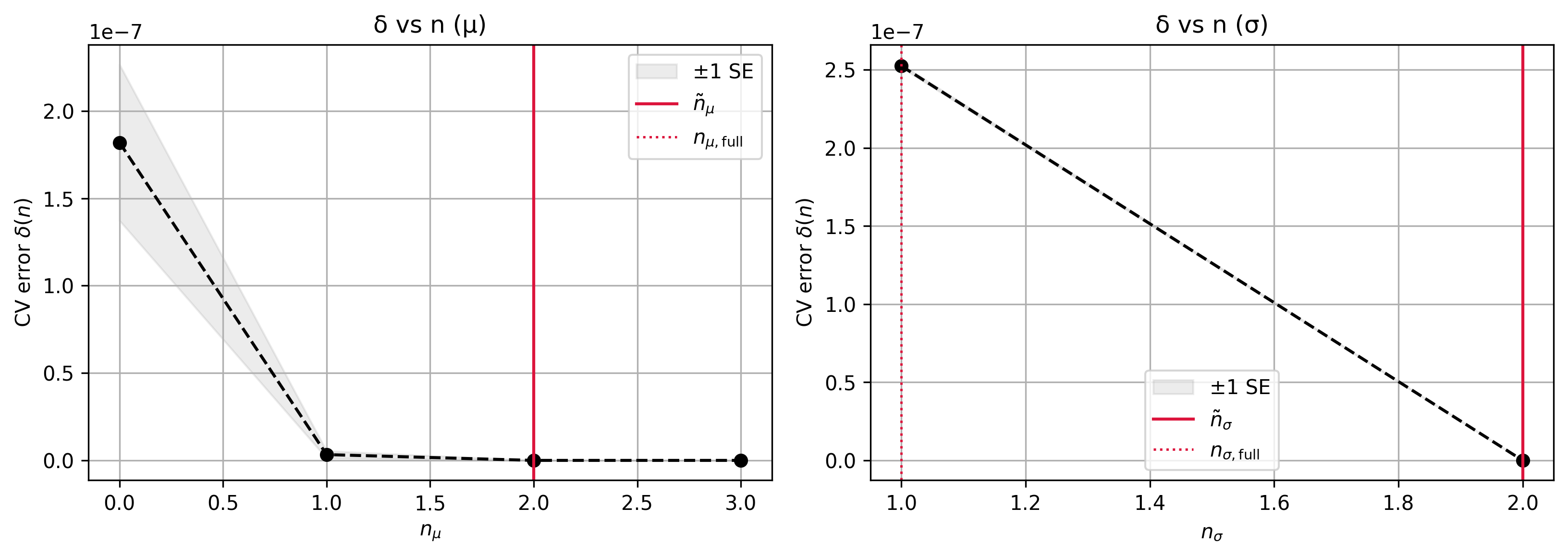}
  \caption{\textit{Validation error \(\delta(n)\) versus support size \(n\) with \(\pm 1\) SE bands for the drift and diffusion blocks. 
  The selected sizes are \(n_\mu=2\) and \(n_\sigma=2\), indicating sparse yet stable models.}}
  \label{fig:size_OU_reg}
\end{figure}

\noindent\textbf{Case 1:} In the first experiment, we use the value of \(\psi_{21}\) obtained by applying \hyperref[psi21_method1]{Method~1} to recover the drift function \(\mu(X_t)\) . 
The CV heatmap and the curves given in Fig.~\ref{fig:cv_OU_reg} show a wide region of small validation error.  The selected pair of hyper-parameters \(\alpha^\dagger\) and \(\rho^\dagger\) lies inside this stable region and is approximately
\[
(\alpha^\dagger,\rho^\dagger) \approx \big(8.7\times 10^{-4},\,0.4\big).
\]
The \(\delta\)-versus-\(n\) plots in Fig.~\ref{fig:size_OU_reg} suggest the estimated model size to be \(n_\mu=2\) and \(n_\sigma=2\). 
This means both the drift and the diffusion blocks keep two terms; adding more terms does not lead to a visible improvement in the validation error. 
With these hyperparameters \(\alpha^\dagger\) and \(\rho^\dagger\), the recovered functions take the form
\[
\widehat{\mu}(x)= -\,0.50031 + 0.00031\,x,
\qquad 
\widehat{\sigma}(x)= 0.29901.
\]

\begin{figure}[H]
  \centering
  \includegraphics[width=.95\linewidth]{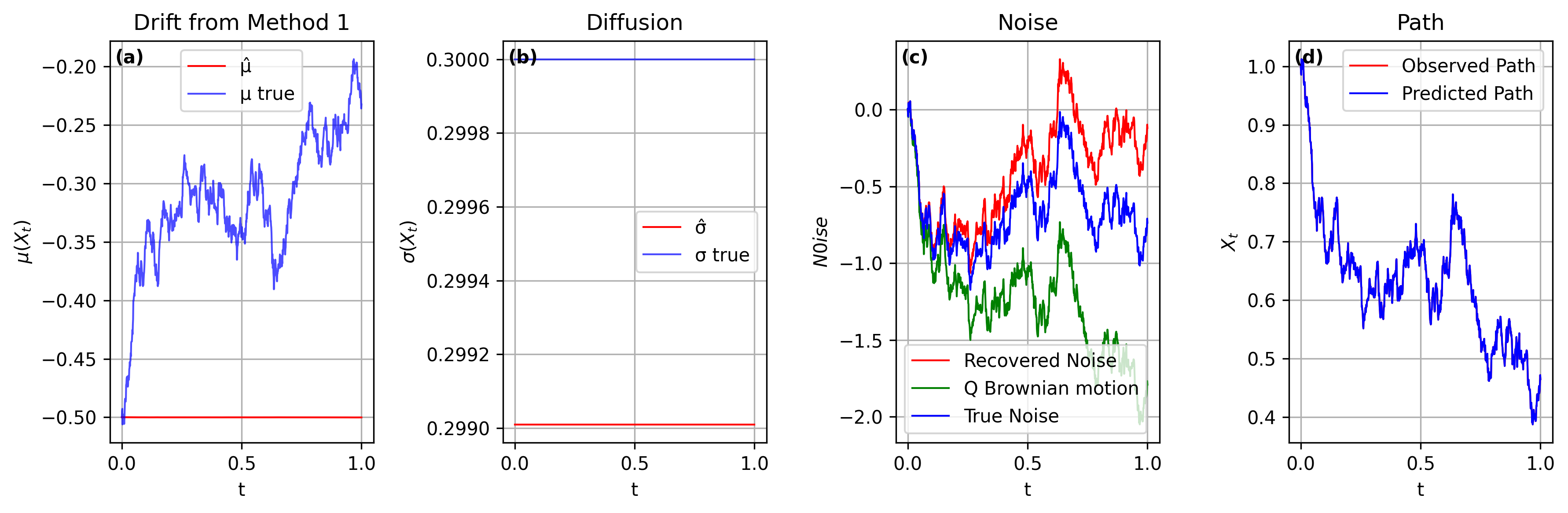}
  \caption{\textit{\textbf{(a)} True \(\mu(X_t)\) and reconstructed \(\widehat{\mu}(X_t)\). 
  \textbf{(b)} True \(\sigma(X_t)\) and reconstructed \(\widehat{\sigma}(X_t)\). 
  \textbf{(c)} True noise, recovered noise, and the \(\mathbb{Q}\)-Brownian motion. 
  \textbf{(d)} Observed path \(X_t\) and the path simulated using the recovered \(\widehat{\mu}(X_t)\) and \(\widehat{\sigma}(X_t)\). 
  Each vector is of length \(M\) with \(M=1000\).}}
  \label{fig:results_OU_reg}
\end{figure}

The estimated diffusion function \(\widehat{\sigma}(x)=0.29901\) is very close to the true function \(\sigma(x)=0.3\). However, the estimated drift function \(\widehat{\mu}(x)=-0.50031+0.00031\,x\) does not capture the linear structure of the true drift \(\mu(x)=-0.5\,x\) well. The coefficient of \(x\) should be negative, but it is positive in the estimated drift. Fig.~\ref{fig:results_OU_reg}(a)--(b) shows that both the true and recovered diffusion functions are constant and close to one another, whereas the drift functions differ substantially. Despite this, the prediction error is very small, with \(\mathrm{MSE}=2.96264\times10^{-16}\).

The recovered noise does not closely follow the driving noise. However, the path generated using \(\widehat{\mu}\) and \(\widehat{\sigma}\) is very close to the observed trajectory; see Fig.~\ref{fig:results_OU_reg}(c)--(d). Overall, these results indicate that the method recovers the diffusion well but not the drift, motivating the use of alternative methods such as \hyperref[psi21_method2]{Method~2} or the \hyperref[psi21_method3]{ODE2 method}.

\begin{figure}[H]
  \centering
  \includegraphics[width=.95\linewidth]{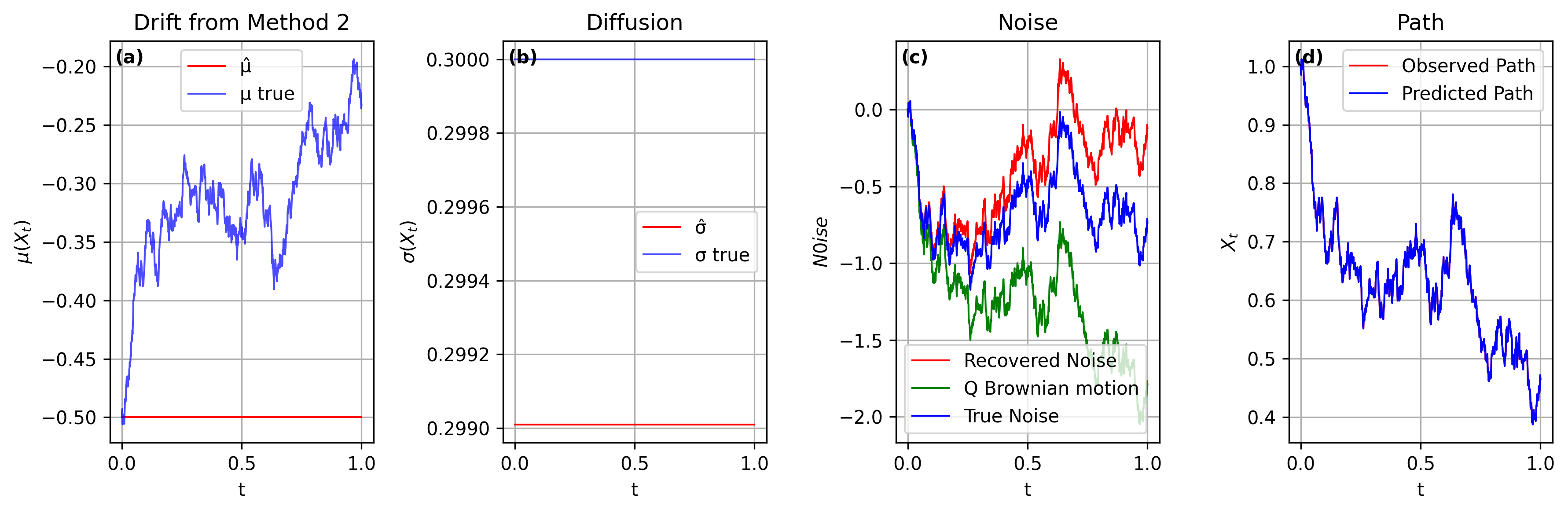}
  \caption{\textit{\textbf{(a)} True \(\mu(X_t)\) and reconstructed \(\widehat{\mu}(X_t)\). 
  \textbf{(b)} True \(\sigma(X_t)\) and reconstructed \(\widehat{\sigma}(X_t)\). 
  \textbf{(c)} True noise, recovered noise, and the \(\mathbb{Q}\)-Brownian motion. 
  \textbf{(d)} Observed path \(X_t\) and the path simulated using the recovered \(\widehat{\mu}(X_t)\) and \(\widehat{\sigma}(X_t)\). 
  Each vector is of length \(M\) with \(M=1000\).}}
  \label{fig:results_OU_true}
\end{figure}

\noindent\textbf{Case 2:} Applying \hyperref[psi21_method2]{Method~2} to obtain \(\psi_{21}\) leads to the recovered functions
\[
\widehat{\mu}(x)=-0.5,\qquad \widehat{\sigma}(x)=0.29901.
\]
Although the estimated diffusion remains very close to the true value \(\sigma(x)=0.3\), the recovered drift is constant and therefore does not capture the linear structure of \(\mu(x)=-0.5\,x\). This difference is visible in Fig.~\ref{fig:results_OU_true}(a)--(b): the diffusion functions are close to one another, whereas the drift functions differ substantially.

\begin{figure}[H]
  \centering
  \includegraphics[width=.78\linewidth]{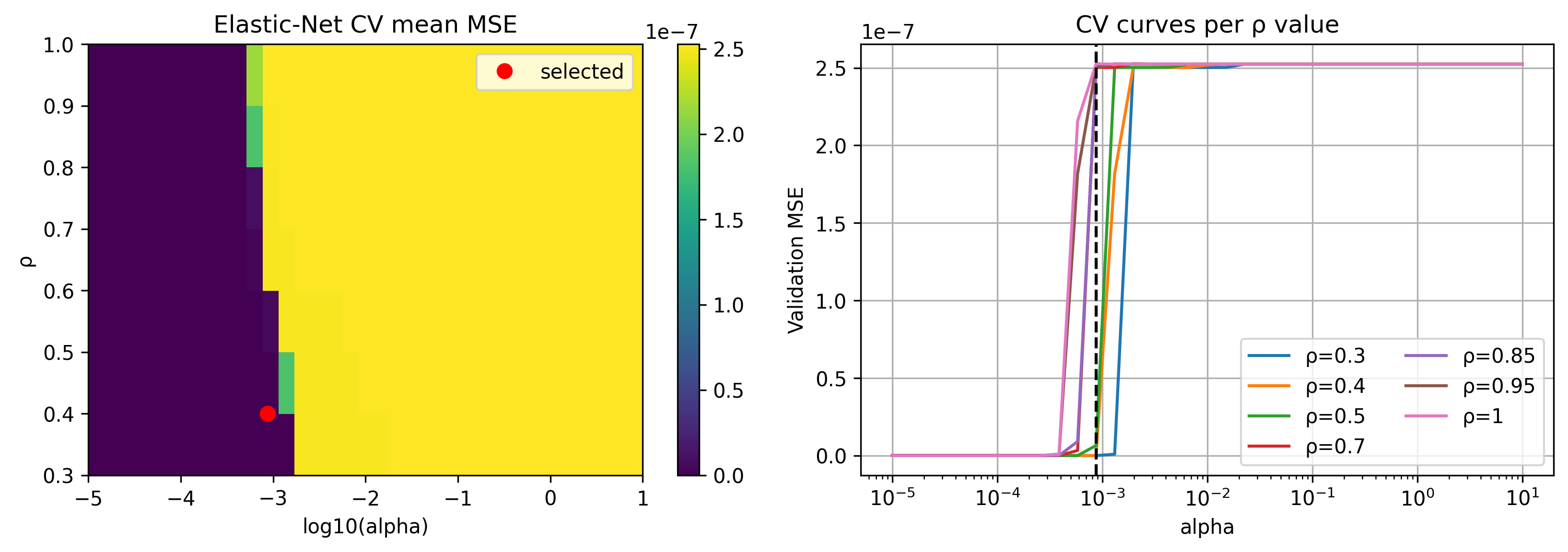}
  \caption{\textit{Time--series CV over \((\alpha,\rho)\) with de-biasing on the active set. The selected point 
  \((\alpha^\dagger,\rho^\dagger)\approx(8.7\times 10^{-4}, 0.4)\) lies in a broad near-optimal region.}}
  \label{fig:cv_OU_true}
\end{figure}

\begin{figure}[H]
  \centering
  \includegraphics[width=.78\linewidth]{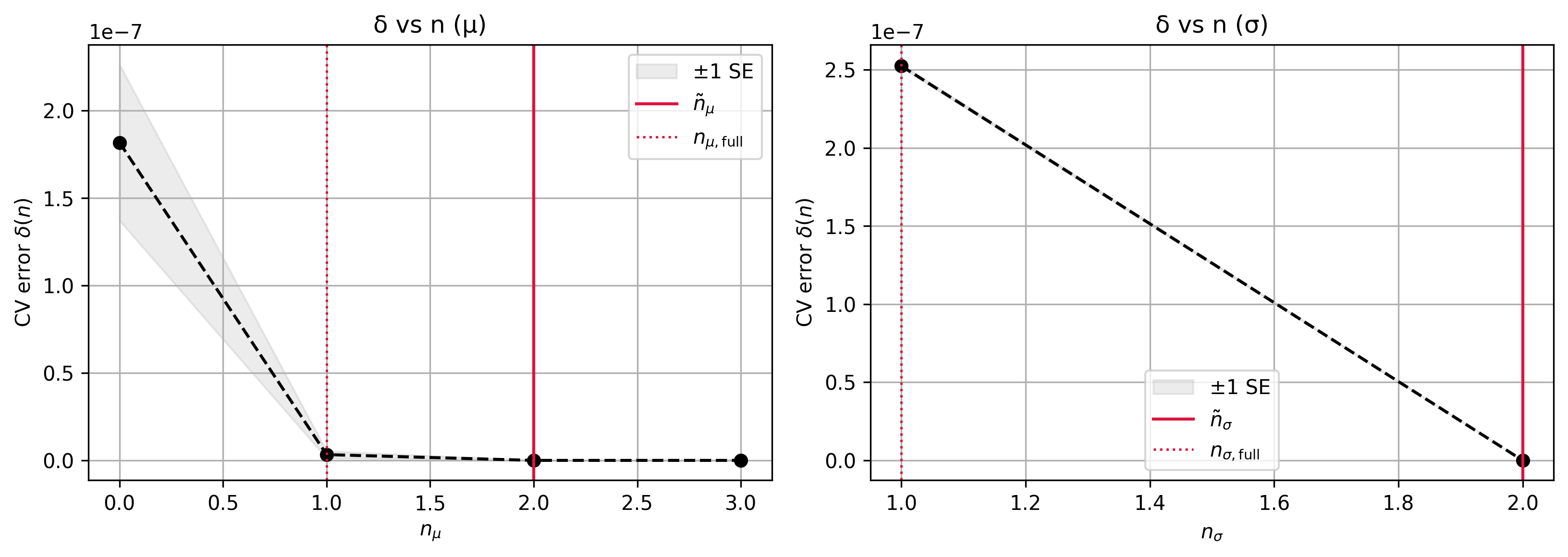}
  \caption{\textit{Validation error \(\delta(n)\) versus support size \(n\) with \(\pm 1\) SE bands for the drift and diffusion blocks. 
  The selected sizes are \(n_\mu=2\) and \(n_\sigma=2\), indicating sparse yet stable models.}}
  \label{fig:size_OU_true}
\end{figure}

For this experiment, the selected hyperparameters are approximately \((\alpha^\dagger,\rho^\dagger)=(8.7\times10^{-4},\,0.4)\), lying within the wide region of small validation error shown in Fig.~\ref{fig:cv_OU_true}. The \(\delta\)-versus-\(n\) plots in Fig.~\ref{fig:size_OU_true} suggest support sizes of two terms for each block, while the final rounded full-data models retain only one constant term in each block.

Despite the discrepancy in the drift, the final prediction error is very small, with \(\mathrm{MSE}=2.15839\times10^{-16}\). The reconstructed path also remains very close to the observed trajectory, even though the recovered noise does not closely follow the driving noise; see Fig.~\ref{fig:results_OU_true}(c)--(d). Thus, the small prediction error does not imply accurate drift recovery. These results motivate examining the \hyperref[psi21_method3]{ODE2 method} as the next alternative.

\begin{figure}[H]
  \centering
  \includegraphics[width=.78\linewidth]{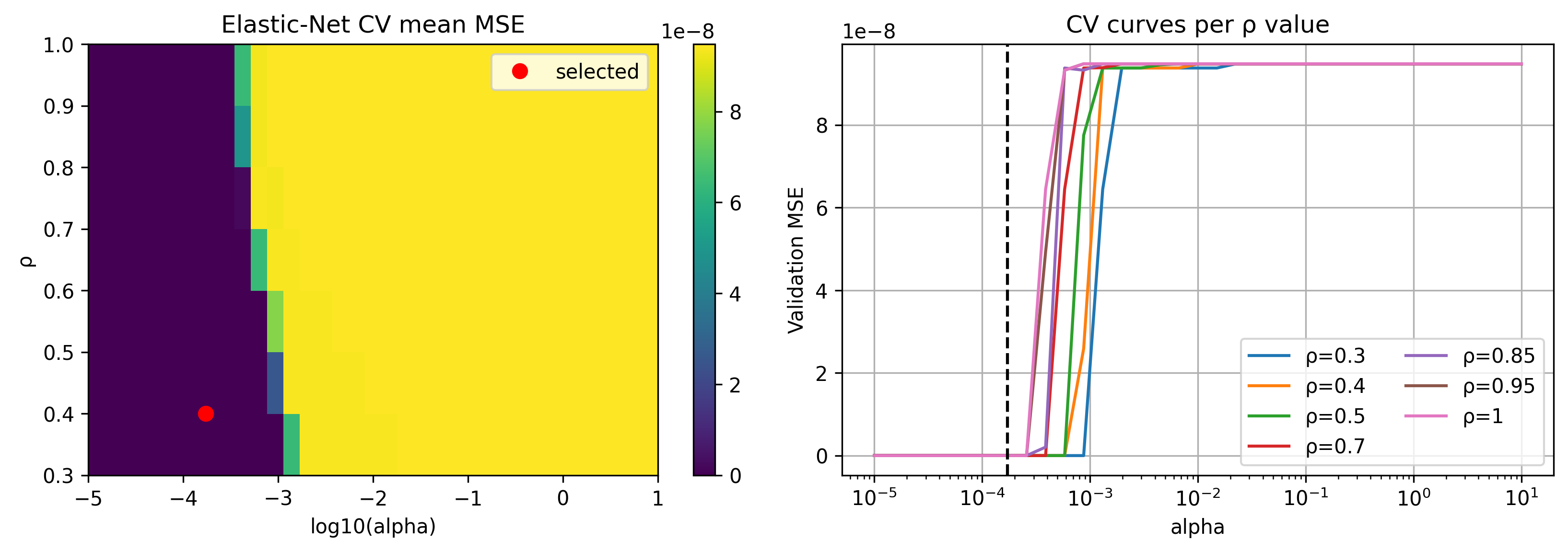}
  \caption{\textit{Time-series CV over \((\alpha,\rho)\) with de-biasing on the active set for the OU experiment using ODE2. The selected point \((\alpha^\dagger,\rho^\dagger)\approx(1.7\times10^{-4},\,0.40)\) lies in a broad region of small validation error.}}
  \label{fig:cv_OU_ode2}
\end{figure}

\begin{figure}[H]
  \centering
  \includegraphics[width=.78\linewidth]{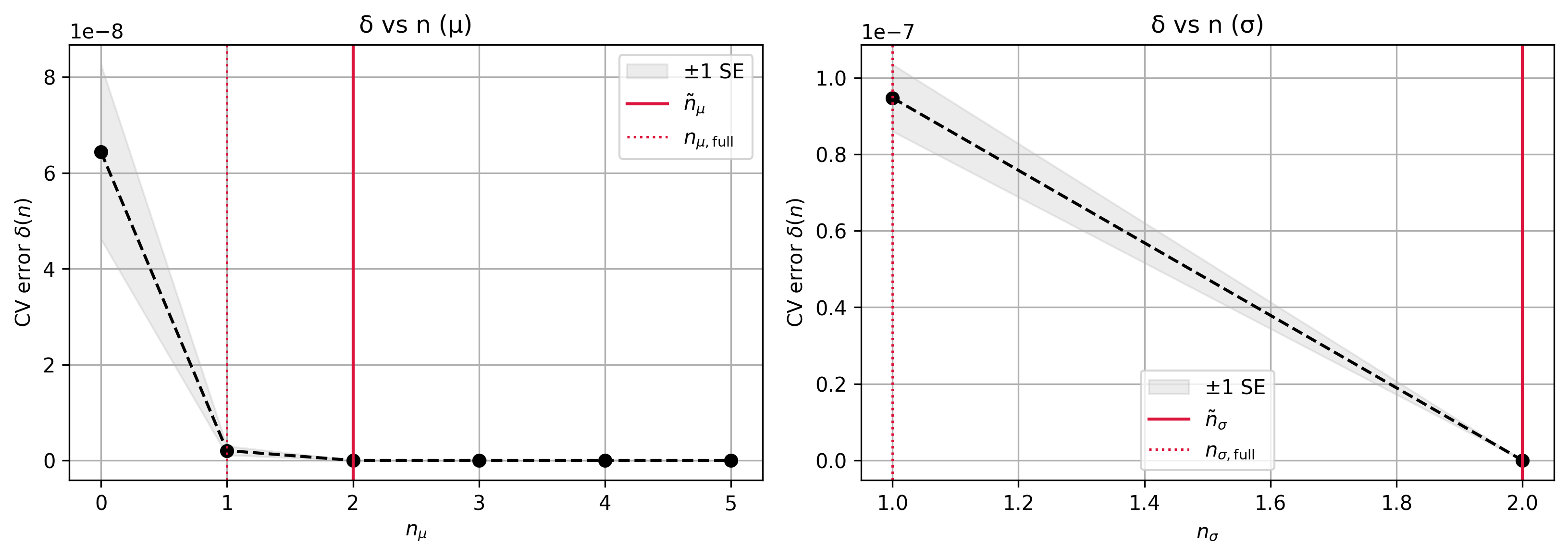}
  \caption{\textit{Validation error \(\delta(n)\) versus support size \(n\), with \(\pm1\)\,SE bands for the drift and diffusion blocks. The solid vertical markers indicate suggested support sizes of two terms in each block, while the dotted markers indicate one retained term in each final rounded full-data model.}}
  \label{fig:size_OU_ode2}
\end{figure}

\noindent\textbf{Case 3:} Finally, we apply the \hyperref[psi21_method3]{ODE2 method} to recover the drift function \(\mu(X_t)\) of the Ornstein--Uhlenbeck SDE and use it to reconstruct the noise and estimate the final drift and diffusion functions. The CV heatmap and curves in Fig.~\ref{fig:cv_OU_ode2} show a broad region of small validation error. The selected hyperparameters are approximately
\[
(\alpha^\dagger,\rho^\dagger)\approx\big(1.7\times10^{-4},\,0.40\big).
\]
Compared with \textbf{Case 2}, the selected \(\alpha^\dagger\) is smaller, while \(\rho^\dagger\) remains unchanged. The \(\delta\)-versus-\(n\) plots in Fig.~\ref{fig:size_OU_ode2} suggest support sizes of two terms for each block. However, the final rounded full-data models retain only one term in each block: a linear drift term and a constant diffusion term.

\begin{figure}[H]
  \centering
  \includegraphics[width=.95\linewidth]{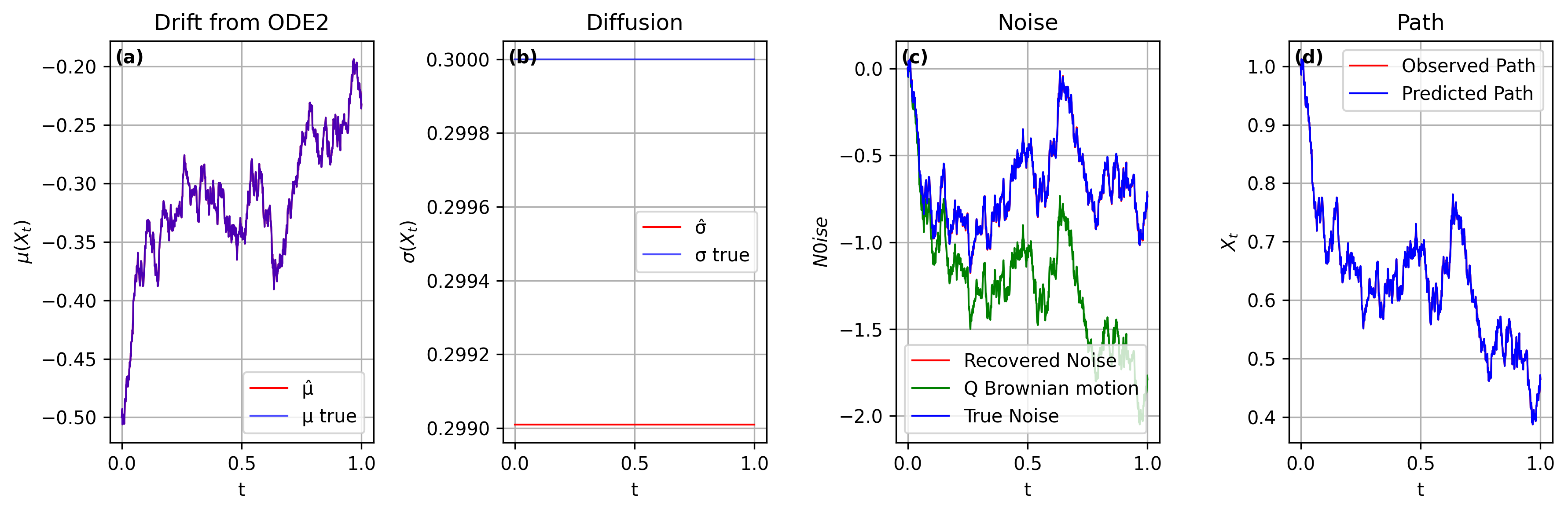}
  \caption{\textit{\textbf{(a)} True \(\mu(X_t)\) and reconstructed \(\widehat{\mu}(X_t)\). 
  \textbf{(b)} True \(\sigma(X_t)\) and reconstructed \(\widehat{\sigma}(X_t)\). 
  \textbf{(c)} True noise, recovered noise, and the \(\mathbb{Q}\)-Brownian motion. 
  \textbf{(d)} Observed path \(X_t\) and the path simulated using the recovered \(\widehat{\mu}(X_t)\) and \(\widehat{\sigma}(X_t)\). 
  Each vector is of length \(M\) with \(M=1000\).}}
  \label{fig:results_OU_ode2}
\end{figure}

The recovered functions are
\[
\widehat{\mu}(x)=-0.50000\,x,\qquad \widehat{\sigma}(x)=0.29901,
\]
with a final prediction error of \(\mathrm{MSE}=2.15707\times10^{-16}\). At the reported precision, the recovered drift matches the true function \(\mu(x)=-0.5\,x\), while the estimated diffusion remains close to \(\sigma(x)=0.3\). Fig.~\ref{fig:results_OU_ode2}(a)--(b) shows this agreement along the observed path. The recovered noise also closely follows the driving noise, and the reconstructed path nearly overlaps the observed trajectory; see Fig.~\ref{fig:results_OU_ode2}(c)--(d). Although the final MSE is similar to those in \textbf{Case 1} and \textbf{Case 2}, the \hyperref[psi21_method3]{ODE2 method} substantially improves drift and noise recovery by correctly identifying the drift function.

\section{Error analysis for Black-Scholes model}
\begin{proposition}\label{prop: step 3}
If $X_t$ follows dynamic \eqref{int_form_stat1}, and the data vector is observed at time grid $0=t_0<t_1<\cdots, <t_N = T$. Denote $L^3(t)$ as the level-$3$ leading expansion term as in  \eqref{step 3 expansion}, then the approximation error of $X_t$ for $t\in [0,T]$ in {\bf Step 3} has the following estimate,
\begin{align}
    \mathbb P\left(\sup_{t\in [t_k,t_{k+1}]} \|X_t-L^3(t)\|\ge \xi |t_{k+1}-t_k|^{2}; \tau< \zeta \right) \le \exp\left(-\frac{c\xi^{\alpha} }{|t_{k+1}-t_k|}\right).
\end{align}
Furthermore, for $t\in [t_k,t_{k+1}]$, we have
\begin{align}
    \mathbb E[\| X_t- L^3(t)\|^2]\le C |t_{k+1}-t_k|^2,
\end{align}
where $C$ depends on the bound of $\mu$, $\sigma$ and their derivatives.
\end{proposition}

\begin{corollary}\label{corollary}
If $X_t$ follows a Black-Scholes dynamics \eqref{eq:BS}, and the data vector is observed at time grid $0=t_0<t_1<\cdots <t_N = T$, then the total approximation error of $X_t$ for $t\in [0,T]$ has the following form,
\begin{align}
    \mathbb E[\| X_t- L^3(t)\|^2]\le C |t_{k+1}-t_k|^2,
\end{align}
where $C$ depends on the bound of $\mu$, $\sigma$ and their derivatives.
\end{corollary}

\bibliographystyle{siam}
\bibliography{ref}
\appendix

\section{Proof of Step 2}\label{section proof step 2}
Recall that
\begin{align}
    X_t &= X_s+A_0(X_s)\int_s^t du +\int_s^t \int_s^u A_0'(X_r) \circ dX_r du + \sigma(X_s)\int_s^t \circ dB_u^{\mathbb Q}\\
    & \quad +\int_s^t\int_s^u \sigma'(X_r)\circ dX_r\circ dB_u^{\mathbb Q}\nonumber.
\end{align}
Plugging in the dynamics of $dX_r$ \ref{martingale SDE stratonovich} and applying the Taylor expansion at the starting point $X_s$, we get 
\begin{align}\label{eq29}
    X_t&= X_s+A_0(X_s)\int_s^t du +\int_s^t \int_s^u A_0'(X_r) (A_0(X_r) dr + \sigma(X_r)\circ dB_r^{\mathbb Q}) du \\
    &+ \sigma(X_s)\int_s^t \circ dB_u^{\mathbb Q} +\int_s^t\int_s^u \sigma'(X_r)  (A_0(X_r) dr + \sigma(X_r)\circ dB_r^{\mathbb Q})\circ dB_u^{\mathbb Q}\nonumber\\
    &=X_s+A_0(X_s)\int_s^t du +A_0'(X_s) A_0(X_s)\int_s^t \int_s^u   dr du +A_0(X_s)\int_s^t du  \nonumber\\
    & +A_0'(X_s) \sigma(X_s)\int_s^t\int_s^u \circ dB_r^{\mathbb Q} du  \nonumber\\
    &+ \sigma(X_s)\int_s^t \circ dB_u^{\mathbb Q}+ \sigma'(X_s)  A_0(X_s)\int_s^t\int_s^u dr \circ dB_u^{\mathbb Q} \nonumber\\
    & +  \sigma'(X_s)  \sigma(X_s)\int_s^t\int_s^u \circ dB_r^{\mathbb Q}\circ dB_u^{\mathbb Q}\nonumber\\
    &+\int_s^t\int_s^u\int_s^r (A_0''(X_v)A_0(X_v)+A_0'(X_v)A_0'(X_v))\circ dX_vdr\circ dB_u^{\mathbb Q} \nonumber\\
    &+\int_s^t\int_s^u\int_s^r(A_0''(X_v)\sigma(X_v)+A_0'(X_v)\sigma'(X_v))\circ dX_v \circ dB_r^{\mathbb Q} du\nonumber\\
    &+\int_s^t\int_s^u\int_s^r(\sigma''(X_v)A_0(X_v)+\sigma'(X_v)A_o'(X_v) )\circ dX_v dr \circ dB_u^{\mathbb Q}\nonumber\\
    &+ \int_s^t\int_s^u\int_s^r (\sigma''(X_v)\sigma(X_v)+\sigma'(X_v)\sigma'(X_v) )\circ dX_v\circ dB_r^{\mathbb Q}\circ dB_u^{\mathbb Q}\nonumber
\end{align}
We collect higher order terms (i.e., with order equal or higher than $|t-s|^{2-\varepsilon}$, for any $\varepsilon>0$,  when $|t-s|$ is small) and denote them as the error term $\varepsilon^{\Psi_2}$. The leading term of $\int_s^t\int_s^u\int_s^r (\sigma''(X_v)\sigma(X_v)+\sigma'(X_v)\sigma'(X_v) )\circ dX_v\circ dB_r^{\mathbb Q}\circ dB_u^{\mathbb Q}$ will contribute one more $|t-s|^{3/2-\varepsilon}$ term, which has the following form,
\begin{align*}
\left(\sigma''(X_s)\sigma(X_s)+\sigma'(X_s)\sigma'(X_s) \right)\sigma(X_s)  \int_s^t\int_s^u\int_s^r \circ dB_v^{\mathbb Q}\circ dB_r^{\mathbb Q}\circ dB_u^{\mathbb Q},
\end{align*}
together with higher-order error terms. By collecting all the second-order iterated integrals, where the integral against Brownian motion has order $\frac{1}{2}-\varepsilon$, we end up with
\begin{align}
    X_t=&X_s+\psi_1^{s,t} \int_s^t du +\psi_2^{s,t} \int_s^t d B_u+ \psi_{12}^{s,t}\int_s^t\int_s^u dr \circ dB_u^{\mathbb Q} \nonumber\\
    &+\psi_{21}^{s,t}\int_s^t\int_s^u\circ  dB^{\mathbb Q}_r du +\psi_{22}^{s,t}\int_s^t\int_s^u \circ d B^{\mathbb Q}_r \circ dB_u^{\mathbb Q} \nonumber \\
    & + \psi_{222}^{s,t}\int_s^t\int_s^u\int_s^r \circ dB_v^{\mathbb Q}\circ dB_r^{\mathbb Q} \circ dB_u^{\mathbb Q} + \mathsf{\varepsilon}^{\Psi_2},\nonumber
\end{align}
where we denote $\Psi_{2}^{s,t}:=\{\psi_1^{s,t},\psi_2^{s,t},  \psi_{12}^{s,t}, \psi_{21}^{s,t}, \psi_{22}^{s,t},\psi_{222}^{s,t}\}$ as the coefficients of the second order expansion of $X_t$, and $\varepsilon^{\Psi_2}(\approx |t-s|^{2-\varepsilon})$ as the error term for the second order expansion. The proof is thus completed. \qed 

\section{Proof of Step 3}\label{section proof step 3}
Applying stochastic Taylor expansion on \(\tilde{\mu}\) and \(\sigma\) around \(X_{s}\), and truncating the Taylor expansion at level 2, we have the following approximation of $\tilde{\mu}$ and $\sigma$, 

\begin{align}\label{expansion tilde u}
\tilde{\mu}(X_u) &= \tilde{\mu}(X_s) + \tilde{\mu}'(X_s) \int_s^u \circ dX_r  + \tilde{\mu}''(X_s) \int_s^u \int_s^r  \circ dX_v \circ dX_r \\
&+\underbrace{\int_s^u \int_s^r \int_s^v \tilde{\mu}'''(\xi_\tau) \circ dX_\tau \circ dX_v \circ dX_r}_{
\varepsilon^{\tilde{\mu}}} \nonumber
\end{align}

\begin{align} \label{expansion sigma}
\sigma(X_u) &= \sigma(X_s) + \sigma'(X_s) \int_s^u \circ dX_r  + \sigma''(X_s) \int_s^u \int_s^r \circ dX_v \circ dX_r\\
&+ \underbrace{\int_s^u \int_s^r \int_s^v \sigma'''(\eta_\tau) \circ dX_\tau \circ dX_v \circ dX_r}_{\varepsilon^{\sigma}} \nonumber
\end{align}

Where we denote
$\varepsilon^{\tilde \mu}$ and $\varepsilon^{\sigma}$ as the higher order approximation error wrt the second order expansion. Under some nominal H\"older assumptions on the underlying process $X$ and smoothness of the functions $\tilde{\mu}$ and $\sigma$, these approximation errors of \eqref{expansion tilde u} and \eqref{expansion sigma} are of order $|t_{i+1}-t_i|^{\frac{3}{2}-\varepsilon}$ on each time grid $[t_i,t_{i+1}]$. For clarity, we highlight the substitution of the expansion into the dynamics in blue. Substituting the second-order Taylor expansions \eqref{expansion tilde u} and \eqref{expansion sigma} into \eqref{int_form_stat1} gives
\begin{align*}
    X_t & = X_s + \int_s^t  \left( \tilde{\mu}(X_s) + \tilde{\mu}'(X_s) \int_s^u \circ dX_r  + \tilde{\mu}''(X_s) \int_s^u \int_s^r { \circ}dX_v \circ dX_r +\varepsilon^{\tilde{\mu}}\right) du\\
    &+ \int_s^t  \left( \sigma(X_s) + \sigma'(X_s) \int_s^u \circ dX_r  + \sigma''(X_s) \int_s^u \int_s^r \circ dX_v \circ dX_r+\varepsilon^{\sigma} \right) \circ dB_u^{\mathbb P}.\\
    \end{align*}
   Plugging in dynamic \eqref{int_form_stat1} for $dX_t$ in the above expansion, we get
    \begin{align*}
   X_t & = X_s + \tilde{\mu}(X_s) \int_s^t du + \sigma(X_s) \int_s^t \circ dB_u^{\mathbb P}\\
   & \quad + \tilde{\mu}'(X_s) \int_s^t \int_s^u   \left(  \tilde{\mu}(X_r)dr+\sigma(X_r) \circ dB_r^{\mathbb P} \right)  du\\
    & \quad + \tilde{\mu}''(X_s) \int_s^t \int_s^u \int_s^r   \left(  \tilde{\mu}(X_v)dv+\sigma(X_v) \circ dB_v^{\mathbb P} \right)  \left(  \tilde{\mu}(X_r)dr+\sigma(X_r) \circ dB_r^{\mathbb P} \right)  du \\
    & \quad +  \sigma'(X_s) \int_s^t \int_s^u   \left(  \tilde{\mu}(X_r)dr+\sigma(X_r) \circ dB_r^{\mathbb P} \right)  \circ dB_u^{\mathbb P} \\
    & \quad + \sigma''(X_s) \int_s^t \int_s^u \int_s^r   \left(  \tilde{\mu}(X_v)dv+\sigma(X_v) \circ dB_v^{\mathbb P} \right) \left(  \tilde{\mu}(X_r)dr+\sigma(X_r) \circ dB_r^{\mathbb P} \right)  \circ dB_u^{\mathbb P}\\
    &\quad+\varepsilon^{\tilde{\mu}} \int_s^t ds+\varepsilon^{\sigma}\int_s^t \circ dB^{\mathbb P}_u\\
    & = X_s + \tilde{\mu}(X_s) \int_s^t du + \sigma(X_s) \int_s^t \circ dB_u^{\mathbb P} + \tilde{\mu}'(X_s) \int_s^t \int_s^u    \tilde{\mu}(X_r)  drdu   \\
    & \quad +  \tilde{\mu}'(X_s) \int_s^t \int_s^u   \sigma(X_r)  \circ dB_r^{\mathbb P} du + \sigma'(X_s) \int_s^t \int_s^u    \tilde{\mu}(X_r)  dr \circ dB_u^{\mathbb P} \\
   &  \quad + \sigma'(X_s) \int_s^t \int_s^u    \tilde{\mu}(X_r)  dr \circ dB_u^{\mathbb P} +  \sigma'(X_s) \int_s^t \int_s^u   \sigma(X_r)  \circ dB_r^{\mathbb P} \circ dB_u^{\mathbb P}\\
    & \quad + \tilde{\mu}''(X_s) \int_s^t \int_s^u \int_s^r   \left(  \tilde{\mu}(X_v)dv+\sigma(X_v) \circ dB_v^{\mathbb P} \right)  \left(  \tilde{\mu}(X_r)dr+\sigma(X_r) \circ dB_r^{\mathbb P} \right)  du \\
    & \quad + \sigma''(X_s) \int_s^t \int_s^u \int_s^r   \left(  \tilde{\mu}(X_v)dv+\sigma(X_v) \circ dB_v^{\mathbb P} \right) \left(  \tilde{\mu}(X_r)dr+\sigma(X_r) \circ dB_r^{\mathbb P} \right)  \circ dB_u^{\mathbb P}\\
    &\quad+\varepsilon^{\tilde{\mu}} \int_s^t du+\varepsilon^{\sigma}\int_s^t \circ dB^{\mathbb P}_u\\
    \end{align*}
For the double integral in the above expansion, apply first-order Taylor expansion for $\tilde{\mu}(X_r)$ and $\sigma(X_r)$ at $X_s$ and plug in the dynamic for $dX_t$ again, we further get 
    \begin{align*}
X_t    & = X_s + \tilde{\mu}(X_s) \int_s^t du + \sigma(X_s) \int_s^t \circ dB_u^{\mathbb P}\\
& + \tilde{\mu}'(X_s) \tilde{\mu}(X_s) \int_s^t \int_s^u drdu +  \tilde{\mu}'(X_s) \sigma(X_s) \int_s^t \int_s^u  \circ dB_r^{\mathbb P} du\\
    & + \sigma'(X_s) \tilde{\mu}(X_s) \int_s^t \int_s^u dr \circ dB_u^{\mathbb P} +  \sigma'(X_s) \sigma(X_s) \int_s^t \int_s^u \circ dB_r^{\mathbb P} \circ dB_u^{\mathbb P}\\
    & + (\tilde{\mu}'(X_s))^2 \int_s^t \int_s^u \int_s^r    \circ dX_v  drdu + \tilde{\mu}'(X_s) \sigma'(X_s) \int_s^t \int_s^u \int_s^r  \circ  dX_v  \circ dB_r^{\mathbb P} du\\
    & + \sigma'(X_s) \tilde{\mu}'(X_s) \int_s^t \int_s^u \int_s^r   \circ dX_v  dr \circ dB_u^{\mathbb P} \\
    & + (\sigma'(X_s))^2 \int_s^t \int_s^u \int_s^r  \circ  dX_v  \circ dB_r^{\mathbb P} \circ dB_u^{\mathbb P}\\
    &+ \tilde{\mu}''(X_s) \int_s^t \int_s^u \int_s^r   \left(  \tilde{\mu}(X_v)dv+\sigma(X_v) \circ dB_v^{\mathbb P} \right)  \left(  \tilde{\mu}(X_r)dr+\sigma(X_r) \circ dB_r^{\mathbb P} \right)  du \\
    & + \sigma''(X_s) \int_s^t \int_s^u \int_s^r   \left(  \tilde{\mu}(X_v)dv+\sigma(X_v) \circ dB_v^{\mathbb P} \right) \left(  \tilde{\mu}(X_r)dr+\sigma(X_r) \circ dB_r^{\mathbb P} \right)  \circ dB_u^{\mathbb P}\\
    &\quad+\varepsilon^{\tilde{\mu}} \int_s^t du+\varepsilon^{\sigma}\int_s^t \circ dB^{\mathbb P}_u\\
    & = X_s + \tilde{\mu}(X_s) \int_s^t du + \sigma(X_s) \int_s^t \circ dB_u^{\mathbb P} + \tilde{\mu}'(X_s) \tilde{\mu}(X_s) \int_s^t \int_s^u drdu   \\
    &+  \tilde{\mu}'(X_s) \sigma(X_s) \int_s^t \int_s^u  \circ dB_r^{\mathbb P} du + \sigma'(X_s) \tilde{\mu}(X_s) \int_s^t \int_s^u dr \circ dB_u^{\mathbb P} \\
    &+  \sigma'(X_s) \sigma(X_s) \int_s^t \int_s^u \circ dB_r^{\mathbb P} \circ dB_u^{\mathbb P}\\
    &+ (\tilde{\mu}'(X_s))^2 \int_s^t \int_s^u \int_s^r    \left(  \tilde{\mu}(X_v)dv+\sigma(X_v) \circ dB_v^{\mathbb P} \right)   drdu \\
    &+ \tilde{\mu}'(X_s) \sigma'(X_s) \int_s^t \int_s^u \int_s^r   \left(  \tilde{\mu}(X_v)dv+\sigma(X_v) \circ dB_v^{\mathbb P} \right)   \circ dB_r^{\mathbb P} du\\
    &+ \sigma'(X_s) \tilde{\mu}'(X_s) \int_s^t \int_s^u \int_s^r    \left(  \tilde{\mu}(X_v)dv+\sigma(X_v) \circ dB_v^{\mathbb P} \right)  dr \circ dB_u^{\mathbb P} \\
    &+ (\sigma'(X_s))^2 \int_s^t \int_s^u \int_s^r   \left(  \tilde{\mu}(X_v)dv+\sigma(X_v) \circ dB_v^{\mathbb P} \right)  \circ dB_r^{\mathbb P} \circ dB_u^{\mathbb P}\\
    &+ \tilde{\mu}''(X_s) \int_s^t \int_s^u \int_s^r   \left(  \tilde{\mu}(X_v)dv+\sigma(X_v) \circ dB_v^{\mathbb P} \right)  \left(  \tilde{\mu}(X_r)dr+\sigma(X_r) \circ dB_r^{\mathbb P} \right)  du \\
    &+ \sigma''(X_s) \int_s^t \int_s^u \int_s^r   \left(  \tilde{\mu}(X_v)dv+\sigma(X_v) \circ dB_v^{\mathbb P} \right) \left(  \tilde{\mu}(X_r)dr+\sigma(X_r) \circ dB_r^{\mathbb P} \right)  \circ dB_u^{\mathbb P}\\
    &+\varepsilon^{\tilde{\mu}} \int_s^t du+\varepsilon^{\sigma}\int_s^t \circ dB^{\mathbb P}_u\\
\end{align*}

Finally, replacing \(\tilde{\mu}\) and \(\sigma\) with their Taylor expansion around \(X_s\) and combining the like terms up to triple integrals, gives the following expansion 
\begin{align}
X_t& = X_s + \tilde{\mu}(X_s) \int_s^t du + \sigma(X_s) \int_s^t \circ dB_u^{\mathbb P} + \tilde{\mu}'(X_s) \tilde{\mu}(X_s) \int_s^t \int_s^u drdu \nonumber\\
    & \quad +  \tilde{\mu}'(X_s) \sigma(X_s) \int_s^t \int_s^u  \circ dB_r^{\mathbb P} du  + \sigma'(X_s) \tilde{\mu}(X_s) \int_s^t \int_s^u dr \circ dB_u^{\mathbb P} \nonumber\\
    & \quad +  \sigma'(X_s) \sigma(X_s) \int_s^t \int_s^u \circ dB_r^{\mathbb P} \circ dB_u^{\mathbb P} \nonumber\\
    &  \quad + \left((\tilde{\mu}'(X_s))^2 \tilde{\mu}(X_s) + \tilde{\mu}''(X_s) (\tilde{\mu}(X_s))^2\right) \int_s^t \int_s^u \int_s^r dvdrdu\nonumber\\
    &  \quad + \left((\tilde{\mu}'(X_s))^2 \sigma(X_s) + \tilde{\mu}''(X_s) \tilde{\mu}(X_s) \sigma(X_s)\right) \int_s^t \int_s^u \int_s^r \circ dB_v^{\mathbb P} drdu\nonumber\\
    &  \quad + \left(\tilde{\mu}'(X_s) \sigma'(X_s)\tilde{\mu}(X_s) + \tilde{\mu}''(X_s) \tilde{\mu}(X_s) \sigma(X_s)\right) \int_s^t \int_s^u \int_s^r dv \circ dB_r^{\mathbb P} du\nonumber\\
    &  \quad + \left(\tilde{\mu}'(X_s) \sigma'(X_s)\tilde{\mu}(X_s) + \sigma''(X_s) (\tilde{\mu}(X_s))^2 \right) \int_s^t \int_s^u \int_s^r dv dr \circ dB_u^{\mathbb P}\nonumber \\
    &  \quad + \left(\tilde{\mu}'(X_s) \sigma'(X_s) \sigma(X_s) + \tilde{\mu}''(X_s) (\sigma(X_s))^2 \right) \int_s^t \int_s^u \int_s^r \circ dB_v^{\mathbb P} \circ dB_r^{\mathbb P} du \nonumber\\
    &  \quad + \left(\tilde{\mu}'(X_s) \sigma'(X_s) \sigma(X_s) + \sigma''(X_s) \sigma(X_s) \tilde{\mu}(X_s) \right) \int_s^t \int_s^u \int_s^r \circ dB_v^{\mathbb P} dr \circ dB_u^{\mathbb P} \nonumber\\
    &  \quad + \left( (\sigma'(X_s))^2 \tilde{\mu}(X_s) + \sigma''(X_s) \sigma(X_s) \tilde{\mu}(X_s) \right) \int_s^t \int_s^u \int_s^r  dv \circ dB_r^{\mathbb P} \circ dB_u^{\mathbb P}\nonumber \\
    &  \quad + \left( (\sigma'(X_s))^2 \sigma(X_s) + \sigma''(X_s) (\sigma(X_s))^2  \right) \int_s^t \int_s^u \int_s^r \circ dB_v^{\mathbb P} \circ dB_r^{\mathbb P} \circ dB_u^{\mathbb P}\nonumber\\
&\quad+\underbrace{\varepsilon^{\tilde{\mu}} \int_s^t du+\varepsilon^{\sigma}\int_s^t \circ dB^{\mathbb P}_u+\tilde \varepsilon^{\tilde{\mu},\sigma}}_{:=\varepsilon^{\tilde{\mu},\sigma}},\label{proof step 3 expansion}
\end{align}
where we denote $\tilde \varepsilon^{\tilde{\mu},\sigma}$ as the collection of all the higher order error terms from double and triple integrals.
We denote $\varepsilon^{\tilde{\mu},\sigma}:=\varepsilon^{\tilde{\mu}} \int_0^t ds+\varepsilon^{\sigma}\int_0^t \circ dB^{\mathbb P}_s+\tilde \varepsilon^{\tilde{\mu},\sigma} $ as the overall error term from the above expansion. This completes the proof.  \qed 

\section{Proof of Step 4}\label{section proof step 4}
Applying Girsanov theorem, plugging the dynamic \(dB_t^{\mathbb P} = dB_t^{\mathbb Q} - \frac{\mu(X_t)}{\sigma(X_t)} dt\) into the expanison \eqref{proof step 3 expansion} from Step 3, we have
\begin{align}\label{X_t expansion}
    X_t & = X_s + \tilde{\mu}(X_s) \int_s^t du + \sigma(X_s) \int_s^t  \circ \left( dB_u^{\mathbb Q} - \frac{\mu(X_u)}{\sigma(X_u)}du \right)   \nonumber \\
    & \quad + \tilde{\mu}'(X_s) \tilde{\mu}(X_s) \int_s^t \int_s^u drdu +  \tilde{\mu}'(X_s) \sigma(X_s) \int_s^t \int_s^u \circ   \left( dB_r^{\mathbb Q} - \frac{\mu(X_r)}{\sigma(X_r)}dr \right)  du \nonumber\\
    & \quad + \sigma'(X_s) \tilde{\mu}(X_s) \int_s^t \int_s^u dr  \circ \left( dB_u^{\mathbb Q} - \frac{\mu(X_u)}{\sigma(X_u)}du \right) \nonumber\\
    & \quad  +  \sigma'(X_s) \sigma(X_s) \int_s^t \int_s^u  \circ \left( dB_r^{\mathbb Q} - \frac{\mu(X_r)}{\sigma(X_r)}dr \right)  \circ  \left( dB_u^{\mathbb Q} - \frac{\mu(X_u)}{\sigma(X_u)}du \right) \nonumber\\
    &  \quad + \left((\tilde{\mu}'(X_s))^2 \tilde{\mu}(X_s) + \tilde{\mu}''(X_s) (\tilde{\mu}(X_s))^2\right) \int_s^t \int_s^u \int_s^r dvdrdu\nonumber\\
    &  \quad + \left((\tilde{\mu}'(X_s))^2 \sigma(X_s) + \tilde{\mu}''(X_s) \tilde{\mu}(X_s) \sigma(X_s)\right) \nonumber\\
    & \quad \quad \times \int_s^t \int_s^u \int_s^r  \circ \left( dB_v^{\mathbb Q} - \frac{\mu(X_v)}{\sigma(X_v)}dv \right)  drdu \nonumber\\
    &  \quad + \left(\tilde{\mu}'(X_s) \sigma'(X_s)\tilde{\mu}(X_s) + \tilde{\mu}''(X_s) \tilde{\mu}(X_s) \sigma(X_s)\right) \nonumber\\
    & \quad \quad \times \int_s^t \int_s^u \int_s^r dv  \circ \left( dB_r^{\mathbb Q} - \frac{\mu(X_r)}{\sigma(X_r)}dr \right)  du \nonumber\\
    &  \quad + \left(\tilde{\mu}'(X_s) \sigma'(X_s)\tilde{\mu}(X_s) + \sigma''(X_s) (\tilde{\mu}(X_s))^2 \right)   \nonumber\\
    & \quad \quad \int_s^t \int_s^u \int_s^r dv dr  \circ \left( dB_u^{\mathbb Q} - \frac{\mu(X_u)}{\sigma(X_u)}du \right) \nonumber\\
    &  \quad + \left(\tilde{\mu}'(X_s) \sigma'(X_s) \sigma(X_s) + \tilde{\mu}''(X_s) (\sigma(X_s))^2 \right)  \nonumber\\
    & \quad \quad \times \int_s^t \int_s^u \int_s^r   \circ \left( dB_v^{\mathbb Q} - \frac{\mu(X_v)}{\sigma(X_v)}dv \right)    \circ \left( dB_r^{\mathbb Q} - \frac{\mu(X_r)}{\sigma(X_r)}dr \right)  du \nonumber\\
    &  \quad + \left(\tilde{\mu}'(X_s) \sigma'(X_s) \sigma(X_s) + \sigma''(X_s) \sigma(X_s) \tilde{\mu}(X_s) \right)  \nonumber\\
    & \quad \quad \times \int_s^t \int_s^u \int_s^r   \circ \left( dB_v^{\mathbb Q} - \frac{\mu(X_v)}{\sigma(X_v)}dv \right)  dr \circ  \left( dB_u^{\mathbb Q} - \frac{\mu(X_u)}{\sigma(X_u)}du \right) \nonumber\\
    &  \quad + \left( (\sigma'(X_s))^2 \tilde{\mu}(X_s) + \sigma''(X_s) \sigma(X_s) \tilde{\mu}(X_s) \right)   \nonumber\\
    &\quad\quad   \times \int_s^t \int_s^u \int_s^r  dv  \circ \left( dB_r^{\mathbb Q} - \frac{\mu(X_r)}{\sigma(X_r)}dr \right)   \circ \left( dB_u^{\mathbb Q} - \frac{\mu(X_u)}{\sigma(X_u)}du \right)  \nonumber\\
    &  \quad + \left( (\sigma'(X_s))^2 \sigma(X_s) + \sigma''(X_s) (\sigma(X_s))^2  \right)\nonumber  \\
    &\quad   \times \int_s^t \int_s^u \int_s^r   \circ \left( dB_v^{\mathbb Q} - \frac{\mu(X_v)}{\sigma(X_v)}dv \right)  \circ \left( dB_r^{\mathbb Q} - \frac{\mu(X_r)}{\sigma(X_r)}dr \right)  \circ \left( dB_u^{\mathbb Q} - \frac{\mu(X_u)}{\sigma(X_u)}du \right) \nonumber \\
&\quad+\underbrace{\varepsilon^{\tilde{\mu}} \int_s^t du+\varepsilon^{\sigma}\int_s^t \circ dB^{\mathbb P}_u+\tilde \varepsilon^{\tilde{\mu},\sigma}}_{:=\varepsilon^{\tilde{\mu},\sigma}},
\end{align}
Note that for the Black-Scholes model, the linear constant drift and diffusion coefficient assumption ensures that $\frac{\mu(X_u)}{\sigma(X_u)} = \frac{\mu(X_s)}{\sigma(X_s)}$ for all $u\in [s,t)$. In the spirit of Black-Sholes, we propose further approximation of \(\frac{\mu(X_u)}{\sigma(X_u)}\) by \(\frac{\mu(X_s)}{\sigma(X_s)} \) for the interval $[s,t)$, and we denote $\varepsilon^{\mu/\sigma}$ as the error term from this approximation.  This ensures some cancellations. Higher approximations of \(\frac{\mu(X_u)}{\sigma(X_u)}\) is also possible here; we leave this for future studies. To derive the third-order approximation, we only collect the integrals with order $|t-s|^{3/2}$ terms in the above expansion, leaving all the higher terms in the error term. With a slight abuse of notation, we denote by $\varepsilon^{\tilde{\mu},\sigma}$  the error term associated with the Taylor expansion for functions $\tilde{\mu}$ and $\sigma$, and by $\varepsilon^{\mu/\sigma}$ the error term associated with the approximation for $\frac{\mu}{\sigma}$.  The final result after truncating to a third-order iterated integral and replacing \(\frac{\mu(X_u)}{\sigma(X_u)}\) by the starting point \(\frac{\mu(X_s)}{\sigma(X_s)}\) will be as follows.

\begin{align*}
    X_t & = X_s - \frac{1}{2} \sigma(X_s) \sigma'(X_s)  \int_s^t du + \sigma(X_s) \int_s^t  \circ dB_u^{\mathbb Q}   \\
    & \quad + \left( \mu'(X_s) \sigma(X_s) - \mu(X_s) \sigma'(X_s) - \frac{1}{2}\sigma(X_s) \sigma'(X_s)^2 - \frac{1}{2}\sigma(X_s)^2 \sigma''(X_s) \right)   \\
    & \qquad \qquad \qquad \qquad \qquad \qquad \qquad \qquad \qquad \qquad \qquad \qquad \qquad  \times \int_s^t \int_s^u   \circ dB_r^{\mathbb Q}   du\\
    & \quad - \frac{1}{2} \sigma(X_s) \sigma'(X_s)^2 \int_0^t \int_0^s   du \circ dB_s^{\mathbb Q}   +  \sigma'(X_s) \sigma(X_s) \int_s^t \int_s^u  \circ dB_r^{\mathbb Q}  
 \circ dB_u^{\mathbb Q} \\
 &+\left( (\sigma'(X_s))^2 \sigma(X_s) + \sigma''(X_s) (\sigma(X_s))^2  \right) \int_s^t\int_s^u\int_s^r  \circ dB_v^{\mathbb Q}\circ d B_r^{\mathbb Q}\circ dB_u^{\mathbb Q} \\
 & \quad +\varepsilon^{\tilde{\mu},\sigma}+\varepsilon^{\mu/\sigma}. 
\end{align*}
The overall error terms $\varepsilon^{\tilde{\mu},\sigma}+\varepsilon^{\mu/\sigma}$ has the same order as $|t-s|^{2-\varepsilon}$, which is small if $|t-s|$ is small.  The proof is thus finished. \qed 

\section{Proof of Step 5}\label{section proof step 5}
Comparing the coefficients of  \eqref{SDE 2 order expansion} from Step 2 and \eqref{expansion step 4} from Step 4, for the expansion of $X_t$ at the initial point $X_s$ on interval $[s,t)$, we have the following equalities,
\begin{align}
    \begin{cases}
   &  - \frac{1}{2} \sigma(X_s) \sigma'(X_s)  = \psi_1^{s,t},\quad 
    \sigma(X_s)  = \psi_2^{s,t}, \quad - \frac{1}{2} \sigma(X_s) \sigma'(X_s)^2  = \psi_{12}^{s,t}\\
      &\mu'(X_s) \sigma(X_s) - \mu(X_s) \sigma'(X_s) - \frac{1}{2}\sigma(X_s) \sigma'(X_s)^2 - \frac{1}{2}\sigma(X_s)^2 \sigma''(X_s)  = \psi_{21}^{s,t},\\
       &\sigma(X_s) \sigma'(X_s)  = \psi_{22}^{s,t}.\\
    \end{cases}
\end{align}
As we already get the values/functions of $\sigma$ from step 1, we only focus on the function $\mu$ here. 
The equation associated with $\psi_{21}^{s,t}$ implies that 
\begin{equation}
    \mu'(X_s) \sigma(X_s) - \mu(X_s) \sigma'(X_s) - \frac{1}{2}\sigma(X_s) \sigma'(X_s)^2 - \frac{1}{2}\sigma(X_s)^2 \sigma''(X_s) = \psi_{21}^{s,t}. \label{B}
\end{equation}
Following the above observation, the parameter $\psi^{t_i,t_{i+1}}_{21}$ can be found corresponding to the initial value \(X_{t_i}, \text{ for }i =0, 1,2, \cdots, n\). With this, the equation \eqref{B} holds true for each \(t_i, i = 0,1,2, \dots, n\). For example, \eqref{B} is equivalently written as
\begin{align}\mu'(X_{t_i}) \sigma(X_{t_i}) - \mu(X_{t_i}) \sigma'(X_{t_i}) - \frac{1}{2}\sigma(X_{t_i}) \sigma'(X_{t_i})^2 - & \frac{1}{2}\sigma(X_{t_i})^2 \sigma''(X_{t_i}) = \psi_{21}^{t_i,t_{i+1}},\\
& \quad \forall \; i=0,1,2,\dots, n. \nonumber \label{C}
\end{align}
Denote \(a(X_r) = \frac{\sigma'(X_r)}{\sigma(X_r)}\) and \(b(X_r) = \frac{2\psi^{t_i,t_{i+1}}_{21}+\sigma(X_{r}) \sigma'(X_{r})^2 +\sigma(X_{r})^2 \sigma''(X_{r})}{2\sigma(X_r)}\), for $r\in [t_i,t_{i+1})$.
Since we know \(\mu(X_0)\), using Eular scheme for each interval $[t_i,t_{i+1})$, we can solve $\mu$ as follows
\begin{equation} \label{Euler_for_mu}
    \mu(X_{t_i}) = \mu(X_{t_{i-1}}) + \left[a(X_{t_{i-1}}) \mu(X_{t_{i-1}}) + b(X_{t_{i-1}} ) \right](X_{t_{i}} - X_{t_{i-1}}), \quad \forall \; i=1,2,\dots, n.
\end{equation}
Thus, we use estimate the values of $\mu(X_{t_i})$ for $i=1,\cdots,N.$ The proof is thus finished. \qed  
\section{Proof of Proposition \ref{prop: step 3}}
The expansion of $X_t$ in {\bf Step 3} is a Taylor expansion for the following SDE,
\begin{align}
    X_t= X_s+\int_s^t V_0(X_r)dr +\int_s^t V_1(X_r)\circ dB_r^{\mathbb P},\label{Taylor SDE}
\end{align}
where we denote the vector field $V_0(X_r)=\tilde{\mu}(X_r)$, and $V_1(X_r)=\sigma(X_r)$ to simplify the notation. According to \cite[Theorem 1.6]{azencott2006formule} (see also \cite[Lemma 3.9]{feng2020taylor} for a more general set-up of this expansion), for the one dimensional $X_t$, we have 
\begin{align}\label{tail expansion}
X_t=X_s+\sum_{k=1}^Ng_k(t,s)+|t-s|^{\frac{N+1}{2}}R^{N+1}(t,s),
\end{align}
where we set $\mathbf B^{0}_t=t$, and $\mathbf B^{1}_t=B^{\mathbb P}_t$, and introduce the following notation, 
\begin{align}
g_k(t,s)=\sum_{|I|=k}P_I(X_s) \int_{s<t_1<t_2\cdots<t_k<t} \circ \; d\mathbf B^{i_1}_{t_1}\cdots \circ d\mathbf B^{i_k}_{t_k},
\end{align}
for a word $I=(i_1,\cdots, i_k)\in \{0,1\}^k$. Here we denote $|I|$ the size of $I$ which equals $k$, and
 and we denote $P_{I}(X_s)=(V_{i_1}\cdots V_{i_k})(X_s)$. In particular,  $V_{i_1}\cdots V_{i_k}:= (V_{i_1}\cdots(V_{i_{k-2}}(V_{i_{k-1}} V_{i_k}))\cdots)$ represents the iterative operation of the vector fields in order. Following from \cite[Proposition 4.3]{azencott2006formule} (see also \cite[Lemma 3.9]{feng2020taylor} with $H=1/2$), there exists a random time $\zeta$, some constants $\alpha, c>0$, for every $\xi\ge 1$, such that 
\begin{align*}
    \label{remainder est}
\mathbb P\left(\sup_{t\in [t_k,t_{k+1}]} \||t-t_k|^{\frac{N+1}{2}}R_{m+1}(t,t_k)\|\ge \xi |t_{k+1}-t_k|^{\frac{N+1}{2}}; \tau< \zeta \right)\\ \le \exp\left(-\frac{c\xi^{\alpha} }{|t_{k+1}-t_k|}\right).
\end{align*}  The random time $\zeta$ is the exit time of $X_t$ in a compact domain $K$. We pick the compact domain $K$ large enough, such that $X_t$ stays inside $K$ for $t\in [t_{k},t_{k+1}]$. Since we take the Taylor expansion at level 3,
by selecting $N=3$, we have 
\begin{align*}
\mathbb P\left(\sup_{t\in [t_k,t_{k+1}]} \|R_{m+1}(t,t_k)\|\ge \xi |t_{k+1}-t_k|^{2}; \tau< \zeta \right) \le \exp\left(-\frac{c\xi^{\alpha} }{|t_{k+1}-t_k|}\right).
\end{align*}
Applying the above estimate,
denote $Z=\| X_t-L^3(t)\|^2$, where $L^3(t)=X_{t_k}+\sum_{k=1}^3g_k(t,t_k)$ as defined in \eqref{tail expansion}, we have 
\begin{align*}
\mathbb E [\| X_t-L^3(t)\|^2]=    \mathbb E[Z] =\int_0^{\infty} \mathbb P(Z>z) dz\le C |t_{k+1}-t_k|^2,
\end{align*}
which completes the proof.
\section{Proof of Corollary \ref{corollary}}

First, recall that the total approximation error of $X_t$ accumulated from \textbf{Step 1-3} is coming from the error terms $\varepsilon^{\tilde{\mu}}$, $\varepsilon^{\sigma}$, $\varepsilon^{\tilde{\mu},\sigma}$ and  $\varepsilon^{\mu/\sigma}$. For the Black-Scholes model since $\sigma(X_t) = \sigma X_t$ and $\tilde{\mu}(X_t) = \mu X_t - \sigma^2 X_t$, the Equations \eqref{expansion tilde u} and \eqref{expansion sigma} simplifies to
\begin{align*}
  \tilde{\mu}(X_u)& = \tilde{\mu}(X_s) + (\mu-\sigma^2) \int_s^u \circ dX_r  + 0  \\
 {\sigma}(X_u)& = \sigma(X_s) + \sigma^2 \int_s^u \circ dX_r  + 0. 
\end{align*} 

This implies $\varepsilon^{\tilde{\mu}} =\varepsilon^{\sigma} = 0$ in \textbf{Step 2}. Furthermore, $\frac{\mu(X_s)}{\sigma(X_s)}=\frac{\mu}{\sigma}$, we get $\varepsilon^{\mu/\sigma}=0$ in {\bf Step 4}.
Thus, we only need to estimate the error $X_t$ in {\bf Step 3}, the proof follows directly from Proposition  \ref{prop: step 3}.

\end{document}